%% file: AandA.tex
\documentclass[longauth]{aa}

\usepackage{graphicx}
\usepackage{natbib}
\usepackage{scalerel}
\usepackage{graphicx}
\usepackage{subcaption}
\usepackage{adjustbox}

\usepackage{txfonts}
\usepackage[pdfencoding=auto,psdextra]{hyperref}
\hypersetup{
    colorlinks=true,
    linkcolor=blue,
    filecolor=magenta,      
    urlcolor=blue,
    citecolor=blue
}
\makeatletter
\renewcommand*\aa@pageof{, page \thepage{} of \pageref*{LastPage}}
\makeatother

\usepackage[utf8]{inputenc}

\usepackage[switch, modulo]{lineno}

\usepackage{euclid}

\renewcommand{\mp}{M^\mathrm{2D}}
\newcommand{\mrat}{\mp/M}

\titlerunning{\Euclid\/ preparation. Projection effects and correlations}
\authorrunning{Euclid Collaboration:  Ragagnin et al.}
\begin{document}
%
%
\title{\Euclid preparation}
\subtitle{The impact of line-of-sight projections on the covariance between galaxy cluster multi-wavelength observable properties -- insights from hydrodynamic simulations}    

     \include{authorlist.tex}

 \date{\bf June 16, 2023}

%
%

  \abstract
  {
Cluster cosmology can benefit from combining multi-wavelength studies, which themselves can benefit from a characterisation of the correlation coefficients between different mass-observable relations.
}   
   {In this work, we aim to provide information on the scatter, the skewness, and the covariance of various mass-observable relations in galaxy clusters in cosmological hydrodynamic simulations. This information will help future analyses to better tackle accretion histories and projection effects and model mass observable relations for cosmology studies.} 
   {We identify galaxy clusters in Magneticum Box2b simulations with mass $M_{\rm 200c}>10^{14} {\rm M}_\odot$ at redshift $z=0.24$ and $z=0.90$. Our analysis includes \Euclid-derived properties such as richness, stellar mass, lensing mass, and concentration. Additionally, we investigate complementary multi-wavelength data, including X-ray luminosity, integrated  Compton-$y$ parameter, gas mass, and temperature. The impact of projection effects on mass-observable residuals and correlations is then examined.}
   {We find that at intermediate redshift ($z=0.24$) projection effects impact lensing concentration, richness, and gas mass the most in terms of scatter and skewness of log-residuals of scaling relations.  The contribution of projection effects can be significant enough to boost a spurious hot- vs. cold-baryons correlation and consequently hide underlying correlations due to halo accretion histories.
   At high redshift ($z=0.9$), the richness has a much lower scatter (of log-residuals), and the quantity that is most impacted by projection effects is the lensing mass.
   Lensing concentration reconstruction, in particular, is affected by deviations of the reduced-shear profile shape from the one derived by an NFW profile rather than interlopers in the line of sight.
  
   }
   {}
   

   \maketitle
%

%
%
%
   \section{Introduction}

Galaxy clusters are the largest gravitationally bound, collapsed, and virialised structures in our Universe and represent unique laboratories for testing cosmological models,  galaxy evolution,  and thermodynamics of the intracluster medium~\citep[ICM, see][for a review on galaxy clusters]{2012ARA&A..50..353Kravtsov}. 
Regarding galaxy cluster cosmology studies~\citep[see, e.g.,][]{2010ApJ...708..645Rozo,2019ApJ...878...55Bocquet}, an accurate characterisation of the selection function and of the mass-observable scaling relations represent the dominant systematic uncertainties \citep[see the review on the cluster mass scale in][]{pratt19}.

Cluster masses cannot be observed directly,  and their reconstruction requires both a number of assumptions and high-quality data \citep[see, e.g.,][]{meneghetti10}. 
This means that precise estimates are rare~\citep{2010ApJ...721..875Okabe,2012MNRAS.427.1298Hoekstra,2015MNRAS.449.2219Melchior,2016MNRAS.459.3251VanUitert,2019MNRAS.485...69Stern,2023PhRvD.108l3521Sugiyama,2024PhRvD.110h3510Bocquet}.
Once a set of highly accurate mass determinations are available, together with other mass proxies recovered from multi-band observations, well-calibrated mass-observable relations (for instance, the mass-richness relation or the mass-temperature relation) can be established and used to estimate galaxy cluster masses for larger samples with known observable properties. For this purpose, it is important to calibrate accurately the mass-observable relations ~\citep{2013SSRv..177..247Giodini,2011ARA&A..49..409Allen,2021MNRAS.505.3923Schrabback}, including proper modelling of their associated scatter~\citep{2005PhRvD..72d3006Lima,2019ApJ...878...55Bocquet}.

This process is complicated by the fact that studies at different wavelengths are biased by various astrophysical processes and projection effects to various degrees.
For instance,  X-ray surveys tend
to favour the selection of clusters with centrally peaked gas distributions~\citep{2007MNRAS.382.1289Pacaud,2010A&A...513A..37Hudson,2011A&A...536A..37Andreon,2016A&A...593A...2Andreon,2018A&A...619A.162Xu} and suffer from AGN contamination~\citep[see, e.g.,][]{Bhargava2023AGNContamiantion}, while projection effects are known to strongly impact weak lensing mass reconstructions~\citep{2014ApJ...797...34Meneghetti,Giocoli-EP30} and richness evaluations \citep[e.g.,][]{Castignani2016richness}.
This is particularly relevant for cluster cosmology studies, 
where the aim is to reduce uncertainty by combining constraints on different mass-observable relations.
For \Euclid~\citep{EuclidSkyOverview}, this will include quantities such as richness, stellar mass, and properties of stacked weak lensing signals~\citep{2020A&A...638A.141Pires}, of the cluster samples detected using tools such as AMICO~\citep{amicoA,amicoB} or PZWav~\citep{Adam-EP3}, possibly together with other multi-wavelength observations~\cite{2011ARA&A..49..409Allen}.
These properties are known to be biased by projection effects~\citep{2014ApJ...797...34Meneghetti}, accretion histories~\citep{2022A&A...666A..22Ragagnin},  mis-centring~\citep{2022MNRAS.509.1127Sommer,Sommer+23subm}, and the fit procedure~\citep{2016JCAP...01..042Sereno}. 
  Projection effects, especially, are expected to generate some covariance between the richness and weak lensing signal, the uncertainty of which may significantly affect the performance of the mission for cluster population analyses.
  This effect is one the major sources of systematics for current optical cluster surveys~\citep{2019MNRAS.488.4779Costanzi,2020PhRvD.102b3509Abbott}, and thus is expected to play an even more critical role for the \Euclid cluster sample.

  Numerical simulations are thus a critical tool to mitigate the impact of the aforementioned biases on cosmological cluster studies. Indeed, the power of observations to constrain them is limited, thus increasing the final uncertainty budget. However, scatter and covariance parameters are also prime sources of uncertainty when aiming to combine information originating from different wavelengths. 
For instance, various observational works hint towards different directions for the hot- vs. cold-baryon covariance~\citep{2019NatCo..10.2504Farahi,2022MNRAS.511.2968Puddu,2022A&A...666A..22Ragagnin}, as different formation times are related with satellite accretion history~\citep{Giocoli2008mass-loss}.

In this context, numerical simulations have proven to be a very powerful tool for helping observational studies in modelling mass-observable relations, which are strongly affected by galaxy cluster accretion histories~\citep{2012MNRAS.427.1322Ludlow,2019MNRAS.490.5693Bose,2020MNRAS.491.4462Davies,2020MNRAS.495..686Anbajagane,2022A&A...666A..22Ragagnin}, projection effects \citep{2014ApJ...797...34Meneghetti}, and deviations~\citep[see, e.g.,][]{Ragagnin2021MC} from the Navarro--Frenk--White density profile~\citep[NFW,][]{1997ApJ...490..493Navarro}, which is often adopted in weak lensing studies.
Thus, simulations can suggest the most suitable functional forms of scaling relations for cosmological studies~\citep[as in the works of][]{2019MNRAS.488.4779Costanzi, 2016MNRAS.456.2361Bocquet, 2019ApJ...878...55Bocquet, Ghirardini2024arXivCosmo}.
They can provide informative priors on their correlation coefficients,  which are among the most difficult parameters to be constrained directly from observed quantities, guiding the forward modelling setup of cluster cosmology studies.

There are various works in the literature that study how simulations can help disentangle physical models~\citep[see e.g.,][]{Cui22GizmoRun,Angelinelli23Redshift} cosmological models~\citep[see e.g.,][]{Bocquet2020MiraTitan,Angulo2021Bacco,VillaescusaNavarro22Cosmology1Galaxy} or dark matter types~\citep[see e.g.,][]{Ragagnin24SIDM,Fischer24Mergers,ContrerasSantos24NoExotic}, and study observable cross-correlations~\citep[see e.g.,][]{2010ApJ...715.1508Stanek,2020MNRAS.495..686Anbajagane}.

In this work, besides focusing on correlations between observable properties of interests for multi-wave length studies, we also study the impact of projection effects. The impact of uncorrelated large-scale structure on the covariance between observable properties can be modelled analytically~\citep{2003MNRAS.339.1155Hoekstra,McClintock2019,2019MNRAS.488.4779Costanzi}, but the covariance of different observable properties below a few tens of Mpc requires dedicated simulations.
At these scales, numerical hydrodynamic simulations, with their self-consistent depiction of the ICM, emerge as an ideal tool for exploring multi-wavelength observable properties since they incorporate the effects of large-scale structures within which clusters are situated. Indeed, baryon feedback influences the ICM not only within cluster virial radii but also beyond~\citep[see, e.g.,][]{Angelinelli2022Mapping,Angelinelli2023Redshift}.

While it is true that cosmological simulations are influenced by the underlying sub-grid prescriptions, and while it is true that these simulations may diverge on small scales, they generally exhibit agreement on quantities integrated up to the sizes of galaxy groups and clusters~\citep[see, e.g.,][]{2020MNRAS.495..686Anbajagane}.
At the same time, different cosmological parameters can affect galaxy cluster properties, such as their masses~\citep{Ragagnin2021MC}, satellite abundance~\citep{VanDenBosch2005Satellites}, and mass-observable relations \citep{2020MNRAS.494.3728Singh}. On the other hand, the qualitative significance of covariances and projection effects on observable properties is not expected to significantly hinge on cosmological parameters \citep{2019ApJ...878...55Bocquet}, and possible deviations from this expectation could be estimated using emulators~\citep[see, e.g.,][]{Bocquet2020MiraTitan,Ragagnin2021MC,Angulo2021Bacco,Ragagnin2023HOD}.



We will study the impact of projection effects using hydrodynamic simulations in order to gain insight into which fraction of the scatter and skewness of scaling relations originates from projection effects (i.e., alignment with filaments and objects) or different accretion histories.

In Sect. \ref{sec:setup}, we present how we set up our \Euclid-like observable properties and the others coming from the other wavelengths. 
In Sect.~\ref{sec:proj}, we study how projection effects impact the scatter and skewness of log-residuals of scaling relations and discuss the impact of projection effects on observable covariance.
In Sect.~\ref{sec:MC}, we focus on the mass-concentration relation and how it is affected by projection effects and deviations from the functional form of profiles and the radial ranges of the fits.
In Sect.~\ref{sec:cova}, we focus on the covariance between observable properties and study how different accretion histories and projection effects impact them.
Finally, we draw our conclusions in Sect.~\ref{sec:conclu}.

\section{Numerical Setup}
\label{sec:setup}

We will conduct our study by analysing clusters obtained from the Magneticum\footnote{\url{http://www.magneticum.org}} hydrodynamic cosmological simulations \citep{2013MNRAS.428.1395Biffi,2014MNRAS.440.2610Saro,2015MNRAS.448.1504Steinborn,2016MNRAS.463.1797Dolag,2015MNRAS.451.4277Dolag, 2015ApJ...812...29Teklu,2016MNRAS.458.1013Steinborn,2016MNRAS.456.2361Bocquet,Ragagnin2019MC}.
They are based on the $N$-body code \texttt{Gadget3}, which is built upon \texttt{Gadget2} \citep{2005Natur.435..629Springel,2005MNRAS.364.1105Springel,2009MNRAS.398.1150Boylan} with an improved smoothed particle hydrodynamics (SPH) solver from \cite{2016MNRAS.455.2110Beck}.
Magneticum initial conditions are generated using a standard $\Lambda$CDM cosmology with \WMAP 7~\citep{2011ApJS..192...18Komatsu} cosmological parameters. 
The large-scale structure evolution in Magneticum simulations includes a treatment of radiative cooling, heating from a uniform redshift-dependent ultraviolet (UV) background,  star formation, and stellar feedback processes as in \cite{2005MNRAS.361..776Springel}. The stellar feedback is then connected to a detailed chemical evolution and enrichment model as in \cite{2007MNRAS.382.1050Tornatore}, which follows  11 chemical  elements \citep[H, He,
C, N, O, Ne, Mg, Si, S, Ca, and Fe, with cooling tables from][]{Wiersma2009Cooling} which are produced with the  \texttt{CLOUDY} photo-ionisation
code \citep{1998PASP..110..761Ferland}. \cite{2010MNRAS.401.1670Fabjan} and \cite{2014MNRAS.442.2304Hirschmann} described prescriptions for black hole growth and feedback from AGNs.
Haloes that host galaxy clusters and groups are identified using the friends-of-friends halo finder~\citep{1985ApJ...292..371Davis}, and subhaloes together with their associated galaxies are identified with an improved version of \texttt{SUBFIND} \citep{2001MNRAS.328..726Springel}, which takes into account the presence of baryons   \citep{2009MNRAS.399..497Dolag}.

We define $r_{\Delta \rm c}$  as the radius that encloses an average density of  $\Delta_{\rm c}\,\rho_{\rm cr}, $ where $\rho_{\rm cr}$ is the critical density of the Universe at a given redshift,
\begin{equation}
\label{eq:mass}
M_{\Delta \rm c} = \frac{4}{3} \pi r_{\Delta \rm c}^3 \Delta_{\rm c}\, \rho_{\rm cr}.
\end{equation}
Throughout this paper, when we omit $\Delta_{\rm c}$ from masses and radii we imply the usage of $\Delta_{\rm c}=200$ (i.e., $M=M_{\rm 200c}).$ 

To disentangle the scatter of the mass-observable relation from projection effects, we compute quantities within a sphere of radius $r_{\rm 200c}$ and as integrated into a cylinder.
Projected quantities will be denoted with the superscript ${\rm 2D}$ (for instance, the total mass inside the cylinder is denoted as $M^{\rm 2D}$). 
We opted to employ a random projection plane for each cluster.
Additionally, we set an integration depth of $20$ comoving $h^{-1} {\rm Mpc}$, corresponding to approximately $23 \, {\rm Mpc}$ at $z=0.24$ and $15 \, {\rm Mpc}$ at $z=0.9$ (with $h=0.704$).
This cylinder depth is smaller than \Euclid's galaxy cluster photo-$z$ equivalent uncertainty~\citep{Desprez-EP10} and, while we exclude some uncorrelated projection effects, they are known not to play an important role~\citep{Sunayama2020ImpactProj,Wu2022OpticalBias}. Thus, we ensure that we do not overestimate any projection effect that could be mitigated using photo-$z$. Consequently, all projection effects examined in this paper hold relevance for interpreting forthcoming \Euclid-based catalogues.

This study is based on the results from Box2b/hr~\citep{2014MNRAS.442.2304Hirschmann} Magneticum simulation,  which covers a length of $900$ comoving Mpc,  with dark matter particle masses $m_\mathrm{DM } = 9.8\times10^8{\rm M}_\odot,$  gas initial particle masses of $m_\mathrm{gas} = 2\times10^8{\rm M}_\odot,$ and a gravitational softening of both gas and dark matter of  $\epsilon = 3.75$ comoving kpc.
Euclid is expected to detect clusters with masses $M>10^{14}{\rm M}_\odot$ up to a redshift of approximately $z\approx2$~\citep{Sartoris2016,Adam-EP3}, where the bulk of the cluster population which will be used for mass-calibration will lie below redshift $z\approx1.$ 
Furthermore, the number of haloes contained in this Magneticum simulation drops significantly beyond the same redshift value.
Therefore, we decided to extract haloes at two representative redshift slices: at an intermediate redshift of approximately $z\approx0.24$, yielding $4300$ objects, and at a higher redshift of approximately $z\approx0.9$, yielding $1300$ objects. These extractions were performed using the web portal\footnote{\url{https://c2papcosmosim.uc.lrz.de/}} introduced in \cite{2017A&C....20...52Ragagnin}.
We focus most of the analyses on the qualitative effect of projection effects at our intermediate redshift slice because of the larger statistics of clusters to help us determine projection effects. 
We stress that this mass threshold is high enough so that all of our galaxy clusters have at least $10^{4}$ particles and, therefore, can be considered well resolved in terms of density profile fitting~\citep{1997ApJ...490..493Navarro}.

\subsection{Observable properties}
\label{sec:obs}

\begin{table*}
\caption{List of observable properties used in this work and presented in Sect.~\ref{sec:obs}}              
\label{table:obs}      
\centering                                      
\begin{tabular}{l l l}          
\hline\hline                        
Quantity & Notation & Comments \\    
\hline         
Stellar mass & $M_\star$ \\
Temperature & $T$ & weighted by X-ray emissivity\\
    Richness & $n$ & satellites with ${\rm log}_{10}(M_\star/{\rm M}_\odot)>10.65,$\\
    & & and background subtraction as in {\cite{2016A&A...593A...2Andreon}} \\      
    X-ray luminosity & $L_X$ & in $[0.5,2]{\rm keV}$ band \\
    Gas mass & $M_{\rm g}$ & \\ 
    Thermal SZ parameter & $Y$ & see Eq.~\eqref{eq:sz} and Eq.~\eqref{eq:sz-sph}\\
    NFW profile fit parameters & $M_{\rm NFW},$ $c_{\rm NFW}$ & in 3D: $100$ log-spaced bins with $60{\rm ckpc}<r<r_{\rm 200c},$ \\
    & & in 2D: reduced shear fit as in Eq.~\eqref{eq:nfw}, and uncertainty as in Eq.~\eqref{eq:shear-error},\\
    & & on $12$ log-spaced bins with $300{\rm kpc}<R<3000{\rm kpc},$ a cylinder depth of $20{\rm cMpc},$\\
    & & and a source distribution in redshift as in~\cite{Giocoli-EP30}\\
    
\hline                                             
\end{tabular}
\end{table*}

{
 We now discuss the properties that we compute for each cluster and report a summary in Table~\ref{table:obs}. 
 We compute the total stellar masses $M_\star$ and $M_\star^{\rm 2D}$ as the sum of all stellar particles within the respective volumes.
 We compute  the richness $n$ with a 
 cut of satellites of ${\rm log}_{10} (M_\star/{\rm M}_\odot) > 10.65,$
 and the projected version $n^{\rm 2D}$ that includes \Euclid-like corrections for projection effects where similarly to \cite{2016A&A...593A...2Andreon}.
 In particular, we compute the average projected richness between $3.5$ and $8$ Mpc radii from the cluster centre, divided the annulus in $8$ slices of equal angles,  excluded the two least dense and most dense slices and removed the average projected number density of the $4$ remaining octants from the projected richness within}
 $r_{\rm 200c};$\footnote{
 Note that \Euclid richness will be based on detection algorithms such as AMICO~\citep{Bellagamba2018Amico} or PZWav, which may provide similar or different removal for projection effects. However, we do not have Magneticum Box2b light cones to feed to these algorithms. We stress that our richness computation anyway is very close to what is used in other observational works as \cite{2016A&A...593A...2Andreon}.
 }
 We compute the X-ray luminosities  $L_X,$ and $L_X^{\rm 2D},$ in the $[0.5,2]\,{\rm keV}$ energy band computed  using the \texttt{APEC} model~\citep{2001ApJ...556L..91Smith},
 using SPH particle temperatures together with the \texttt{XSPEC} package\footnote{\url{https://heasarc.gsfc.nasa.gov/xanadu/xspec/}}~\citep{1996ASPC..101...17Arnaud}, which considers the emission of a collisionally-ionized, chemically-enriched plasma implemented with metallicity values taken from the simulated particles~\footnote{We credit~\cite{Biffi17Xray} and~\cite{2018MNRAS.474.4089Truong} for the routines and cooling tables we used.}.
 We compute the temperature $T$ and $T^{\rm 2D}$ as weighted by the X-ray emissivity of gas particles.
We compute the hot gas mass $M_{\rm g}$ and $M_{\rm g}^{\rm 3D}$, computed as the sum of the mass of SPH particles with cold gas fraction greater than $0.1$ and $T>3\times\,10^4\,{\rm K}$ in order to filter out cold gas.

 Note that the projected gas mass is not to be confused with the one inferred from X-ray observations, as X-ray observational works typically perform a de-projection of the surface brightness $\propto n_{\rm e}^2,$ which provides a gas-mass estimate that is closer to the spherical  $M_{\rm g},$ with the addition of some possible alignment effects coming from the central region of clusters. 
 {
Moreover, observational works have the capability to mask possible bright substructures, thus minimising the presence of interlopers. Consequently, we can conceptualise the observed projected gas mass as an intermediate value between our $M_{\rm g}$ and $M_{\rm g}^{\rm 2D}$.
 }

 We estimate the integrated Compton-$y$  parameter produced by thermal Sunyaev--Zeldovich ~\citep[SZ,][]{1972CoASP...4..173Sunyaev}. The Compton-$y$  parameter is defined as 
 \begin{equation}
 \label{eq:sz}
y=\frac{k_{\rm B}\sigma_{\rm T}}{m_{\rm e} c^2}\int {T \,n_{\rm e}\,{\rm d}l},
\end{equation}
where $T$ is the temperature,   $n_{\rm e}$ the number density of the electrons, $k_{\rm B}$ the Boltzmann constant, $\sigma_{\rm T}$ the Thomson cross-section, $c$ the speed of light, and $m_{\rm e}$ the electron rest mass.
We compute the integrated Compton-$y$ parameter $Y=\int y \,  {\rm d}\Omega,$ both within the volume of sphere of ($Y$) and a cylinder ($Y^{\rm 2D}$).  
We estimate the integral in Eq. \eqref{eq:sz} as
\begin{equation}
\label{eq:sz-sph}
\int {T \, n_{\rm e} \, {\rm d}l }\approx \frac{1}{\pi R ^2} \sum_i { T_i  \  f_{{\rm e},i}  \ \frac{m_i}{m_{\rm p}}},
\end{equation}
where the sum runs over all SPH particles,  $m_i$ is the $i-$th SPH particle mass, $T_i$ its temperature and $f_{{\rm e},i}$ is its electron fraction, expressed as local electron number density normalised to the hydrogen number density, and $m_{\rm p}$ is the proton mass.

 For each halo, we also perform fits of the NFW profile $\rho_{\rm NFW}$, defined   as
 \begin{equation}
 \label{eq:nfw}
 \rho_{\rm NFW}\left(r\right) = \frac{\rho_0}{r/r_{\rm s}\left(1+r/r_{\rm s}\right)^2},
 \end{equation}
 where the scaling density $\rho_0$ and the scale radius $r_{\rm s}$ (that is the radius where the density log-slope equals $-2$) are free parameters.
We perform this fit on the total matter density profile on $100$ log-spaced radial bins between $75\,{\rm ckpc}$ (which corresponds to $60\,{\rm kpc}$ at $z=0.24,$ and to $40\,{\rm kpc}$ at $z=0.9$; as it is enough to exclude the deep central potential of baryons) and $r_{\rm 200c},$ and define the corresponding NFW masses and concentration parameters as $M_{\rm NFW}$ and $c_{\rm NFW}$ respectively, and the concentration as $c_{\rm NFW}=r_{\rm NFW}/r_{\rm s},$  where $r_{\rm NFW}$ is obtained from  $M_{\rm NFW}$ via the Eq. \eqref{eq:mass}.
 
 The projected version of the mass and concentrations are obtained by mimicking a lensing reconstruction procedure by fitting the corresponding reduced shear.
 We define the derived masses and concentrations as  $M_{\rm NFW}^{\rm 2D}$ and $c_{\rm NFW}^{\rm 2D},$ where $c^{\rm 2D}_{\rm NFW}=R^{\rm 2D}_{\rm NFW}/r_{\rm s}, $ where $R_{\rm NFW}^{\rm 2D}$ is obtained from  $M_{\rm NFW}^{\rm 2D}$ via the Eq. \eqref{eq:mass}.
Note that the "2D" here, as for the other quantities, means that the quantity is computed in projection, however, a correct fit of the mass from reduced shear NFW profile (namely, our $M_{\rm NFW}^{\rm 2D}$) should provide an estimate of the same NFW halo mass $M_{\rm NFW}$ that would be recovered from a 3D fit.

The fit is computed within the cylinder described above, with a projected radial range of $[300, 3000]\,{\rm kpc}$ at $z=0.2.$
We performed the analyses at  $z=0.9$ by rescaling that range with $H^{-2/3}(z),$ where $H(z)$ is the Hubble parameter, in order to retain the same fractional distances from the virial radius (at fixed mass), which resulted in a range of $[234,2300]\,{\rm kpc}.$  
Note that in this work we are not interested in estimating the contribution of the uncorrelated large-scale structure in the reduced shear reconstruction, therefore we limit our density projection to a cylinder of the depth of $20\,{\rm cMpc},$~\citep[see, e.g.,][]{Giocoli-EP30,Becker11}.

The signal from the source-averaged excess surface mass density $\Delta\Sigma_{\rm gt},$ averaged over circular radii $R$ and a population of sources distributed in redshift, {
can be written as
  \begin{equation}
  \begin{split}
 \ave{
 \Delta\Sigma_{\rm gt}
 }\left(R\right) 
 \simeq 
 \frac{ 
 \ave{
 \Delta\Sigma_{\rm t}
 }\left(R\right) 
 }
 {1 - 
 \ave{\Sigma_{\rm cr}^{-1}}
 \ave{
 \Sigma
 }\left(R\right) 
 }. 
 \label{eq:shear}
 \end{split}
 \end{equation}
   Here $\Sigma$ denotes the surface mass density. The symbol $\ave{...}$ denotes an average over radial bins and redshift lens sources, where we used a redshift distribution as proposed in \cite{Ajani-EP29}, and \cite{Giocoli-EP30}.
   }
 The quantity $\ave{\Delta\Sigma_{\rm t}}$
 is the excess of surface mass density, averaged over polar coordinates and defined as 
\begin{equation}
 \label{eq:shear-error}
\ave{\Delta\Sigma_{\rm t}}\left(R\right) = \frac{1}{\pi\,R^2}\int_0^R2\,\pi\, r\,\ave{\Sigma}(r)\,{\rm d}r - \ave{\Sigma}(R).
\end{equation}
The symbol $\Sigma_{\rm cr}$ in Eq. \eqref{eq:shear} is the  critical surface mass density, that for a given redshift source equals to 
\begin{equation}
    \Sigma_{\rm cr} = \frac{c^2\, {D}_{\rm s}}{4\pi G\,{D}_{\rm ds} {D}_{\rm d}},
\end{equation}
where $G$ the universal gravity constant, ${D}_{\rm s}$ the angular diameter distance to the source, ${D}_{\rm d}$ the angular distance to the lens, and ${D}_{\rm ds}$ the angular distance between the source and the lens.
Similarly to~\cite{Giocoli-EP30},
 we define
 the error associated with each radial bin of the profile in Eq. \eqref{eq:shear} as
\begin{equation}
   \begin{split}
     \delta \ave{
 \Delta\Sigma_{\rm gt}
 } 
 = \ave{\Sigma_{\rm cr}} \frac{\sigma_\epsilon}{ \sqrt{\pi\,n_{\rm g} \left( R_2^2 - R_1^2 \right)}}, 
    \label{eq:deltag}
    \end{split}
\end{equation}
where $\sigma_\epsilon=0.3$~\citep{2012MNRAS.427.1298Hoekstra,Blanchard-EP7} is the dispersion of the total intrinsic ellipticity $\epsilon=(1-q)/(1+q),$ where $q$ is the axis ratio, $R_1$  and $R_2$ represent the inner and outer radius of a bin, and $n_{\rm g}$ is the number density of galaxies.
{
For our redshift source distribution~\cite[we assume the same as in][]{Giocoli-EP30}, we find that
$n_{\rm g} \approx 28\,{\rm arcmin}^{-2}$ for lenses at redshift $z=0.24$ and $n_{\rm g} \approx  14\,{\rm arcmin}^{-2}$  for lenses at redshift $z=0.9.$
}

\subsection{Scaling relations}
\label{sec:cumpa}

\begin{figure}
\includegraphics[width=\linewidth]{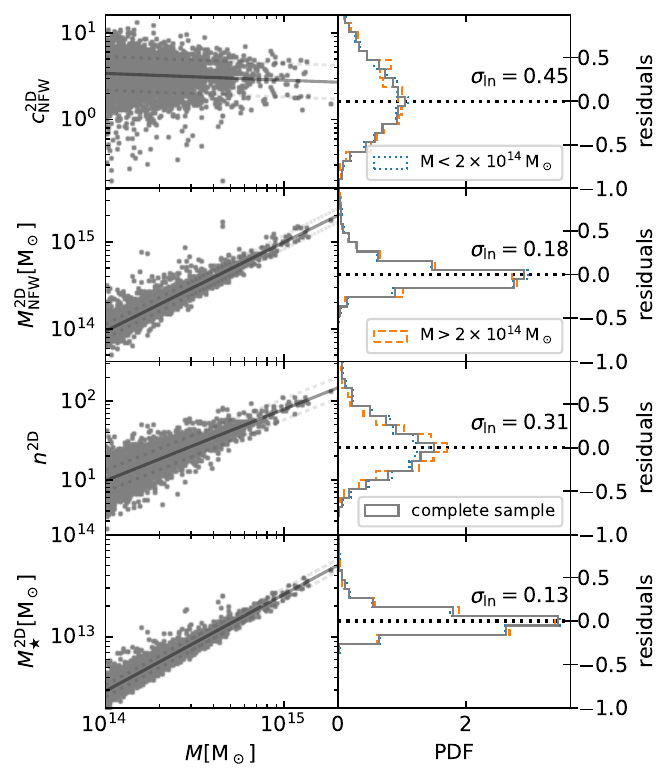}\\
\caption[h]{Magneticum mass-observable relation for \Euclid-like derived quantities. The left column shows scaling relations, relative fit (solid grey line), and a corridor corresponding to one standard deviation (dashed grey line). The right column shows the residual PDF and scatter of log-residuals $\sigma_{\rm ln}.$ We report the following properties: lensing concentration $c^{\rm 2D}_{\rm NFW}$ (first row), lensing mass  $M^{\rm 2D}_{\rm NFW}$ (second row), projected  richness $n^{\rm 2D}$ (third row),  and projected stellar mass $M_\star^{\rm 2D}$ (fourth row). The histogram of residuals for haloes with $M<2\times10^{14}\,{\rm M}_\odot$ is in blue dotted lines, for haloes with $M>2\times10^{14}\,{\rm M}_\odot$ is in orange dashed lines, and for the complete sample is in solid grey lines. Note that the three histograms almost overlap. Each distribution panel reports the value of the natural log scatter $\sigma_{\rm ln}$ for the complete sample.
}
\label{fig:compare_obs_eucl}
\end{figure}

\begin{figure}
\includegraphics[width=\linewidth]{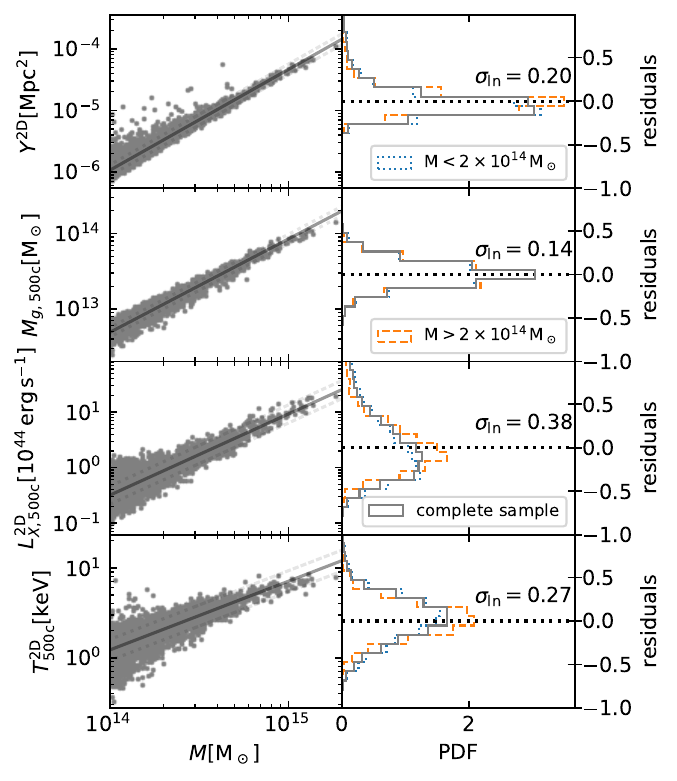}\\
\caption[h]{As Fig.\,\ref{fig:compare_obs_eucl}, but for the following multi-wavelength observable properties: projected integrated Compton-$y$ parameter $Y^{\rm 2D}$ (first row),  the  3D gas mass $M_{\rm g, 500c}$ (second row), the projected X-ray luminosity in the soft band (in range $[0.5,2]\,{\rm keV}$) $L^{\rm 2D}_{\rm X, 500c}$ (third row),    and the projected temperature $T^{\rm 2D}_{\rm 500c}$ (fourth row). The values of X-ray luminosity, gas mass, and temperatures are reported within an overdensity of $r_{\rm 500c}$ as this definition is a typical choice in X-ray-based observations.
}
\label{fig:compare_obs_oth}
\end{figure}

In  Fig.\,\ref{fig:compare_obs_eucl} we show the observable properties vs. true mass $M$ of clusters, derived from Magneticum Box2b/hr simulation for properties that could be derived by \Euclid-like catalogues, such as lensing concentration (first row from top), lensing mass (second row), projected richness (third row), and projected stellar mass (last row), as presented in Sect.~\ref{sec:obs}.
For each property, we fit a scaling relation performed using a linear regression in the log-log space. 
We utilise a log-log linear regression because a single power law proves to be effective in modelling our scaling relations.

In the right panel of Fig.\,\ref{fig:compare_obs_eucl},   we show the log-residual distribution for both low-mass haloes ($M<2\times10^{14}\,{\rm M}_\odot$), high-mass haloes ($M>2\times10^{14}\,{\rm M}_\odot$), and for the complete sample of the log-residuals $\sigma_{{\rm ln},i}$, defined as the logarithmic ratio between the $i-$th cluster property and the corresponding scaling relation value at its mass.
 In the second column, we also report the log-scatter  $\sigma_{\rm ln}$ defined here as the corresponding standard deviation of the log-residual, namely
 \begin{equation}
 \sigma_{\rm ln} = \operatorname{E}\left[\left(\sigma_{{\rm ln},i} - \operatorname{E}\left[\sigma_{{\rm ln},i}\right]\right)^2\right]^{1/2},    
 \label{eq:sigmaln}
 \end{equation}
 where $\operatorname{E}$ is the expectation operator that averages over our catalogue data.
 We note that the concentration has a scatter of $0.45$ which is higher than theoretical expectations~\citep[see, e.g.,][]{Child2018MCscatter}.
 Throughout this paper, we will show that this is due to projection effects; in fact, the 3D concentration has a scatter of $\approx0.33.$

  {
Note that our scatter in temperature exceeds that reported in the theoretical work by \cite{2018MNRAS.474.4089Truong}. We verified that, if we compute mass-weighted temperature, which is known to behave very well in a power law scaling relation, reveals a log scatter of $0.07,$ in agreement with their work. The additional scatter that we see may be due to different X-ray temperature computations (they use core-excised temperature while we take the contribution of the core into account).
 }

 In Fig.\,\ref{fig:compare_obs_oth} we show the mass-observable relations of quantities that could potentially be obtained from various multi-wavelength observations that could enrich studies based on \Euclid-like data products:  the integrated Compton-$y$ parameter, the gas mass $M_{\rm g,500c}$,   the X-ray luminosity $L_{\rm X,500c}^{\rm 2D}$ converted in the soft band $[0.5,2]\,{\rm keV},$ and the temperature $T_{\rm 500c}.$
 We decided to plot the X-ray luminosity, gas mass, and temperature within $r_{\rm 500c}$  because this radius is typically used in various X-ray observations~\cite[see, e.g.,][]{2006ApJ...640..691Vikhlinin,2009ApJ...693.1142Sun}.
 
 {
The typical \Euclid cluster cosmology analysis will therefore rely on mass-observable relations calibrated within $r_{200c}$ (e.g., richness, weak-lensing mass), and follow-up observations calibrated within $r_{500c}$ (e.g., X-ray and SZ mass-proxies). We will thus need to take into account the covariance between observable properties extracted at different radii. 
}.
{
Finally, we note that generally, X-ray observations are expected to align more closely with the 3D mass rather than the projected one~\citep{Ettori2013MassProfiles}, although several details adopted to analyse the X-ray observations (e.g., including masking of substructures, deprojection procedures, etc.) can significantly impact the final result. These choices critically depend on the quality of the observations themselves. Dedicated mocks will thus be required to properly take into account all these effects and are, therefore, beyond the purpose of this work.
 For simplicity, in this work, we consider an X-ray-derived gas mass as close to $M_{\rm g},$ while an SZ-derived gas mass closer to $M_{\rm g}^{\rm 2D}$.}

\section{Projection effects}
\label{sec:proj}

\begin{figure*}
\includegraphics[width=\linewidth]
{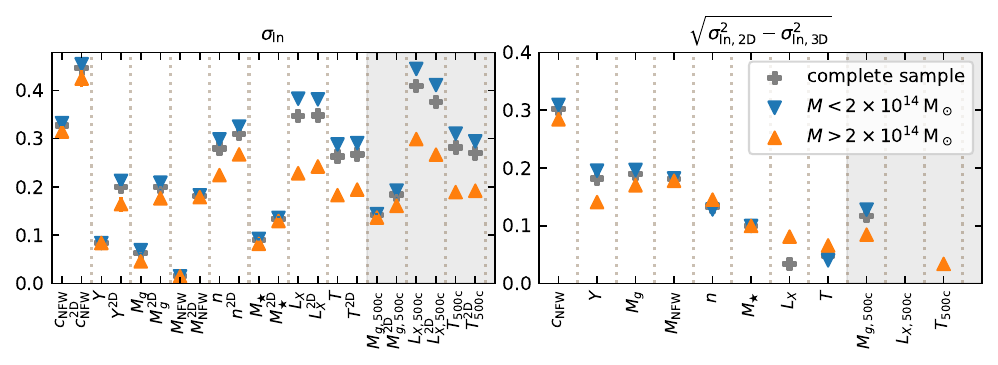}
\caption[h]{Scatter of our mass-observable relations, paired with their projected version: concentration,  integrated Compton-$y$ parameter, gas mass, NFW mass, richness, stellar mass,  X-ray luminosity, and temperature, within a radius of $r_{\rm 200c};$ the grey band reports the gas mass, X-ray luminosity, and temperature within $r_{\rm 500c}.$ 
The left panel reports the fractional scatter of both 3D and 2D quantities.
Each dotted vertical line separates the regions that report a given quantity and its 2D version.
The right panel reports the contribution of projection effects.
Points are coloured by their mass range as in Fig.\,\ref{fig:compare_obs_eucl}, blue down-triangles represent the low mass bin, orange up-triangles represent the high mass bin, and grey crosses represent the complete sample.
Note that we lack the value of projection effects for $L_{\rm X, 500c}$ (see discussion). Points are ordered according to the value of the second panel. Error bars are computed using the jackknife method. }
\label{fig:sigma}
\end{figure*}

\begin{figure}
\includegraphics[width=\linewidth]
{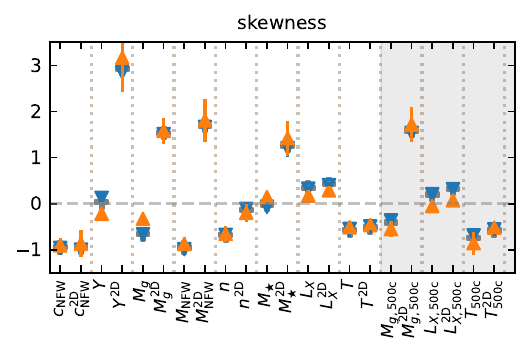}
\caption[h]{Skewness parameters for our cluster properties. Data, line styles and colours are as in Fig.\,\ref{fig:sigma}. Error bars are computed using the jackknife method. }
\label{fig:skew}
\end{figure}

\begin{figure}
\includegraphics[width=0.9\linewidth]{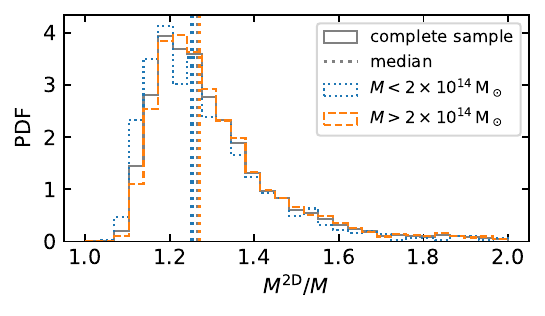}
\caption[h]{Probability density distribution of $M^{\rm 2D}/M$ at different mass bins:  for haloes with $M<2\times10^{14}\,{\rm M}_\odot$ as a blue dotted line,  for haloes with $M>2\times10^{14}\,{\rm M}_\odot$ as a dashed orange line and for the complete sample as a grey solid line. For each mass bin, we report a vertical line with the median values (note that the three lines are very close together).}
\label{fig:M2DPDF}
\end{figure}

\begin{figure*}
\begin{subfigure}[b]{0.03\linewidth} 
{
    \hspace{2.2cm}{
                {
                 \tiny{$\frac{M^{\rm 2D}}{M}=$}
                }
    }
    \rotatebox[origin=l]{90}{\hspace{2.1cm}{
            {
                \tiny{ \texttt{23\,} ${\rm Mpc}$ }
            }
        }
        \hspace{1.5cm}{}
    }
}
    \end{subfigure}
    \hfill
    \begin{subfigure}[b]{0.98\linewidth} 
        \includegraphics[width=\linewidth]{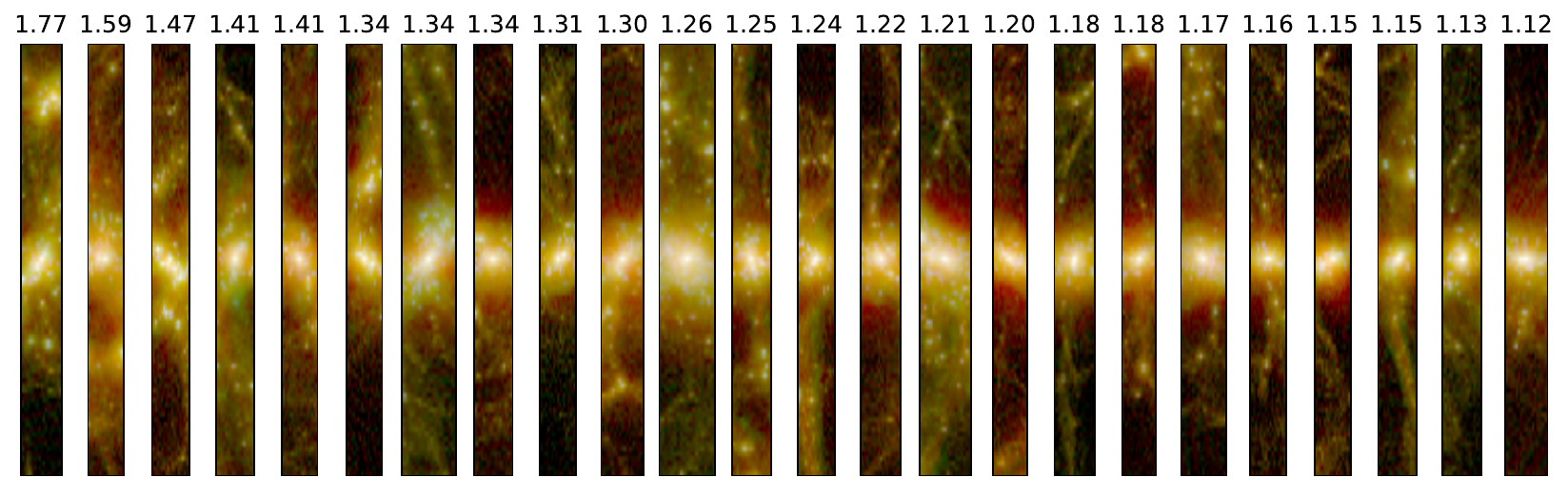}
    \end{subfigure}
\caption[h]{Projected maps along a cylinder of length $23\,{\rm Mpc},$ with radius $r_{\rm 200c}$  and centred on a random sample of our galaxy clusters, ordered by their  $M^{\rm 2D}/M$ (over-plotted above each map) values decreasing from left to right. The pixel red, green, and blue channels are used as follows: the red channel maps the gas projected mass, the green channel maps the dark matter projected mass, and the blue channel maps the stellar projected mass. Columns widths are proportional to the cluster radii.}
\label{fig:map_random}
\end{figure*}

\begin{figure*}
\includegraphics[width=\linewidth]{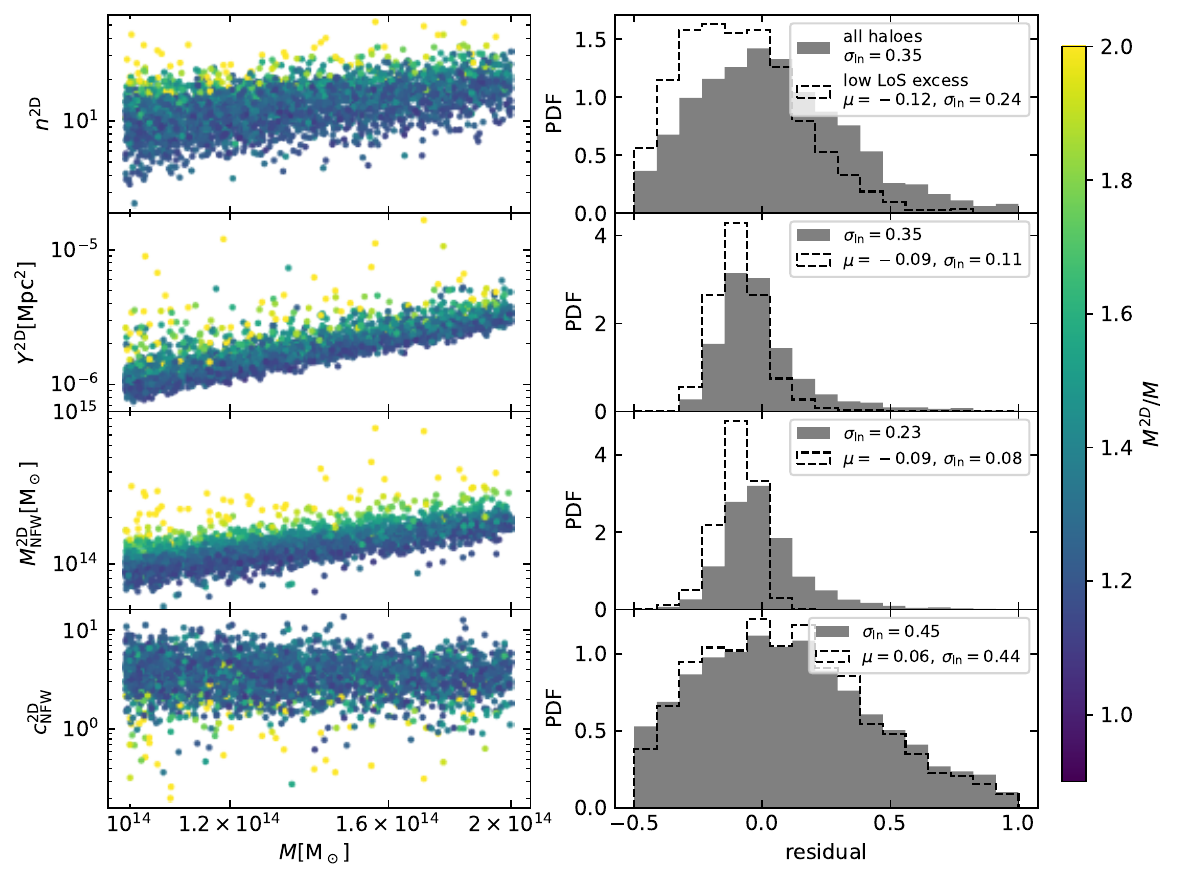}
\caption[h]{Impact of LoS contamination in scaling relations. We show halo properties  as a function of halo mass $M$
in the left column, and colour-coded by the fractional amount of mass in a cylinder ($M^{\rm 2D}/M$), and the residuals PDFs in the right column (grey shaded histogram).
Rows correspond to richness, the integrated Compton-$y$ parameter, lensing mass, and lensing concentration. We also show the residuals of a subset of haloes with low LoS contamination (in particular $M^{\rm 2D}/M<1.26,$ dashed line histogram). Each panel reports the scatter of the residuals $\sigma$ and the mean $\mu$ of the low LoS residuals. }
\label{fig:residuals_2d}
\end{figure*}

\begin{figure}
\includegraphics[width=\linewidth]{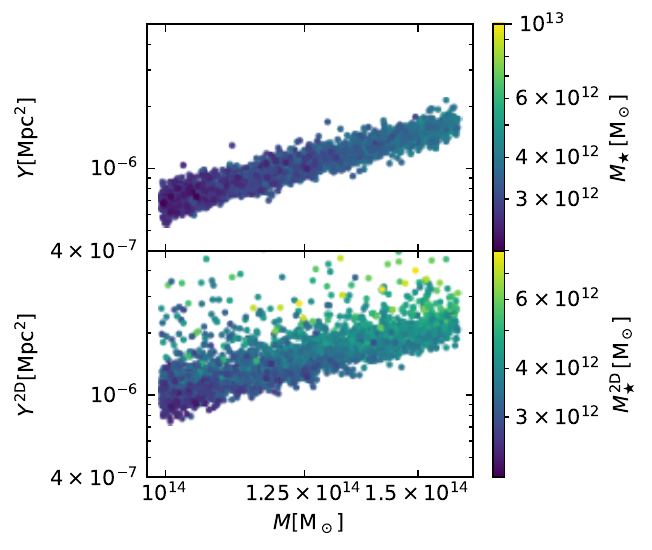}
\caption[h]{Integrated Compton-$y$ parameter vs halo mass, colour-coded by a stellar mass fraction. The top panel shows quantities computed within a sphere of radius $r_{\rm 200c}$, while on the bottom panel, they are computed within a cylinder (of radius $r_{\rm 200c}$ and length $23\,{\rm Mpc}$ as already presented in Sect. \ref{sec:setup}). We limit the plot in the mass range $M\in[1,2]\,10^{14}\,{\rm M}_\odot$.}
\label{fig:3d_vs_2d}
\end{figure}

The main objective of this work is to disentangle the amount of scatter and skewness in scaling relations that is purely due to projection effects.
Note that  observational data are also affected by measurement errors that we do not tackle in this work 
(as, for instance, the Poisson error of the limited number of galaxies used to infer the richness). 
In this Section, we discuss the scatter and skewness of mass-observable relation qualitatively and limit the discussion for the data at redshift $z=0.24,$ as we have a larger sample of galaxy clusters, and the results are qualitatively similar to the ones at $z=0.9$. We stress that we leave the quantitative discussion of the scatter and correlation coefficients for both redshifts on Sect.~\ref{sec:cova}.

To assess the role of projection effects,    Fig.\,\ref{fig:sigma}   reports the 3D and 2D scatter of our cluster properties.
In the left panel of Fig.\,\ref{fig:sigma} we show 
the value of the scatter (see Eq.~\ref{eq:sigmaln}) of log-residuals $\sigma_{\rm ln}$ for all our mass-observable relations.
In the shaded region, we also report the values computed within $r_{\rm 500c}$ for the X-ray luminosity, gas mass, and temperature values because this is 
 the characteristic overdensity used in X-ray analyses.

For each  observable, in Fig.\,\ref{fig:sigma}, we report (with different symbols) both the scatter of the complete sample as well as the one of two 
separate mass range of $10^{14}<M<2\times10^{14}\, {\rm M}_\odot$ and $M>2\times10^{14}\, {\rm M}_\odot$ respectively.
We note that the lower-mass bin ($M<2\times10^{14}\, {\rm M}_\odot$) is the one with the largest scatter because, for a given external object in the line-of-sight (LoS), the profile of a small cluster will be more perturbed with respect to a cluster.

  In the left panel of Fig.\,\ref{fig:sigma}, we see that some quantities have a low scatter in the 3D space and do gain a large amount of scatter once they are seen in projection. 
To better quantify  what is the actual impact of projection effects, in the right panel of Fig.~\ref{fig:sigma} we present the metric $\sqrt{\sigma_{\rm ln, 2D}^2 - \sigma_{\rm ln, 3D}^2}.$  {
This metric shows that the quantities that are most affected by projection effects are the weak lensing concentration, integrated Compton-$y$ parameter, gas mass, and NFW profile lensing mass.
This is expected as all these observable properties (except for weak lensing concentration and X-ray luminosity) scale linearly with the respective observable mass.
Further, we note that the scatter in the richness agrees to the theoretical predictions from~\cite{Castignani2016richness}.
}

{
We observe that X-ray luminosity and temperature are the least affected by projection effects. This is attributed to the fact that X-ray luminosity is contingent upon the square of gas density, thereby being primarily influenced by the most bright regions within an image. Similarly, the temperature is predominantly influenced by the innermost regions of a cluster.
}
Note that we lack the value of projection effects for $L_{\rm X, 500c}$ because the 2D scatter is slightly smaller than the 3D one. This happened because projection effects impact under-luminous haloes more strongly than overly luminous haloes (at a fixed mass bin), with the consequence of the projected X-ray luminosity having a higher normalisation and a lower scatter (see Fig. ~\ref{fig:LX3D_vs_2D}).

{
Some mass-observable relations have a large skewness,  to aid observational works in modelling these relations, we will estimate their skewness.
}
Therefore, we also quantify deviations of residuals from a symmetrical distribution by means of the   Fisher-Pearson coefficient of skewness ${m_3}/{m_2^{3/2}},$ where $m_k$ is the sample  $k$th  central moment.\footnote{
 Where the central moment $m_k$ for a random variable $X$ is defined as $\operatorname{E}\left[ ( X - \operatorname{E[X]})^k \right],$ where a skewness of zero implies a symmetric distribution, a positive/negative value implies a distribution skewed towards the right/left part.
}
We report its dependency on projection effects in  Fig.\,\ref{fig:skew}, showing that the properties whose skewness is most impacted by projection effects are the integrated Compton-$y$ parameter, the gas mass, and the lensing NFW mass.
We also notice that some scaling relation residuals move from having a negative skewness (for the NFW concentration, for instance, due to un-relaxed and merging clusters) to a positive one once projection effects are taken into account (namely, from an asymmetry towards negative residuals towards an asymmetry towards positive residuals).

To assess the impact of projection effects, we introduce a variable to quantify the amount of additional matter in the line of sight that can skew our observable properties. We define it as the ratio of the mass within the cylinder and the mass within a sphere $\mrat,$ both of radius $r_{\rm 200c},$ where the length of the cylinder is described in Sect. \ref{sec:setup}.
We present the distribution of $\mrat$ in Fig.\,\ref{fig:M2DPDF}, where we can see that this quantity is strongly skewed and its median value is $M^{\rm 2D}/M\approx1.26$ 
{
(note that for a NFW profile with $c=4,$ the corresponding analytical cylinder vs. spherical mass  is $1.25$).
}
Although this quantity is not directly observable, we will use it to assess the contribution of LoS objects in the scatter of scaling relations. Note that besides objects in the LoS, different fitting procedures may impact the scatter of projection effects, as we will see in Sect. \ref{sec:MC}.

In Fig.\,\ref{fig:map_random}, we show a random selection of clusters, ordered by decreasing $\mrat$  from left to right.
Objects with high $\mrat$ (the objects in the left-most panels) include clusters that are merging, elongated, or in the LoS.
In the rest of this paper, we will refer to the objects having $\mrat$ greater than the median of the distribution $1.26$ as having LoS excess.

To study the impact of projection effects in scaling relations, in Fig.\,\ref{fig:residuals_2d} we show the scaling relations of the following projected quantities: richness, integrated Compton-$y$ parameter, lensing mass, and concentrations, that are the ones that are most affected by projection effects.
We colour-code these points by   $\mrat$ and focus on a narrow mass range of $M\in[1,2]\times10^{14}\,{\rm M}_\odot$ in order to visualise better how LoS excess impacts these scaling relations.
On the left column, we can visually see that, except for concentration, they strongly correlate with $\mrat,$ as the upper points of the scatter plot tend to have higher values of $\mrat.$
 
We quantify this finding in  the right column by 
comparing the residual distributions (of the power-law fit over the complete mass range presented in Sect.~\ref{sec:cumpa}) of the complete sample with the distribution of objects with low LoS contamination only (we adopt the criteria of $\mrat<1.26$  being the median of the $\mrat$ distribution, as shown in Fig.\,\ref{fig:M2DPDF}),  and report the respective scatter $\sigma$ and average value $\mu$ of the residual distributions (for the complete sample we have that $\mu=0$).
Except for the concentration, residuals of haloes with low $\mrat$ (see dashed histogram) significantly shift towards negative values of $\mu$ and $\sigma.$
For instance, when we consider only objects with a low LoS excess, the scatter of $Y^{\rm 2D}$ decreases from $0.35$ down to $0.11,$ and the $M^{\rm 2D}_{\rm NFW}$ scatter goes from $0.23$ to $0.9$. 
{
The lensing mass is, in fact, well known to be affected by projection effects~\citep{2014ApJ...797...34Meneghetti,Giocoli-EP30}.

In this paper, we refer to projection effects as all effects that take place when going from 3D to projection; they include both LoS effects and model uncertainties.
This definition becomes relevant when dealing with concentration, which is not impacted only by LoS objects but also by the NFW profile fitting procedure. 
We stress that in Fig.\,\ref{fig:sigma} we proved that our projected concentration is actually highly impacted by projection effects, yet only weakly affected by LoS effects. }
We show in the next Section the reason for concentration being strongly affected by projection effects is that their reduced shear profile deviates strongly from the one produced by NFW profile~\citep[see, e.g.,][]{Ragagnin2021MC}, which is used to reconstruct the reduced shear profile.

To conclude this section, we now study how correlations between cluster observable properties can be affected by projection effects.
To this end, we take the case of a possible hot- vs. cold-baryon correlation by studying the stellar mass vs. integrated Compton-$y$ parameter (as the latter should strongly correlate with the gas mass).
In Fig.\,\ref{fig:3d_vs_2d}, we show the integrated Compton-$y$ parameter scaling relation, colour-coded by stellar mass for the 3D quantities (top panel) and projected quantities (bottom panel).

Examining the correlations at a constant halo mass among the computed quantities within spheres (as depicted in the top panel of Fig. \ref{fig:3d_vs_2d}), we find no discernible weak anti-correlation between stellar mass and the integrated Compton-$y$ parameter (which is defined in Sect. \ref{sec:cova}). Conversely, when investigating the properties in the projected space, a more pronounced correlation becomes evident.
This implies that projection effects can strongly impact the correlation between observable properties.
While this analysis is purely qualitative, we will quantify the impact of these projection effects in Sect.~\ref{sec:cova}, where we will compute the correlation coefficients for both 3D quantities and 2D quantities.

\section{Projection effects on lensing concentration}
\label{sec:MC}

\begin{figure}
\includegraphics[width=\linewidth]{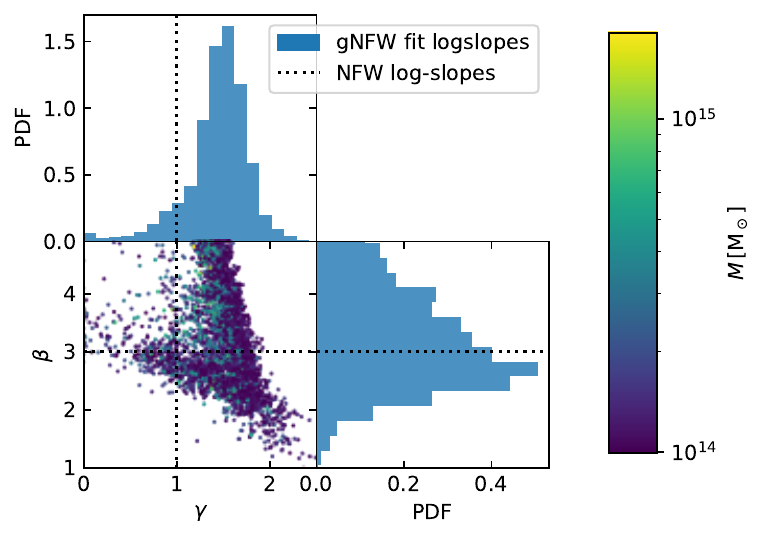}
\caption[h]{Probability density distribution of the parameters $\gamma$ (inner slope, upper panel) and $\beta$ (outer slope, right panel) of Eq. \eqref{eq:gnfw} of the successful gNFW profile fits.
The central panel shows the scatter plot between the two parameters colour-coded by $M$. The dotted lines show the NFW parameters $\gamma=1$ and $\beta=3.$}
\label{fig:gamma_beta}
\end{figure}

\begin{figure}
\includegraphics[width=0.95\linewidth]{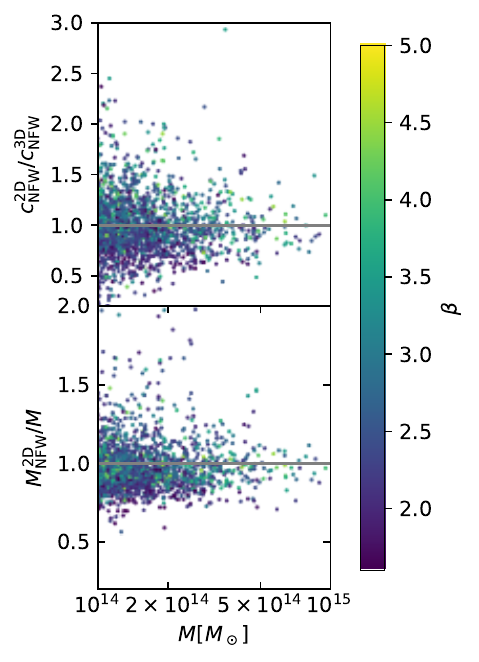}
\caption[h]{Ratio between 2D NFW profile fit parameters and 3D parameters for haloes with a successful 3D gNFW profile fit.
Upper panel: the ratio between concentrations; lower panel: the ratio between halo masses.
Points are colour-coded by the external log-slope $\beta$ of the 3D fit of gNFW.   }
\label{fig:c_ratios}
\end{figure}

\begin{figure}
\includegraphics[width=0.9\linewidth]{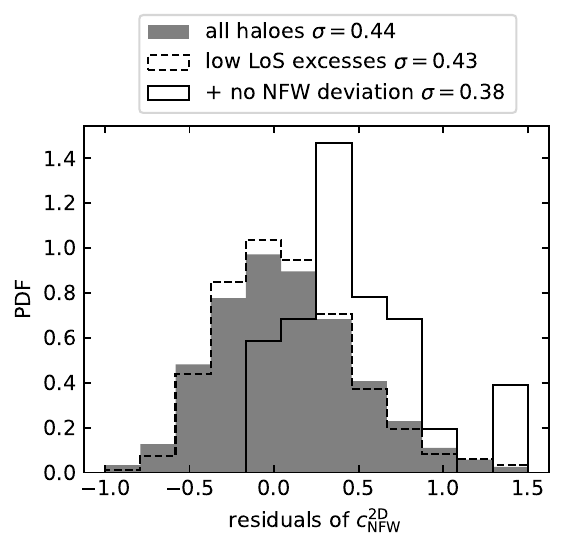}
\caption[h]{Residuals of lensing concentrations with respect to the power-law fit. As in Fig.\,\ref{fig:residuals_2d}, the dashed line histogram indicates the residuals for objects with low LoS effects (low value of $\mrat$).  
The solid line histogram contains the additional constraints of haloes with $\beta$  and $\gamma$ parameters close to the ones of an NFW profile ($2.8<\beta<3.2$ and $0.8<\gamma<1.2$). Each histogram label reports the scatter $\sigma$ of the residuals.}
\label{fig:residuals_2d_mc}
\end{figure}

\begin{figure}
\includegraphics[width=0.9\linewidth]{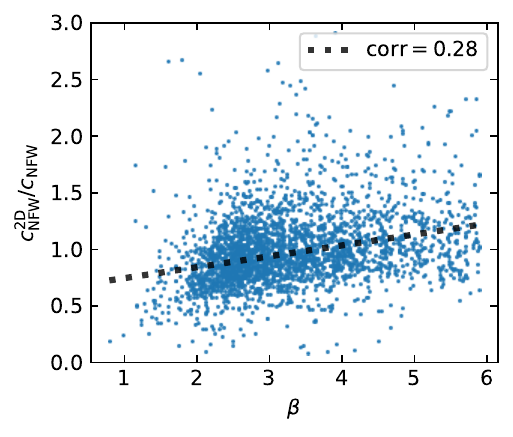}
\caption[h]{Scatter plot of ratio between 2D concentration and 3D concentration against $\beta,$ namely the outer slope of Eq. \eqref{eq:gnfw}, of well behaving 3D gNFW profile fits in a narrow mass range of $M\in[1,2]\times10^{14}\,{\rm M}_\odot.$ We also report the correlation coefficient.}
\label{fig:c_ratio_beta}
\end{figure}

As we found in the previous section,  projection effects significantly increase the scatter and skewness in the scaling of lensing concentration with mass.
However, this scatter increase is not related to external objects along the LoS. Now, we assess if the high scatter of lensing concentration is due to deviations of the reduced shear profile from the one induced by an NFW profile.

In this work, we will not delve into the origin of this deviation as it falls beyond the scope of this paper. Such deviation may arise due to halo elongations, suggesting that alternative profiles such as truncated NFW profiles may better suit galaxy clusters~\citep{Oguri+2011truncatedNFW}. Alternatively, it could stem from the expectation that the NFW profile is intended to describe stacked haloes rather than individual objects. Our focus in this paper is to understand the impact of assuming an NFW profile for each of our haloes. We emphasize that these NFW deviations only affect weak lensing signal reconstruction, as the NFW profile is highly effective in recovering halo mass in 3D.

To study deviations  from the NFW profile of haloes we fit a generalised NFW profile~\citep{Nagai2007gNFW}, hereafter gNFW, in spherical coordinates over the same radial range as our previous NFW profile (described in Sect. \ref{sec:obs}), where the density profile  $\rho_{\rm gNFW}\left( r \right)$ is defined as
\begin{equation}
\rho_{\rm gNFW}\left( r \right) = \frac{\rho_0}{\left(r/r_{\rm s}\right)^\gamma\left(1+r/r_{\rm s}\right)^{\beta-\gamma}},
\label{eq:gnfw}
\end{equation}
where $\gamma$ and $\beta$ are respectively the internal and external log-slopes of the total matter density profiles.
The case $\gamma=1$ and $\beta=3$ produces the NFW profile as in Eq. \eqref{eq:nfw}.
{
Note that the} \cite{Nagai2007gNFW} gNFW profile also depends on the parameter $\alpha$ that we fix to $\alpha=1$ in this work in order to explore internal and external log-slope variations only.

We present the PDF for the gNFW profile parameters  $\gamma$ and $\beta$ in Fig.\,\ref{fig:gamma_beta}, where the fit was performed in 3D with a flat priors for $\gamma\in[0,3]$ and $\beta\in[0,6]$. 
The data points are colour-coded according to the variable $M$, revealing no discernible strong trend with respect to the fitted parameters.
For $~19\%$ of the objects, the resulting best-fit parameters hit the boundaries of hard-cut priors.  Upon visual inspection, these objects are characterised by a very steep matter density profile at large cluster-centric distances, possibly suggesting that a truncated NFW profile might be a better model choice.
As our objective is to examine the effects of deviations from the generalised NFW profile, we omit these objects from the subsequent analysis in this Section.
{
Given the substantial deviation of these objects from NFW profile, they could potentially offer additional insights for our analyses. However, incorporating them would necessitate the use of a profile more general than Eq.~\ref{eq:gnfw}. Therefore, we excluded them in order to make our analysis clearer.}

We observe that the external logslope of Magneticum profiles appears to be slightly flatter than $-3.$ While we emphasize that this discrepancy does not affect the accurate recovery of mass and concentration parameters in 3D NFW fits (such fits can still yield precise estimates of halo mass and concentrations). However, these deviations in the NFW profiles may affect the reduced shear fit, particularly when observed over large radii (remember that in this work, we use $3\,{\rm Mpc}$).

Furthermore, we observe a degeneracy between the $\beta$ and $\gamma$ parameters, indicating that our profiles deviating from NFW profile tend to exhibit a flatter profile compared to NFW profile (as illustrated in Fig.~\ref{fig:example}). However, investigating this discrepancy is beyond the scope of this paper, as the internal log slope of clusters is not currently captured by existing weak lensing studies.

In Fig.\,\ref{fig:c_ratios}, we plot the values of concentration and mass obtained from reduced shear fit, divided by the corresponding 3D quantities and colour-coded by the external 3D gNFW slope $\beta$ for our intermediate redshift haloes. 
As we can see,  haloes with large values of $\beta$ have a projected concentration that is significantly higher than the 3D one (see upper panel).
In Appendix \ref{ap:gnfw} we report the example of a simulated halo with low LoS excess (see Fig.\,\ref{fig:example}) and an analytical one (see Fig.\,\ref{fig:colossus}), both with a flat external log-slope, and we show how the under-estimation of the concentration is caused by the fact that the NFW profile fit on the reduced shear is weighting too much the external part of the profile, that deviates from an NFW profile.

In Fig.\,\ref{fig:residuals_2d_mc}, we show the concentration residual distribution and report their scatter.
We note that the projected concentration scatter is not affected by external material along the LoS (dashed line and shaded histograms match). However, if one restricts our sample to objects having NFW-like profile log-slopes (we used criteria of $2.8<\beta<3.2$ and $0.8<\gamma<1.2$), then the scatter distribution changes drastically.
The concentration residuals decrease from $0.43$ to $0.38$, and the residuals shift towards higher values, suggesting that these objects are more relaxed. Such effect is well known, as studied for instance in~\cite{Maccio07}.

We also show how the external log-slope of the halo profile affects the lensing reconstruction by plotting the ratio between the projected and 3D concentration (i.e., $c^{\rm 2D}_{\rm NFW}/c_{\rm NFW}$) value versus the 3D log-slope $\beta$ in the narrow mass bin of $M\in[1,2]\times10^{14}\,{\rm M}_\odot$ in Fig.~\ref{fig:c_ratio_beta}, where we find a positive correlation coefficient of $\approx0.28,$ in agreement with a shift of residuals we showed in Fig.~\ref{fig:residuals_2d_mc}.

\section{Correlations between properties}
\label{sec:cova}

\begin{figure*}
{\centering 
\includegraphics[width=0.95\linewidth]{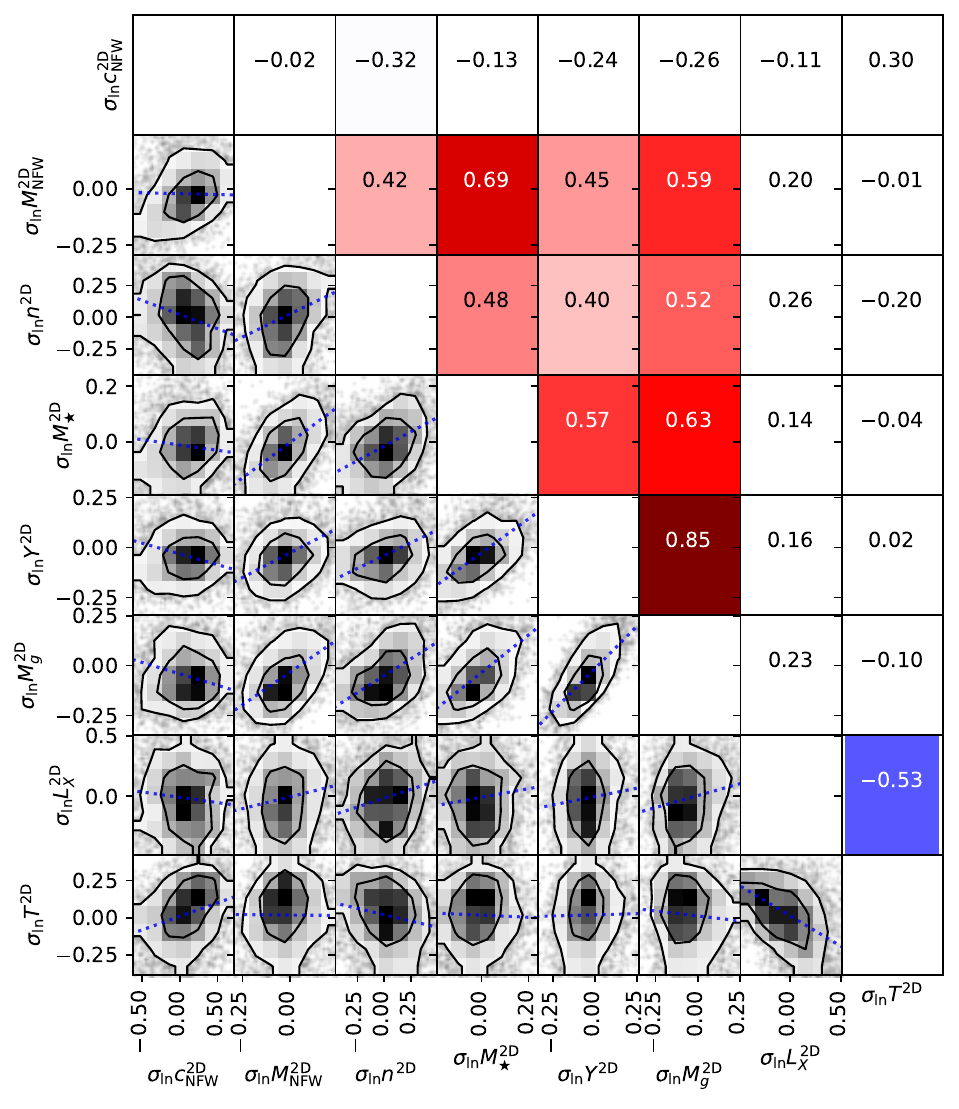}
}
\caption[h]{Correlation coefficients matrix \emph{(upper-right triangle)} and scatter plot \emph{(bottom-left triangle)} of power-law log-residuals of \Euclid-data (lensing concentration, lensing mass, richness, and stellar mass, respectively) and possible outcomes from multi-wavelength observations (integrated Compton-$y$ parameter, gas mass, X-ray luminosity, and temperature, respectively). Cell colouring goes from blue (negative correlation coefficients) to red (positive correlation coefficients) and is white in the interval $[-0.35,0.35]$ in order to enhance the visibility of the most significant coefficients.} 
\label{fig:covarma2D}
\end{figure*}

\begin{figure*}
{\centering 
\includegraphics[width=0.95\linewidth]{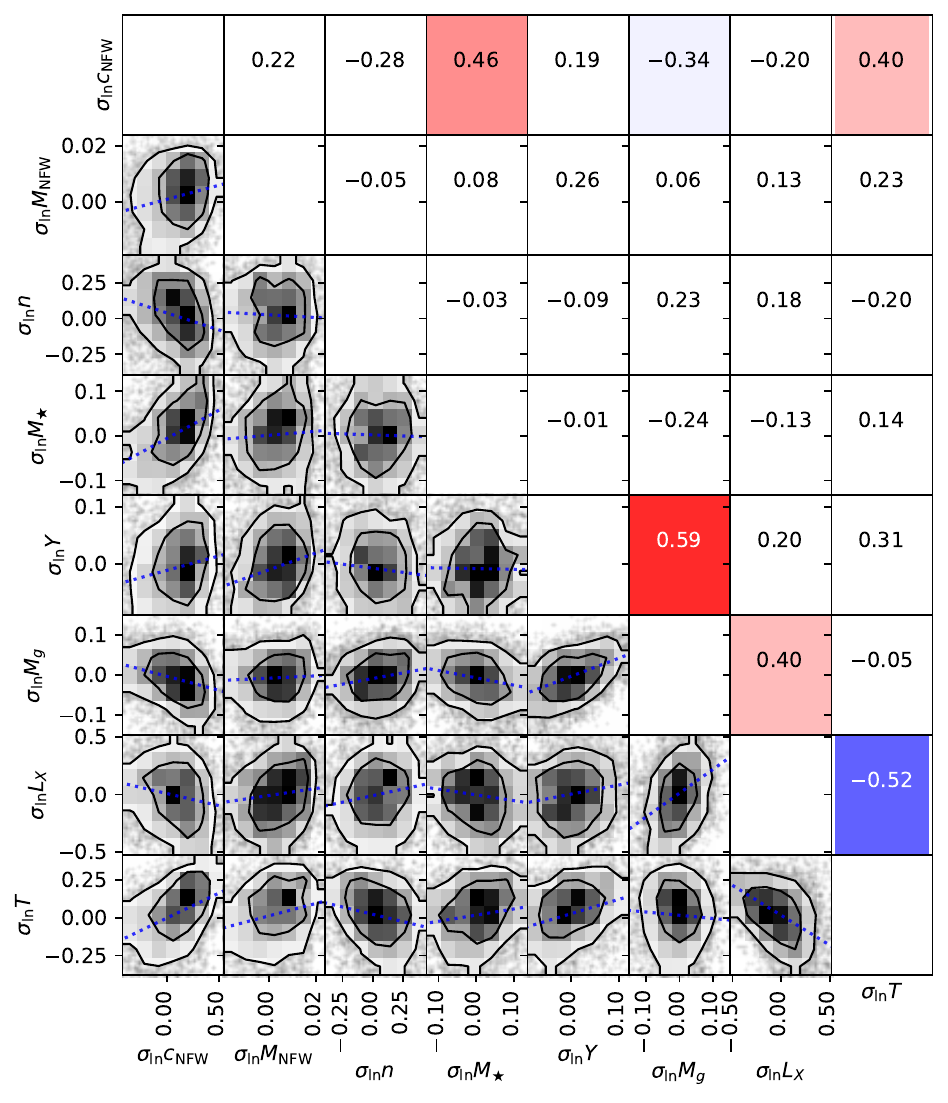}
}
\caption[h]{As Fig.\,\ref{fig:covarma2D}, here we show the quantities computed in the 3D space.}
\label{fig:covarma3D}
\end{figure*}

\begin{figure*}
\includegraphics[width=\linewidth]{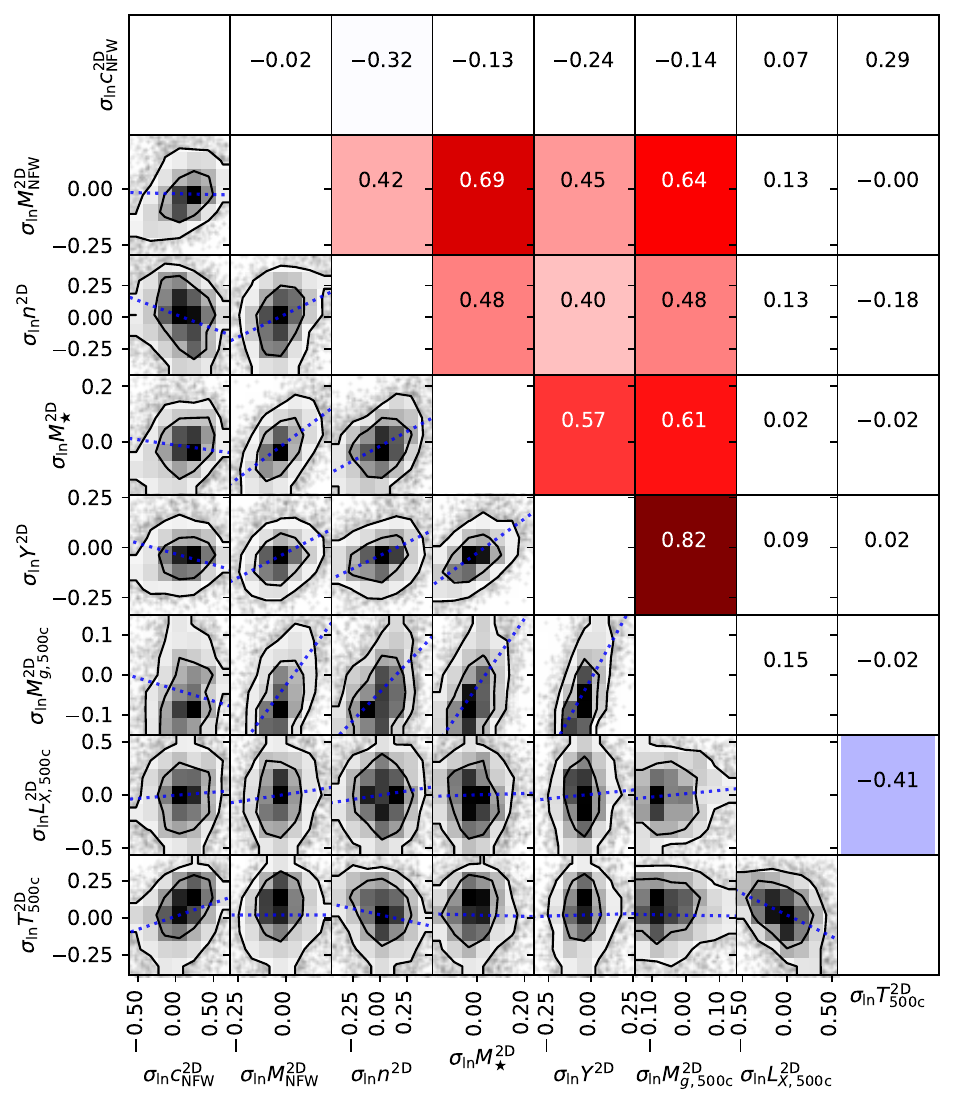}
\caption[h]{As Fig.\,\ref{fig:covarma2D},  but we show the projected X-ray luminosity, the projected gas mass, and the projected temperature computed within the overdensity of $\Delta_{\rm c}=500$ instead of the respective quantities within $\Delta_{\rm c}=200.$}
\label{fig:covarma2D500c}
\end{figure*}

While in the last sections, we investigated the origin of the impact of projection effects in the scatter and skewness of observable properties, we will now quantify how projection effects impact the correlation between observable properties.
To this end, we quantify the  Pearson correlation coefficients between their log-residuals (as defined in Sect.~\ref{sec:cumpa}).
{
We adopt the standard error associated with the Pearson coefficient $\rho$ as derived from two normal distributions, given by $\sigma_\rho = \sqrt{1-\rho^2}/\sqrt{N-2}$~\citep[see Eq. 12-93 in][]{pugh1966analysis}, where $N$ represents the number of objects. This corresponds to a maximum error of $0.015$ (for $\rho=0$) for the sample size at $z=0.24$ and a maximum error of $0.028$ for the sample size at $z=0.90$. It is worth noting that in the correlation coefficient matrices generated in subsequent analyses, we will only colour values with correlation coefficients $|\rho|>0.3$, aiming to highlight strongly correlating properties. We define mild correlation as $0.2<|\rho|<0.3$, as we choose to exercise caution. Correlation coefficients with $|\rho|<0.1$ are disregarded.
}

\subsection{Analysis at $z=0.24$}
\label{sec:z1}

In this Section we focus on the haloes at intermediate redshift $z=0.24.$
In Fig.\,\ref{fig:covarma2D} we show the correlation coefficient matrix between log-residuals at fixed halo mass of our projected observable both from \Euclid-like data (lensing concentration, lensing mass, richness, and stellar mass, respectively) and possible outcomes from multi-wavelength observations (integrated Compton-$y$ parameter, gas mass, X-ray luminosity, and temperature) for intermediate-redshift objects. 

In the lower triangle, we present scatter plots alongside the slope derived from the correlation coefficient. This visualization allows for the identification of instances where the correlation coefficient slope accurately captures the trend of the residuals. Typically, this alignment occurs for quantities that exhibit strong correlation coefficients. For example, our data points show a robust correlation between certain hot baryon tracers ($M_{\rm g}^{\rm 2D}$ and $Y^{\rm 2D}$), some cold baryon components ($n^{\rm 2D}$ and $M_\star^{\rm 2D}$), and weak lensing mass $M_{\rm NFW}^{\rm 2D}$. The underlying reason for these strong correlations lies in projection effects: the greater the amount of matter along the line of sight, the higher the observed values. We will further demonstrate this in the subsequent section by presenting the correlation coefficient matrix for 3D quantities, where many of these correlations diminish. This can be anticipated by observing that $L_x^{\rm 2D}$ and $c_{\rm NFW}^{\rm 2D}$ do not exhibit this positive trend of correlations.

Notably, we observe that the correlation between richness and stellar mass ($\rho=0.48$) is not exceptionally high. Moreover, the stellar mass appears to be more influenced by projection effects compared to richness (evident in their correlations with $M_{\rm NFW}^{\rm 2D}$, where they exhibit $\rho=0.69$ and $\rho=0.42$, respectively). We can speculate on two potential causes: firstly, unlike stellar mass, our richness computation incorporates some observationally-motivated background subtraction; alternatively, since stellar mass encompasses all stellar particles (while richness involves a luminosity-motivated galaxy stellar-mass cut), it is plausible that small subhaloes are influencing the projected stellar mass.

We note that concentration anti-correlates with gas-mass (and integrated Compton-$y$ parameter).
This is in agreement with recent analyses of simulations. In fact, richness at fixed mass anti-correlates with concentration~\citep{2019MNRAS.490.5693Bose}; low concentration is an index of the system being perturbed~\citep{2012MNRAS.427.1322Ludlow}; and un-relaxed systems tend to be gas-rich ~\citep{2020MNRAS.491.4462Davies}.
We refer to~\cite{2022A&A...666A..22Ragagnin} for a more comprehensive study on low luminous groups.
Moreover, at fixed halo mass, the lensing mass correlates strongly with total projected stellar mass ($\rho=0.69$) and projected gas mass ($\rho=0.59$), which may be due to the fact that both correlate strongly with LoS contamination. 
The same holds for the correlation among richness, gas mass, and stellar mass. This is due to projection effects, where LoS excess amplifies all these quantities, as discussed in Section \ref{sec:cumpa}.
We note that the 2D lensing mass and projected X-ray luminosity have a slight positive  ($\rho=0.20$) correlation, in agreement with the observational work of ~\cite{2020MNRAS.492.4528Sereno}.

In Fig.\,\ref{fig:covarma3D} we show the covariance matrix of non-projected quantities for intermediate redshift objects.  
We see that as opposed to Fig.\,\ref{fig:covarma2D}, the 3D covariance matrix shows a mild yet negative covariance between gas mass and stellar mass ($\rho=-0.24$), and a 
positive correlation between richness and gas mass ($\rho=0.23$) because most of their correlations in the projection are due to Line of sights excess, which significantly increases the values of the gas mass, the richness, and the stellar mass.
In Fig.\,\ref{fig:covarma2D500c}  we report the correlation matrix as in Fig.\,\ref{fig:covarma2D} where we present X-ray luminosity, gas mass, and temperature, 
as computed within $r_{\rm 500c},$ which shows an anti-correlation between the gas mass and the concentration residuals ($\rho=-0.14$) that is significantly lower than the one found in  Fig.\,\ref{fig:covarma2D} and Fig.\,\ref{fig:covarma3D}  ($\rho$ equals to $-0.26$ and $-0.34$ respectively).
One possibility is that this change in sign of the correlation is caused by the fact that mixing overdensities (concentration is within $\Delta_{\rm c}= 200$ and gas-mass is within  $\Delta_{\rm c}= 500$) does introduce an additional correlation with the sparsity~\citep{Balmes2014Sparsity,Corasaniti2022Sparsity} that itself correlates with the concentration (see Appendix \ref{ap:500}).

 For completeness, we report the correlation coefficient matrix and the scatter of log-residuals of all quantities in Table~\ref{table:z0}.
 There we also added the core-excised projected X-ray luminosity $L_{\rm X,ce500c}^{\rm 2D}$, as it is typically used in X-ray-based observational studies, where we can see that the scatter and most of the correlation coefficients are smaller than $L_{\rm ce500c}^{\rm 2D},$ while the correlations with the concentration and gas mass increase.
 Note that we do not report the 3D NFW mass ($M_{\rm NFW}$) because it has an extremely low intrinsic scatter $\sigma_{\rm ln} \left(M_{\rm NFW}\right)\approx0.01$ and its correlation coefficients are not meaningful.

\subsection{Analysis at $z=0.9$}
\label{sec:z1}

\begin{figure*}
\includegraphics[width=\linewidth]{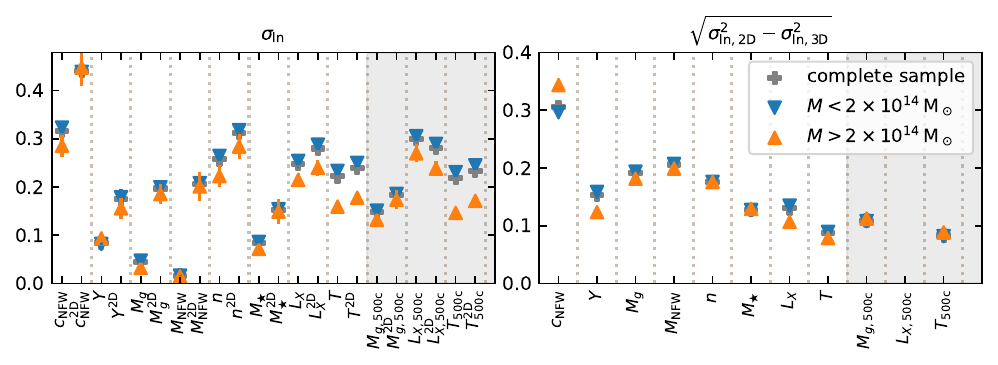}
\caption[h]{Same as Fig.\,\ref{fig:sigma} for the data at $z=0.9.$}
\label{fig:sigmaz1}
\end{figure*}

\begin{figure}
\includegraphics[width=\linewidth]{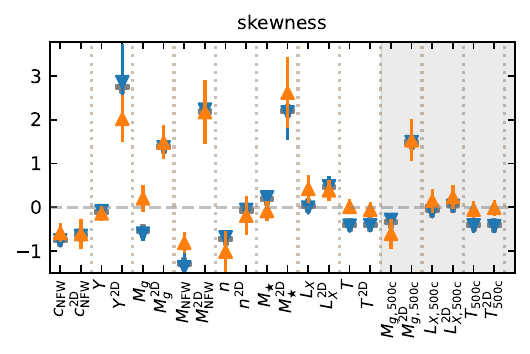}
\caption[h]{Same as Fig.\,\ref{fig:skew} for the data at $z=0.9.$ We do not report the value for the concentration because in our fit radial range the lensing one does not correlate with the 3D one.}
\label{fig:skewz1}
\end{figure}


In this Section,  we focus on observational property covariance matrixes of our haloes $z=0.9$.
At this redshift, we computed projected quantities within a cylinder depth of $35\,{\rm Mpc}$ in order to retain the same relative ratio as the photo-$z$ uncertainty of the low-redshift analysis (it scales with $1+z$).
 For what concerns the cylinder used to integrate $\Delta\Sigma_{\rm gt}$, we rescaled so as to keep it constant in comoving units with the low-redshift analyses. We rescaled the 3D NFW profile minimum radius to $40\,{\rm kpc}$ while we kept the maximum radius at $r_{\rm 200c}.$
For what concerns the radial range of the lensing fit, we rescaled it with $H^{-2/3}(z)$, therefore performing it in the range of $[234,2300]\,{\rm kpc}.$ 

{
We stress that we do not model observational uncertainty. Therefore, the decrease in background source count with redshift does not impact our best fits.
However, it still impacts the fact that we weigh external radial bins more than internal ones.
}
We report the values of the scatter and the projection contribution at $z=0.9$ in Fig.\,\ref{fig:sigmaz1}, while we report the log-residuals and the skewness for each property in Fig.\,\ref{fig:skewz1}.
 
 In particular, the quantities most affected by projection effects are the lensing mass and concentration, whereas the temperature is the lowest.
 These results are qualitatively similar to the low redshift analyses, with the Compton-$Y$ parameter and gas mass being slightly less affected by projection effects.
  Note that since the virial radius is smaller at higher redshift values, our radial range of the reduced shear is closer to the NFW scale radius; therefore, the weak lensing reconstruction is more effective in capturing the scale radius and more sensitive to deviations from an NFW profile.
   As a consequence, we found that the increase of scatter going from $c_{\rm NFW}$ to $c^{\rm 2D}_{\rm NFW}$ compared to the low redshift analyses. 

 We report the correlation coefficient matrix and the scatter log-residuals of the quantities at $z=0.9$ in Table~\ref{table:z1}.
 As for the case at $z=0.24,$ note that we do not report the 3D NFW mass ($M_{\rm NFW}$) because it has an extremely small intrinsic scatter of  $0.01,$  and thus its correlation coefficients have no impact in our study.

\section{Conclusions}
\label{sec:conclu}

In this work, we analysed a number of galaxy clusters from Magneticum hydrodynamic simulation Box2b/hr.
We did so in a mass range, tailored for \Euclid-like data products~\citep[see][]{Sartoris2016,Adam-EP3}, namely with a mass of $M_{\rm 200c}>10^{14}\,{\rm M}_\odot$. 
To this end, we computed properties that could come from \Euclid catalogues of galaxy clusters, such as richness, stellar mass, and lensing masses and concentration, and possible properties coming from multi-wavelength studies such as X-ray luminosity, integrated Compton-$y$ parameter, gas mass, and temperature.
All these properties were computed both within a sphere and within a cylinder (both with radius $r_{\rm 200c}$) to account for projection effects.
Our study considers the remarkable capabilities of \Euclid photo-$z$ measurements in identifying interlopers. However, their importance decreases significantly at scales as small as a few tens Mpc.
This depth is still long enough to contain multiple haloes along the LoS.  Hence, we studied the projection effects on a scale that is significantly smaller than the \Euclid photo-$z$ uncertainty.

We then studied how the scatter and skewness change when one measures quantities in 3D space or in projection. 
Below, we summarise our findings:

\begin{itemize}
\item The properties that are most affected by projection effects are the mass and concentration from lensing, the integrated Compton-$y$ parameter, and the gas mass. In contrast, temperature and X-ray luminosity are the quantities least affected by projection effects.
\item In both redshift slices ($z=0.24$ and $z=0.9$), the influence of LoS effects is substantial and potentially leads to a spurious correlation between gas and stellar masses. These projection effects have the capacity to markedly enhance correlations between gas and stellar mass (they go from a negative value of $-0.24$ to a significantly high value of $0.57$), effectively masking the intrinsic correlation (for instance, driven by distinct accretion histories) beneath.
\item The lensing concentration, on the other hand, is mainly affected by the fact that the profile outskirt of reduced shear deviates from the one coming from an NFW profile (which is the profile typically used in WL analyses). We found that deviations from an ideal NFW profile increase the skewness from $0.6$ to $2.5$ and increase the scatter of log-residuals from $0.33$ (in agreement with theoretical works) up to $0.46.$
\end{itemize}

The analysis presented here has been carried out using a single suite of hydrodynamic simulations.
Regarding weak lensing masses and concentration, since in this work, we did not consider the profile noise due to the finite number of background galaxies, future studies are needed to improve our estimations. 

Some works show that both scatter and correlation coefficients vary between cosmological simulations with different cosmologies~\citep{Ragagnin2023HOD}, the presence of feedback schemes~\citep{2010ApJ...715.1508Stanek}, and different cosmological simulation suite in the market~\citep[see Fig. 7 in ][]{2020MNRAS.495..686Anbajagane}. So, while simulations can provide directions on how to model correlation coefficients, it is possible that when using this kind of data, one needs to allow for variation due to the different baryon physics.

Furthermore, when striving for even more precise results, it is important to acknowledge that mass-observable relations are not exact power laws. Therefore, employing more generic fitting techniques, such as a running median, could yield improvements. Additionally, there is room for enhancement in how we compute correlation coefficients in future studies. One potential approach could involve simultaneously fitting both the mass-observable relation scatter and the correlation coefficients by maximizing multivariate likelihoods. We anticipate that future studies combining \Euclid data with multi-wavelength observations may encounter challenges in shedding light on currently puzzling residual correlations, primarily dominated by projection effects.

\begin{acknowledgements}
We thanks the anonymous referee for the useful comments.
The \textit{Magneticum Pathfinder} simulations were partially performed at the Leibniz-Rechenzentrum with CPU time assigned to the Project `pr86re'.  AR and LM acknowledge support from the grant PRIN-MIUR 2017 WSCC32 and acknowledges the usage of the INAF-OATs IT framework~\citep{2020ASPC..527..307Taffoni,2020ASPC..527..303Bertocco}, and the space filling curve improvement on  \texttt{Gadget3}~\citep{2016pcre.conf..411Ragagnin}. 
Antonio Ragagnin thanks Veronica Biffi and Elena Rasia for the X-ray computation routines and tables.
KD acknowledges support by the COMPLEX project from the European Research Council (ERC) under the European Union’s Horizon 2020 research and innovation program grant agreement ERC-2019-AdG 882679. LM acknowledges the financial contribution from the grant
PRIN-MUR 2022 20227RNLY3 “The concordance cosmological model: stress-tests with galaxy clusters” supported by Next Generation EU. 
CG and LM acknowledge support from the grant ASI n.2018-23-HH.0. 
AR and CG acknowledge funding from INAF theory Grant 2022: Illuminating Dark Matter using Weak Lensing by Cluster Satellites, PI: Carlo Giocoli. SB acknowledges partial financial support from the INFN InDark grant. AMCLB was supported by a fellowship of PSL University hosted by the Paris Observatory.
We used the package \texttt{colossus}  \citep[see][]{2018ApJS..239...35Diemer} for computing $\Sigma$ and $\Delta\Sigma$ as expected from NFW profiles. AR and FC acknowledge co-funding by the European Union
– NextGenerationEU within PRIN 2022 project n.20229YBSAN
- Globular clusters in cosmological simulations and in lensed
fields: from their birth to the present epoch.
\AckEC
\end{acknowledgements}

\section*{Data Availability}
Raw simulation data were generated at C$^2$PAP/LRZ cosmology simulation web portal \url{https://c2papcosmosim.uc.lrz.de/}. Derived data supporting the findings of this study are available from the corresponding author AR on request.

 \bibliography{referenze,Euclid} 

 \begin{appendix}

\section{Fit of gNFW}
\label{ap:gnfw}

\begin{figure}
\scalebox{.85}{\includegraphics{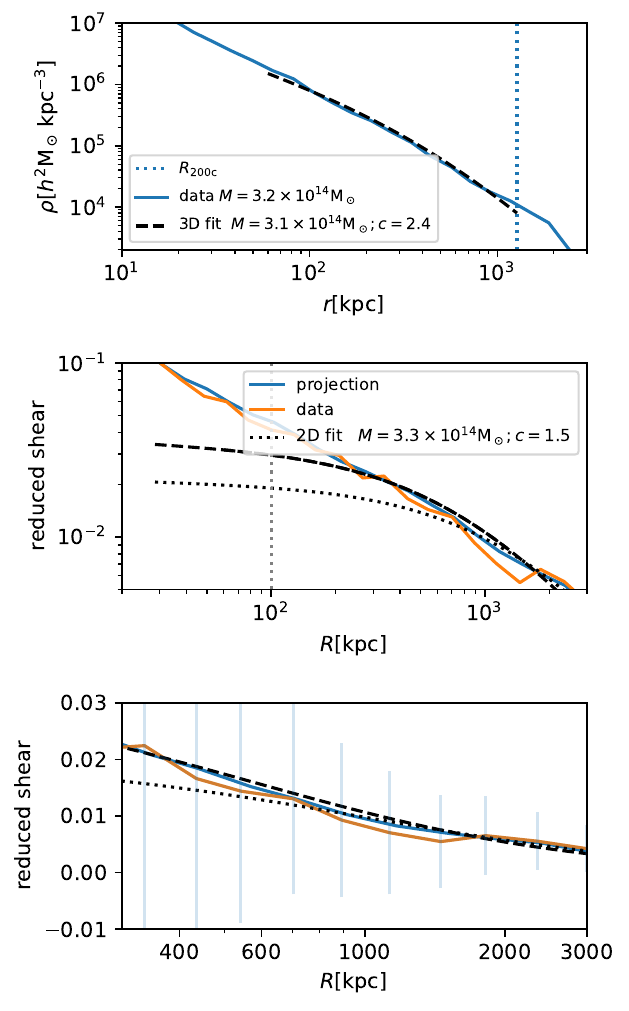}}
\caption[h]{Density profiles of a simulated halo and the corresponding  NFW profile fit. Upper panel: total matter density profile (blue solid line) and the respective NFW profile fit profile (black dashed line). Central panel: reduced shear from simulated particles (orange solid line), and from the analytical projection of the density profile $\Sigma_{\rm conv.}$ presented in Eq. \eqref{eq:sigmaconv} and performed in the radial range $[60,3000]\,{\rm kpc}$, in the blue solid line. The dashed vertical line indicates the minimum radius of the shear fit, and the fit profiles (black lines) are extrapolated down to $10\,{\rm kpc}$ to enhance the central densities predicted by the two fits. The bottom panel shows the same as the central panel but focuses on the radial range of the fit. The error bars indicate the uncertainty for each radial bin, as defined in Eq.  \eqref{eq:deltag}.}
\label{fig:example}
\end{figure}

In Fig.\,\ref{fig:example},  we present the density profile of a halo that deviates from NFW profile and has no  LoS contamination. In particular, it has $\beta\approx1.8$ and $\gamma=1.5.$
We show its NFW profile fit profile on the 3D density in the upper panel of Fig.\,\ref{fig:example}, where we can see that 3D NFW profile (performed on radial bins in a sphere) is capable of capturing the shape of the halo and to estimate its mass with high accuracy (within $\approx5\%$).
In the central panel, we show the reduced shear profile and best fits, where we can see that the fit performed on the reduced shear underestimated the concentration and is not able to capture the more internal part of the shear profile, as it is done by the profile that was fit in 3D.

We first exclude this mismatch as being due to projection effects by showing that both the reduced shear from the particle data (orange line) matches the one recovered by performing an analytical projection of the 3D profile (blue solid line). 
In particular we project the density profile $\rho(r)$ and derive the surface mass density $\Sigma_{\rm conv.},$ as follows
\begin{equation}
\label{eq:sigmaconv}
\Sigma_{\rm conv.}(R) = \int_{-\infty}^{+\infty} \rho\left(\sqrt{R^2+z^2}\right)\, {\rm d}z=2\,\int_{R}^{\infty}\rho(r)\,\frac{r}{\sqrt{r^2 - R^2}}\, {\rm d}r \ .
\end{equation}
 What we find is that the shear obtained with the aid of an analytical projection $\Sigma_{\rm conv.}$ matches very well the real one (i.e., orange and blue lines do match).
 This hints that for this cluster there are no strong LoS effects.
 
To understand why the fit on the shear is not able to capture the concentration of the original halo, we zoom our fit in the bottom panel of Fig.\,\ref{fig:example},  where it looks like the fit is very good in capturing the final part of the profile and not able to capture the internal.
It is crucial to emphasise that we did not include observational uncertainties in these analyses. Therefore, the uncertainty outlined in Eq.\,\eqref{eq:deltag} affects the fit by assigning more weight to external radial bins compared to internal ones. It is worth noting that the proportionality factors in Eq.\,\eqref{eq:deltag} will not affect our best fit.

\begin{figure}
\scalebox{.85}{\includegraphics{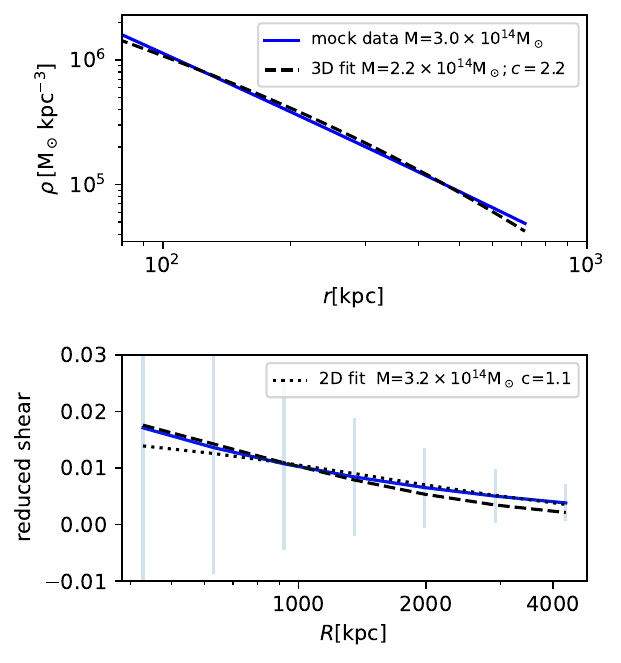}}
\caption[h]{
Density profiles of a mock halo that deviates from NFW and the corresponding  NFW profile fit.
The mock halo mass, concentration parameter and gNFW log-slopes are chosen to match the ones of the simulated halo presented in Fig.\,\ref{fig:example}.
The upper panel reports the total matter density profile of the mock halo (solid blue line) and the profile from the corresponding NFW profile fit (dashed black line). The bottom panel shows the reduced shear and the profile from the corresponding NFW profile fit (dotted black line). 
The error bars indicate the uncertainty for each radial bin, as defined in Eq. \eqref{eq:deltag}.}
\label{fig:colossus}
\end{figure}

To validate this point in Fig.\,\ref{fig:colossus} we study the bias on fitting an NFW profile on a mock gNFW profile that has $\beta=1.8$ and $\gamma=1.5,$ a mass of $3\times10^{14}M_\odot$ and a concentration $c=2.4$ (as the halo presented in Fig.\,\ref{fig:example}).  We see that the 3D NFW profile is capable of estimating both its mass and its gNFW concentration with high accuracy (see top panel match between blue and dashed black lines).
On the other hand the fit of the shear (we report in the bottom panel of Fig.  \ref{fig:colossus})  has the same problems as the one on the cluster in Fig.\,\ref{fig:example}: it recovers a low concentration (with a value of $1.5$).
This may be because, at outer radii, the model fits the data.
It is possible that the under-estimation of concentration at low radii is caused by the combination of two factors: the fit under-estimates the shear at lower radii (with the result of under-estimating the lensing concentration), or the fact that $\gamma$ is different than $3$ induces an NFW profile fit with a low concentration.

We then performed the experiment of fitting the analytical profile with constant (yet unrealistic) error bars.
While the fit was able to capture the shape of the profile, it recovered a concentration of $1.6,$  implying that there is indeed a degeneracy between the shear of low-concentrated NFW profiles and steeper-NFW profiles.

\section{Correlations with different overdensities}
\label{ap:500}

\begin{figure}
\includegraphics[width=\linewidth]{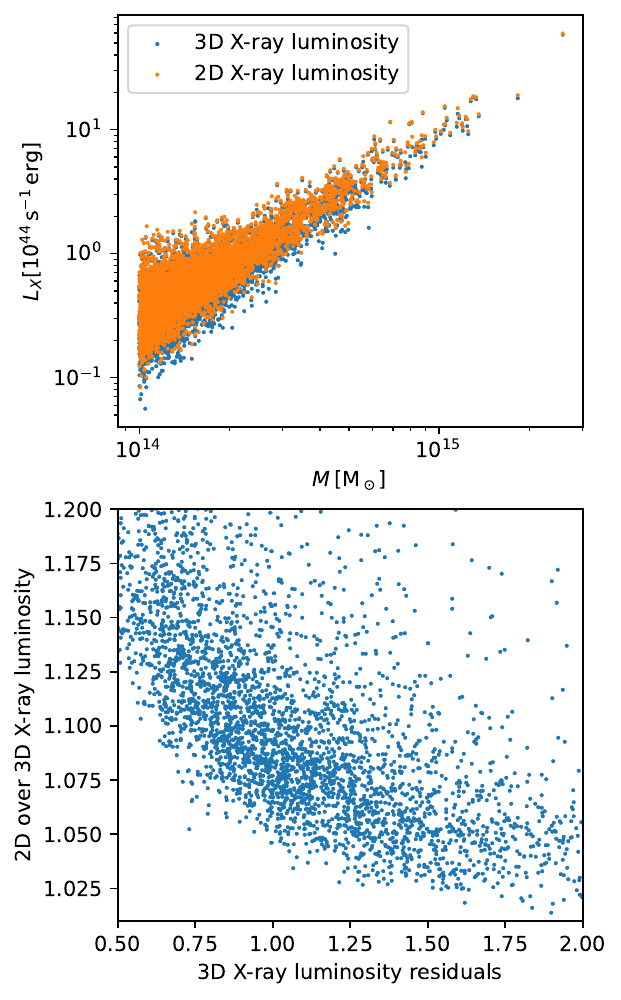}
\caption[h]{Comparison between 3D and projected X-ray luminosities. Top panel shows a scatter plot of the two mass-observable relations, while the bottom panel shows their ratio as a function of the 3D X-ray scaling relation residuals.}
\label{fig:LX3D_vs_2D}
\end{figure}

\begin{figure}
\includegraphics[width=\linewidth]{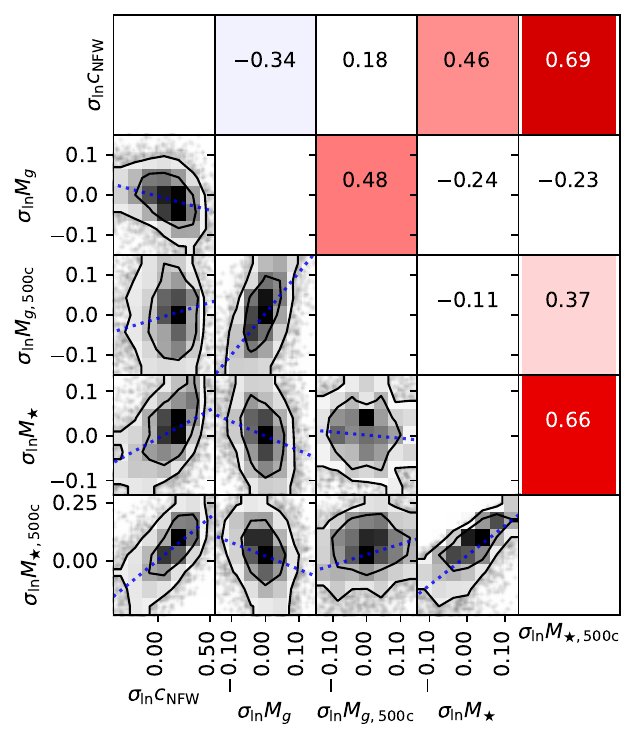}
\caption[h]{We report the correlation coefficient between concentration, gas mass and stellar mass computed at both $r_{\rm 500c}$ and $r_{\rm 200c}.$}
\label{fig:covarma_500_vs_200}
\end{figure}

In this appendix, we discuss the differences between scaling relation scatters and covariance values at different overdensities. 
First of all, we tackle the fact that when we compute X-ray luminosity within $r_{\rm 500c}$ (instead of $r_{\rm 200c}$), we find that the scatter of the scaling relation of the projected quantity is larger than the 3D one.

To investigate this feature, we will focus on the bolometric X-ray luminosity. We report the 3D and projected bolometric X-ray luminosity in Fig. ~\ref{fig:LX3D_vs_2D} (top panel), where it is visually clear that the projected X-ray luminosity is (as expected) always larger than the 3D one. One can also notice that the increase in X-ray luminosity depends on the fact that a halo is over-luminous or not: the increase of luminosity growing from the 3D to 2D is larger for under-luminous haloes than for over-luminous haloes. We prove this point in the bottom panel of Fig.~\ref{fig:LX3D_vs_2D} where we show the ratio between the 2D and 3D luminosity as a function of their residual of the 3D scaling relation (the higher the value of the $x$ axis, the more over-luminous is the object for its mass bin), where we can see a strong anti-correlation: overly luminous objects (for a given mass bin) are not going to be affected much by the fact that their luminosity is computed in 3D or 2D. The possible cause is that an interloper in the LoS will not affect much an overly luminous object.  

For completeness, in Fig.\,\ref{fig:covarma_500_vs_200} we show the correlation coefficients between the gas mass and stellar mass computed within both $r_{\rm 500c}$ and $r_{\rm 200c}$ and the concentration.
Here we can see a change of sign between $M_{\star,\rm 500c}$-$M_{g,\rm 500c}$ correlations and $M_{\star,\rm 500c}$-$M_{\rm g}$ correlations and a change in the sign between $c_{\rm NFW}$-$M_{\rm g}$ correlations and $c_{\rm NFW}$-$M_{g,\rm500c}$  correlations.

\begin{figure*}
\includegraphics[width=\linewidth]{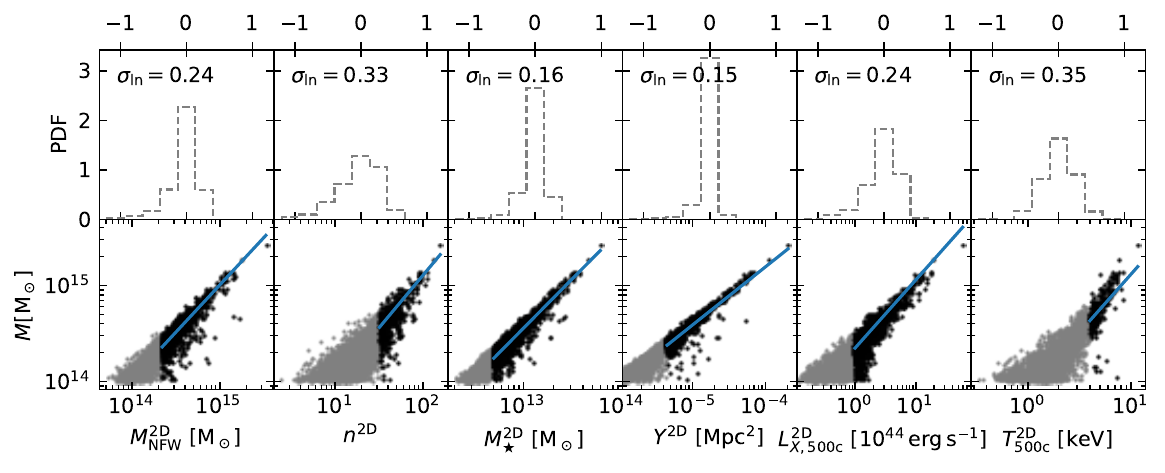}
\caption[h]{Halo masses at fixed observable properties.
We report the lensing mass, lensing richness, projected stellar mass, projected integrated Compton-$y$, projected X-ray luminosity,  and projected temperature in each column, respectively. The top panel shows residuals of the observable-mass relations and respective scatter of log-residuals $\sigma_{\rm ln},$ and its axes are on the upper part of the plot. The bottom panel shows the scaling relation fit (blue solid line); the data used to perform the fit (black data points) over-plotted on top of the total sample (grey data points) of the mass $M$ as a function of the observable properties. 
}
\label{fig:obs_scatter}
\end{figure*}

Finally, in Fig.\,\ref{fig:obs_scatter} we report the scatter of observable properties at fixed mass for both \Euclid-like quantities (lensing mass, richness, and projected stellar mass), and possible multi-wavelength properties (integrated Compton-$y$ parameter, X-ray luminosity, and temperature), where we compute X-ray luminosity and temperature within $r_{\rm 500c}$ as they are typically derived within this overdensity.
The upper panel of Fig.\,\ref{fig:obs_scatter} shows the residuals of the log-log linear regression where we see that in terms of 2D scatter, the properties with the lowest scatter are the stellar mass and the temperature.
The bottom panel shows the data points used to perform the fit (in black) where we used a visually-inspected cut on the halo mass values in order to ensure that mass values are complete for a given observable value.

\begin{table*}
\caption{Scatter and correlation coefficient matrix between $z=0.24$ log-residuals of scaling relations.}              
\label{table:z0}      
\centering   
\small
\setlength\tabcolsep{1.4pt}
\begin{tabular}{l  r  r  r  r  r  r  r  r  r  r  r  r  r  r  r  r  r  r  r  r  r  r }
 \hline\hline & 
\rule{0pt}{1\normalbaselineskip} $c_{\rm NFW}$ & $c^{\rm 2D}_{\rm NFW}$ & $M^{\rm 2D}_{\rm NFW}$ & $n$ & $n^{\rm 2D}$ & $M_{\bigstar}$ & $M_{\bigstar}^{\rm 2D}$ & $Y$ & $Y^{\rm 2D}$ & $M_g$ & $M_g^{\rm 2D}$ & $M_{g,{\rm 500c}}$ & $M^{\rm 2D}_{g,{\rm 500c}}$ & $L_X$ & $L_X^{\rm 2D}$ & $L_{X,{\rm 500c}}$ & $L^{\rm 2D}_{X,{\rm 500c}}$ & $L^{\rm 2D}_{X,ce{\rm 500c}}$ & $T$ & $T^{\rm 2D}$ & $T_{\rm 500c}$ & $T^{\rm 2D}_{\rm 500c}$ \\ \hline   
$c_{\rm NFW}$ &  $0.33$&     &     &     &     &     &     &     &     &     &     &     &     &     &     &     &     &     &     &     &     &      \\ 
$c^{\rm 2D}_{\rm NFW}$ &  $0.60$& $0.45$&     &     &     &     &     &     &     &     &     &     &     &     &     &     &     &     &     &     &     &      \\ 
$M^{\rm 2D}_{\rm NFW}$ &    $--$  &   $--$  & $0.18$&     &     &     &     &     &     &     &     &     &     &     &     &     &     &     &     &     &     &      \\ 
$n$ & $-0.28$&$-0.22$&   $--$  & $0.28$&     &     &     &     &     &     &     &     &     &     &     &     &     &     &     &     &     &      \\ 
$n^{\rm 2D}$ &    $--$  &$-0.32$& $0.42$& $0.71$& $0.31$&     &     &     &     &     &     &     &     &     &     &     &     &     &     &     &     &      \\ 
$M_{\bigstar}$ &  $0.46$& $0.23$&   $--$  &   $--$  &   $--$  & $0.09$&     &     &     &     &     &     &     &     &     &     &     &     &     &     &     &      \\ 
$M_{\bigstar}^{\rm 2D}$ &  $0.23$&   $--$  & $0.69$&   $--$  & $0.48$& $0.59$& $0.13$&     &     &     &     &     &     &     &     &     &     &     &     &     &     &      \\ 
$Y$ &    $--$  &   $--$  &   $--$  &   $--$  &   $--$  &   $--$  &   $--$  & $0.09$&     &     &     &     &     &     &     &     &     &     &     &     &     &      \\ 
$Y^{\rm 2D}$ &    $--$  &$-0.24$& $0.45$&   $--$  & $0.40$&   $--$  & $0.57$& $0.44$& $0.20$&     &     &     &     &     &     &     &     &     &     &     &     &      \\ 
$M_g$ & $-0.34$&$-0.21$&   $--$  & $0.23$& $0.23$&$-0.24$&   $--$  & $0.59$& $0.26$& $0.07$&     &     &     &     &     &     &     &     &     &     &     &      \\ 
$M_g^{\rm 2D}$ &    $--$  &$-0.26$& $0.59$&   $--$  & $0.52$&   $--$  & $0.63$& $0.21$& $0.85$& $0.21$& $0.20$&     &     &     &     &     &     &     &     &     &     &      \\ 
$M_{g,{\rm 500c}}$ &    $--$  & $0.35$&   $--$  &   $--$  &   $--$  &   $--$  &   $--$  & $0.54$&   $--$  & $0.48$&   $--$  & $0.14$&     &     &     &     &     &     &     &     &     &      \\ 
$M^{\rm 2D}_{g,{\rm 500c}}$ &    $--$  &   $--$  & $0.64$&   $--$  & $0.48$&   $--$  & $0.61$& $0.28$& $0.82$& $0.22$& $0.97$&   $--$  & $0.18$&     &     &     &     &     &     &     &     &      \\ 
$L_X$ &    $--$  &   $--$  &   $--$  &   $--$  &   $--$  &   $--$  &   $--$  & $0.20$&   $--$  & $0.40$&   $--$  & $0.36$&   $--$  & $0.35$&     &     &     &     &     &     &     &      \\ 
$L_X^{\rm 2D}$ &    $--$  &   $--$  & $0.20$&   $--$  & $0.26$&   $--$  &   $--$  & $0.20$&   $--$  & $0.37$& $0.23$& $0.31$& $0.26$& $0.91$& $0.35$&     &     &     &     &     &     &      \\ 
$L_{X,{\rm 500c}}$ &    $--$  &   $--$  &   $--$  &   $--$  &   $--$  &   $--$  &   $--$  & $0.29$&   $--$  & $0.33$&   $--$  & $0.55$&   $--$  & $0.91$& $0.82$& $0.41$&     &     &     &     &     &      \\ 
$L^{\rm 2D}_{X,{\rm 500c}}$ &    $--$  &   $--$  &   $--$  &   $--$  &   $--$  &   $--$  &   $--$  & $0.28$&   $--$  & $0.35$&   $--$  & $0.47$&   $--$  & $0.93$& $0.91$& $0.94$& $0.38$&     &     &     &     &      \\ 
$L^{\rm 2D}_{X,ce{\rm 500c}}$ & $-0.28$&   $--$  &   $--$  &   $--$  &   $--$  &$-0.24$&   $--$  & $0.28$&   $--$  & $0.49$&   $--$  & $0.56$&   $--$  & $0.59$& $0.62$& $0.52$& $0.66$& $0.33$&     &     &     &      \\ 
$T$ &  $0.40$& $0.30$&   $--$  &$-0.20$&   $--$  &   $--$  &   $--$  & $0.31$&   $--$  &   $--$  &   $--$  & $0.28$&   $--$  &$-0.52$&$-0.48$&$-0.32$&$-0.40$&   $--$  & $0.26$&     &     &      \\ 
$T^{\rm 2D}$ &  $0.39$& $0.30$&   $--$  &   $--$  &   $--$  &   $--$  &   $--$  & $0.31$&   $--$  &   $--$  &   $--$  & $0.30$&   $--$  &$-0.44$&$-0.53$&$-0.26$&$-0.37$&   $--$  & $0.94$& $0.27$&     &      \\ 
$T_{\rm 500c}$ &  $0.38$& $0.27$&   $--$  &   $--$  &   $--$  &   $--$  &   $--$  & $0.26$&   $--$  &   $--$  &   $--$  & $0.21$&   $--$  &$-0.53$&$-0.49$&$-0.44$&$-0.46$&   $--$  & $0.94$& $0.88$& $0.28$&      \\ 
$T^{\rm 2D}_{\rm 500c}$ &  $0.38$& $0.29$&   $--$  &   $--$  &   $--$  &   $--$  &   $--$  & $0.30$&   $--$  &   $--$  &   $--$  & $0.28$&   $--$  &$-0.47$&$-0.52$&$-0.29$&$-0.41$&$-0.21$& $0.95$& $0.98$& $0.91$& $0.27$ \\ 
\hline\end{tabular}
\tablefoot{Diagonal terms report the scatter of the log-residuals of each quantity, namely $\sigma_{\rm ln}$ of Eq.~\eqref{eq:sigmaln}, while the off-diagonal terms report the correlation coefficient between the log-residuals. We do not report values of the correlation coefficient below $0.20$ because they are not significant. We do not report the values for the 3D NFW mass $M_{\rm 200c}$ because it has a very low scatter of log-residuals ($\approx0.01$) and its correlation coefficients are not meaningful.
Note that in this table we also added the core-excised X-ray luminosity.}
\end{table*}

 \begin{table*}
\caption{Scatter and correlation coefficient matrix between $z=0.9$ log-residuals of scaling relations.}              
\label{table:z1}      
\centering   
\small
\setlength\tabcolsep{1.4pt}
\begin{tabular}{l  r  r  r  r  r  r  r  r  r  r  r  r  r  r  r  r  r  r  r  r  r  r }
 \hline\hline & 
\rule{0pt}{1\normalbaselineskip} $c_{\rm NFW}$ & $c^{\rm 2D}_{\rm NFW}$ & $M^{\rm 2D}_{\rm NFW}$ & $n$ & $n^{\rm 2D}$ & $M_{\bigstar}$ & $M_{\bigstar}^{\rm 2D}$ & $Y$ & $Y^{\rm 2D}$ & $M_g$ & $M_g^{\rm 2D}$ & $M_{g,{\rm 500c}}$ & $M^{\rm 2D}_{g,{\rm 500c}}$ & $L_X$ & $L_X^{\rm 2D}$ & $L_{X,{\rm 500c}}$ & $L^{\rm 2D}_{X,{\rm 500c}}$ & $L^{\rm 2D}_{X,ce{\rm 500c}}$ & $T$ & $T^{\rm 2D}$ & $T_{\rm 500c}$ & $T^{\rm 2D}_{\rm 500c}$ \\ \hline   
$c_{\rm NFW}$ &  $0.32$&     &     &     &     &     &     &     &     &     &     &     &     &     &     &     &     &     &     &     &     &      \\ 
$c^{\rm 2D}_{\rm NFW}$ &  $0.51$& $0.44$&     &     &     &     &     &     &     &     &     &     &     &     &     &     &     &     &     &     &     &      \\ 
$M^{\rm 2D}_{\rm NFW}$ &    $--$  &   $--$  & $0.21$&     &     &     &     &     &     &     &     &     &     &     &     &     &     &     &     &     &     &      \\ 
$n$ & $-0.28$&   $--$  &   $--$  & $0.26$&     &     &     &     &     &     &     &     &     &     &     &     &     &     &     &     &     &      \\ 
$n^{\rm 2D}$ & $-0.22$&$-0.34$& $0.51$& $0.70$& $0.31$&     &     &     &     &     &     &     &     &     &     &     &     &     &     &     &     &      \\ 
$M_{\bigstar}$ &  $0.37$&   $--$  &   $--$  &   $--$  &   $--$  & $0.08$&     &     &     &     &     &     &     &     &     &     &     &     &     &     &     &      \\ 
$M_{\bigstar}^{\rm 2D}$ &    $--$  &$-0.31$& $0.78$&   $--$  & $0.57$& $0.47$& $0.15$&     &     &     &     &     &     &     &     &     &     &     &     &     &     &      \\ 
$Y$ &  $0.34$& $0.27$&   $--$  &$-0.23$&   $--$  &   $--$  &   $--$  & $0.08$&     &     &     &     &     &     &     &     &     &     &     &     &     &      \\ 
$Y^{\rm 2D}$ &    $--$  &$-0.30$& $0.63$&   $--$  & $0.32$&   $--$  & $0.64$& $0.41$& $0.17$&     &     &     &     &     &     &     &     &     &     &     &     &      \\ 
$M_g$ & $-0.29$&   $--$  &   $--$  & $0.23$& $0.21$&$-0.30$&   $--$  & $0.40$& $0.20$& $0.05$&     &     &     &     &     &     &     &     &     &     &     &      \\ 
$M_g^{\rm 2D}$ &    $--$  &$-0.38$& $0.78$&   $--$  & $0.53$&   $--$  & $0.71$&   $--$  & $0.79$&   $--$  & $0.20$&     &     &     &     &     &     &     &     &     &     &      \\ 
$M_{g,{\rm 500c}}$ &    $--$  & $0.41$&   $--$  &   $--$  &   $--$  &$-0.28$&$-0.25$& $0.52$&   $--$  & $0.44$&   $--$  & $0.15$&     &     &     &     &     &     &     &     &     &      \\ 
$M^{\rm 2D}_{g,{\rm 500c}}$ &    $--$  &$-0.24$& $0.80$&   $--$  & $0.49$&   $--$  & $0.67$&   $--$  & $0.77$&   $--$  & $0.96$&   $--$  & $0.18$&     &     &     &     &     &     &     &     &      \\ 
$L_X$ & $-0.20$&   $--$  &   $--$  & $0.26$&   $--$  &   $--$  &   $--$  &   $--$  &   $--$  & $0.51$&   $--$  & $0.51$&   $--$  & $0.25$&     &     &     &     &     &     &     &      \\ 
$L_X^{\rm 2D}$ &    $--$  &   $--$  & $0.42$& $0.26$& $0.48$&   $--$  & $0.32$&   $--$  & $0.22$& $0.44$& $0.39$& $0.35$& $0.44$& $0.79$& $0.28$&     &     &     &     &     &     &      \\ 
$L_{X,{\rm 500c}}$ &    $--$  & $0.26$&   $--$  &   $--$  &   $--$  &   $--$  &   $--$  & $0.32$&   $--$  & $0.45$&   $--$  & $0.73$&   $--$  & $0.89$& $0.67$& $0.30$&     &     &     &     &     &      \\ 
$L^{\rm 2D}_{X,{\rm 500c}}$ &    $--$  &   $--$  & $0.23$& $0.23$& $0.29$&$-0.23$&   $--$  & $0.23$&   $--$  & $0.50$&   $--$  & $0.59$& $0.22$& $0.87$& $0.86$& $0.87$& $0.28$&     &     &     &     &      \\ 
$L^{\rm 2D}_{X,ce{\rm 500c}}$ & $-0.31$&   $--$  & $0.25$& $0.22$& $0.31$&$-0.30$&   $--$  &   $--$  &   $--$  & $0.49$&   $--$  & $0.58$& $0.24$& $0.69$& $0.75$& $0.65$& $0.85$& $0.31$&     &     &     &      \\ 
$T$ &  $0.52$& $0.40$&   $--$  &$-0.22$&   $--$  &   $--$  &   $--$  & $0.52$&   $--$  &   $--$  &   $--$  & $0.33$&   $--$  &   $--$  &   $--$  &   $--$  &   $--$  &   $--$  & $0.22$&     &     &      \\ 
$T^{\rm 2D}$ &  $0.47$& $0.43$&   $--$  &$-0.22$&$-0.28$&   $--$  &   $--$  & $0.50$&   $--$  &   $--$  &$-0.20$& $0.37$&   $--$  &   $--$  &$-0.28$&   $--$  &   $--$  &   $--$  & $0.90$& $0.24$&     &      \\ 
$T_{\rm 500c}$ &  $0.52$& $0.35$&   $--$  &$-0.21$&   $--$  &   $--$  &   $--$  & $0.45$&   $--$  &   $--$  &   $--$  & $0.21$&   $--$  &   $--$  &   $--$  &   $--$  &   $--$  &   $--$  & $0.95$& $0.84$& $0.22$&      \\ 
$T^{\rm 2D}_{\rm 500c}$ &  $0.48$& $0.41$&   $--$  &$-0.21$&$-0.27$&   $--$  &   $--$  & $0.49$&   $--$  &   $--$  &   $--$  & $0.34$&   $--$  &   $--$  &$-0.27$&   $--$  &   $--$  &   $--$  & $0.90$& $0.98$& $0.86$& $0.23$ \\ 
\hline\end{tabular}
\tablefoot{Rows and columns are as in Table~\ref{table:z0}.}
\end{table*}
 \end{appendix}
\end{document}

%% file: authorlist.tex
\newcommand{\orcid}[1]{} 
\author{Euclid Collaboration: A.~Ragagnin\orcid{0000-0002-8106-2742}\thanks{\email{antonio.ragagnin@unibo.it}}\inst{\ref{aff1},\ref{aff2},\ref{aff3},\ref{aff4}}
\and A.~Saro\orcid{0000-0002-9288-862X}\inst{\ref{aff5},\ref{aff2},\ref{aff6},\ref{aff7},\ref{aff4}}
\and S.~Andreon\orcid{0000-0002-2041-8784}\inst{\ref{aff8}}
\and A.~Biviano\orcid{0000-0002-0857-0732}\inst{\ref{aff6},\ref{aff2}}
\and K.~Dolag\inst{\ref{aff9}}
\and S.~Ettori\orcid{0000-0003-4117-8617}\inst{\ref{aff1},\ref{aff10}}
\and C.~Giocoli\orcid{0000-0002-9590-7961}\inst{\ref{aff1},\ref{aff11}}
\and A.~M.~C.~Le~Brun\orcid{0000-0002-0936-4594}\inst{\ref{aff12}}
\and G.~A.~Mamon\orcid{0000-0001-8956-5953}\inst{\ref{aff13},\ref{aff14}}
\and B.~J.~Maughan\orcid{0000-0003-0791-9098}\inst{\ref{aff15}}
\and M.~Meneghetti\orcid{0000-0003-1225-7084}\inst{\ref{aff1},\ref{aff16}}
\and L.~Moscardini\orcid{0000-0002-3473-6716}\inst{\ref{aff3},\ref{aff1},\ref{aff16}}
\and F.~Pacaud\orcid{0000-0002-6622-4555}\inst{\ref{aff17}}
\and G.~W.~Pratt\inst{\ref{aff18}}
\and M.~Sereno\orcid{0000-0003-0302-0325}\inst{\ref{aff1},\ref{aff16}}
\and S.~Borgani\orcid{0000-0001-6151-6439}\inst{\ref{aff5},\ref{aff2},\ref{aff6},\ref{aff7}}
\and F.~Calura\orcid{0000-0002-6175-0871}\inst{\ref{aff1}}
\and G.~Castignani\orcid{0000-0001-6831-0687}\inst{\ref{aff1}}
\and M.~De~Petris\orcid{0000-0001-7859-2139}\inst{\ref{aff19}}
\and D.~Eckert\orcid{0000-0001-7917-3892}\inst{\ref{aff20}}
\and G.~F.~Lesci\orcid{0000-0002-4607-2830}\inst{\ref{aff3},\ref{aff1}}
\and J.~Macias-Perez\orcid{0000-0002-5385-2763}\inst{\ref{aff21}}
\and M.~Maturi\orcid{0000-0002-3517-2422}\inst{\ref{aff22},\ref{aff23}}
\and A.~Amara\inst{\ref{aff24}}
\and N.~Auricchio\orcid{0000-0003-4444-8651}\inst{\ref{aff1}}
\and C.~Baccigalupi\orcid{0000-0002-8211-1630}\inst{\ref{aff2},\ref{aff6},\ref{aff7},\ref{aff25}}
\and M.~Baldi\orcid{0000-0003-4145-1943}\inst{\ref{aff26},\ref{aff1},\ref{aff16}}
\and S.~Bardelli\orcid{0000-0002-8900-0298}\inst{\ref{aff1}}
\and D.~Bonino\orcid{0000-0002-3336-9977}\inst{\ref{aff27}}
\and E.~Branchini\orcid{0000-0002-0808-6908}\inst{\ref{aff28},\ref{aff29},\ref{aff8}}
\and M.~Brescia\orcid{0000-0001-9506-5680}\inst{\ref{aff30},\ref{aff31},\ref{aff32}}
\and J.~Brinchmann\orcid{0000-0003-4359-8797}\inst{\ref{aff33}}
\and S.~Camera\orcid{0000-0003-3399-3574}\inst{\ref{aff34},\ref{aff35},\ref{aff27}}
\and V.~Capobianco\orcid{0000-0002-3309-7692}\inst{\ref{aff27}}
\and C.~Carbone\orcid{0000-0003-0125-3563}\inst{\ref{aff36}}
\and J.~Carretero\orcid{0000-0002-3130-0204}\inst{\ref{aff37},\ref{aff38}}
\and S.~Casas\orcid{0000-0002-4751-5138}\inst{\ref{aff39}}
\and M.~Castellano\orcid{0000-0001-9875-8263}\inst{\ref{aff40}}
\and S.~Cavuoti\orcid{0000-0002-3787-4196}\inst{\ref{aff31},\ref{aff32}}
\and A.~Cimatti\inst{\ref{aff41}}
\and C.~Colodro-Conde\inst{\ref{aff42}}
\and G.~Congedo\orcid{0000-0003-2508-0046}\inst{\ref{aff43}}
\and C.~J.~Conselice\orcid{0000-0003-1949-7638}\inst{\ref{aff44}}
\and L.~Conversi\orcid{0000-0002-6710-8476}\inst{\ref{aff45},\ref{aff46}}
\and Y.~Copin\orcid{0000-0002-5317-7518}\inst{\ref{aff47}}
\and F.~Courbin\orcid{0000-0003-0758-6510}\inst{\ref{aff48}}
\and H.~M.~Courtois\orcid{0000-0003-0509-1776}\inst{\ref{aff49}}
\and A.~Da~Silva\orcid{0000-0002-6385-1609}\inst{\ref{aff50},\ref{aff51}}
\and H.~Degaudenzi\orcid{0000-0002-5887-6799}\inst{\ref{aff20}}
\and G.~De~Lucia\orcid{0000-0002-6220-9104}\inst{\ref{aff6}}
\and J.~Dinis\orcid{0000-0001-5075-1601}\inst{\ref{aff50},\ref{aff51}}
\and F.~Dubath\orcid{0000-0002-6533-2810}\inst{\ref{aff20}}
\and X.~Dupac\inst{\ref{aff46}}
\and M.~Farina\orcid{0000-0002-3089-7846}\inst{\ref{aff52}}
\and S.~Farrens\orcid{0000-0002-9594-9387}\inst{\ref{aff18}}
\and S.~Ferriol\inst{\ref{aff47}}
\and M.~Frailis\orcid{0000-0002-7400-2135}\inst{\ref{aff6}}
\and E.~Franceschi\orcid{0000-0002-0585-6591}\inst{\ref{aff1}}
\and M.~Fumana\orcid{0000-0001-6787-5950}\inst{\ref{aff36}}
\and K.~George\orcid{0000-0002-1734-8455}\inst{\ref{aff9}}
\and B.~Gillis\orcid{0000-0002-4478-1270}\inst{\ref{aff43}}
\and A.~Grazian\orcid{0000-0002-5688-0663}\inst{\ref{aff53}}
\and F.~Grupp\inst{\ref{aff54},\ref{aff9}}
\and S.~V.~H.~Haugan\orcid{0000-0001-9648-7260}\inst{\ref{aff55}}
\and W.~Holmes\inst{\ref{aff56}}
\and I.~Hook\orcid{0000-0002-2960-978X}\inst{\ref{aff57}}
\and F.~Hormuth\inst{\ref{aff58}}
\and A.~Hornstrup\orcid{0000-0002-3363-0936}\inst{\ref{aff59},\ref{aff60}}
\and K.~Jahnke\orcid{0000-0003-3804-2137}\inst{\ref{aff61}}
\and E.~Keih\"anen\orcid{0000-0003-1804-7715}\inst{\ref{aff62}}
\and S.~Kermiche\orcid{0000-0002-0302-5735}\inst{\ref{aff63}}
\and A.~Kiessling\orcid{0000-0002-2590-1273}\inst{\ref{aff56}}
\and M.~Kilbinger\orcid{0000-0001-9513-7138}\inst{\ref{aff18}}
\and B.~Kubik\orcid{0009-0006-5823-4880}\inst{\ref{aff47}}
\and M.~K\"ummel\orcid{0000-0003-2791-2117}\inst{\ref{aff9}}
\and M.~Kunz\orcid{0000-0002-3052-7394}\inst{\ref{aff64}}
\and H.~Kurki-Suonio\orcid{0000-0002-4618-3063}\inst{\ref{aff65},\ref{aff66}}
\and S.~Ligori\orcid{0000-0003-4172-4606}\inst{\ref{aff27}}
\and P.~B.~Lilje\orcid{0000-0003-4324-7794}\inst{\ref{aff55}}
\and V.~Lindholm\orcid{0000-0003-2317-5471}\inst{\ref{aff65},\ref{aff66}}
\and I.~Lloro\inst{\ref{aff67}}
\and D.~Maino\inst{\ref{aff68},\ref{aff36},\ref{aff69}}
\and E.~Maiorano\orcid{0000-0003-2593-4355}\inst{\ref{aff1}}
\and O.~Mansutti\orcid{0000-0001-5758-4658}\inst{\ref{aff6}}
\and O.~Marggraf\orcid{0000-0001-7242-3852}\inst{\ref{aff17}}
\and K.~Markovic\orcid{0000-0001-6764-073X}\inst{\ref{aff56}}
\and M.~Martinelli\orcid{0000-0002-6943-7732}\inst{\ref{aff40},\ref{aff70}}
\and N.~Martinet\orcid{0000-0003-2786-7790}\inst{\ref{aff71}}
\and F.~Marulli\orcid{0000-0002-8850-0303}\inst{\ref{aff3},\ref{aff1},\ref{aff16}}
\and R.~Massey\orcid{0000-0002-6085-3780}\inst{\ref{aff72}}
\and S.~Maurogordato\inst{\ref{aff73}}
\and E.~Medinaceli\orcid{0000-0002-4040-7783}\inst{\ref{aff1}}
\and S.~Mei\orcid{0000-0002-2849-559X}\inst{\ref{aff74}}
\and Y.~Mellier\inst{\ref{aff13},\ref{aff14}}
\and G.~Meylan\inst{\ref{aff48}}
\and M.~Moresco\orcid{0000-0002-7616-7136}\inst{\ref{aff3},\ref{aff1}}
\and E.~Munari\orcid{0000-0002-1751-5946}\inst{\ref{aff6},\ref{aff2}}
\and C.~Neissner\orcid{0000-0001-8524-4968}\inst{\ref{aff75},\ref{aff38}}
\and S.-M.~Niemi\inst{\ref{aff76}}
\and J.~W.~Nightingale\orcid{0000-0002-8987-7401}\inst{\ref{aff77},\ref{aff72}}
\and C.~Padilla\orcid{0000-0001-7951-0166}\inst{\ref{aff75}}
\and S.~Paltani\orcid{0000-0002-8108-9179}\inst{\ref{aff20}}
\and F.~Pasian\orcid{0000-0002-4869-3227}\inst{\ref{aff6}}
\and K.~Pedersen\inst{\ref{aff78}}
\and V.~Pettorino\inst{\ref{aff76}}
\and G.~Polenta\orcid{0000-0003-4067-9196}\inst{\ref{aff79}}
\and M.~Poncet\inst{\ref{aff80}}
\and L.~A.~Popa\inst{\ref{aff81}}
\and L.~Pozzetti\orcid{0000-0001-7085-0412}\inst{\ref{aff1}}
\and F.~Raison\orcid{0000-0002-7819-6918}\inst{\ref{aff54}}
\and A.~Renzi\orcid{0000-0001-9856-1970}\inst{\ref{aff82},\ref{aff83}}
\and J.~Rhodes\inst{\ref{aff56}}
\and G.~Riccio\inst{\ref{aff31}}
\and E.~Romelli\orcid{0000-0003-3069-9222}\inst{\ref{aff6}}
\and M.~Roncarelli\orcid{0000-0001-9587-7822}\inst{\ref{aff1}}
\and E.~Rossetti\orcid{0000-0003-0238-4047}\inst{\ref{aff26}}
\and R.~Saglia\orcid{0000-0003-0378-7032}\inst{\ref{aff9},\ref{aff54}}
\and Z.~Sakr\orcid{0000-0002-4823-3757}\inst{\ref{aff22},\ref{aff84},\ref{aff85}}
\and A.~G.~S\'anchez\orcid{0000-0003-1198-831X}\inst{\ref{aff54}}
\and D.~Sapone\orcid{0000-0001-7089-4503}\inst{\ref{aff86}}
\and B.~Sartoris\orcid{0000-0003-1337-5269}\inst{\ref{aff9},\ref{aff6}}
\and R.~Scaramella\orcid{0000-0003-2229-193X}\inst{\ref{aff40},\ref{aff70}}
\and P.~Schneider\orcid{0000-0001-8561-2679}\inst{\ref{aff17}}
\and T.~Schrabback\orcid{0000-0002-6987-7834}\inst{\ref{aff87}}
\and A.~Secroun\orcid{0000-0003-0505-3710}\inst{\ref{aff63}}
\and E.~Sefusatti\orcid{0000-0003-0473-1567}\inst{\ref{aff6},\ref{aff2},\ref{aff7}}
\and G.~Seidel\orcid{0000-0003-2907-353X}\inst{\ref{aff61}}
\and S.~Serrano\orcid{0000-0002-0211-2861}\inst{\ref{aff88},\ref{aff89},\ref{aff90}}
\and C.~Sirignano\orcid{0000-0002-0995-7146}\inst{\ref{aff82},\ref{aff83}}
\and G.~Sirri\orcid{0000-0003-2626-2853}\inst{\ref{aff16}}
\and L.~Stanco\orcid{0000-0002-9706-5104}\inst{\ref{aff83}}
\and J.~Steinwagner\inst{\ref{aff54}}
\and P.~Tallada-Cresp\'{i}\orcid{0000-0002-1336-8328}\inst{\ref{aff37},\ref{aff38}}
\and I.~Tereno\inst{\ref{aff50},\ref{aff91}}
\and R.~Toledo-Moreo\orcid{0000-0002-2997-4859}\inst{\ref{aff92}}
\and F.~Torradeflot\orcid{0000-0003-1160-1517}\inst{\ref{aff38},\ref{aff37}}
\and I.~Tutusaus\orcid{0000-0002-3199-0399}\inst{\ref{aff84}}
\and L.~Valenziano\orcid{0000-0002-1170-0104}\inst{\ref{aff1},\ref{aff10}}
\and T.~Vassallo\orcid{0000-0001-6512-6358}\inst{\ref{aff9},\ref{aff6}}
\and G.~Verdoes~Kleijn\orcid{0000-0001-5803-2580}\inst{\ref{aff93}}
\and A.~Veropalumbo\orcid{0000-0003-2387-1194}\inst{\ref{aff8},\ref{aff29}}
\and Y.~Wang\orcid{0000-0002-4749-2984}\inst{\ref{aff94}}
\and J.~Weller\orcid{0000-0002-8282-2010}\inst{\ref{aff9},\ref{aff54}}
\and G.~Zamorani\orcid{0000-0002-2318-301X}\inst{\ref{aff1}}
\and E.~Zucca\orcid{0000-0002-5845-8132}\inst{\ref{aff1}}
\and M.~Bolzonella\orcid{0000-0003-3278-4607}\inst{\ref{aff1}}
\and A.~Boucaud\orcid{0000-0001-7387-2633}\inst{\ref{aff74}}
\and E.~Bozzo\orcid{0000-0002-8201-1525}\inst{\ref{aff20}}
\and C.~Burigana\orcid{0000-0002-3005-5796}\inst{\ref{aff95},\ref{aff10}}
\and M.~Calabrese\orcid{0000-0002-2637-2422}\inst{\ref{aff96},\ref{aff36}}
\and D.~Di~Ferdinando\inst{\ref{aff16}}
\and J.~A.~Escartin~Vigo\inst{\ref{aff54}}
\and R.~Farinelli\inst{\ref{aff1}}
\and J.~Gracia-Carpio\inst{\ref{aff54}}
\and N.~Mauri\orcid{0000-0001-8196-1548}\inst{\ref{aff41},\ref{aff16}}
\and V.~Scottez\inst{\ref{aff13},\ref{aff97}}
\and M.~Tenti\orcid{0000-0002-4254-5901}\inst{\ref{aff16}}
\and M.~Viel\orcid{0000-0002-2642-5707}\inst{\ref{aff2},\ref{aff6},\ref{aff25},\ref{aff7},\ref{aff4}}
\and M.~Wiesmann\orcid{0009-0000-8199-5860}\inst{\ref{aff55}}
\and Y.~Akrami\orcid{0000-0002-2407-7956}\inst{\ref{aff98},\ref{aff99}}
\and V.~Allevato\orcid{0000-0001-7232-5152}\inst{\ref{aff31}}
\and S.~Anselmi\orcid{0000-0002-3579-9583}\inst{\ref{aff83},\ref{aff82},\ref{aff12}}
\and M.~Ballardini\orcid{0000-0003-4481-3559}\inst{\ref{aff100},\ref{aff1},\ref{aff101}}
\and P.~Bergamini\orcid{0000-0003-1383-9414}\inst{\ref{aff68},\ref{aff1}}
\and A.~Blanchard\orcid{0000-0001-8555-9003}\inst{\ref{aff84}}
\and L.~Blot\orcid{0000-0002-9622-7167}\inst{\ref{aff102},\ref{aff12}}
\and S.~Bruton\orcid{0000-0002-6503-5218}\inst{\ref{aff103}}
\and R.~Cabanac\orcid{0000-0001-6679-2600}\inst{\ref{aff84}}
\and A.~Calabro\orcid{0000-0003-2536-1614}\inst{\ref{aff40}}
\and G.~Canas-Herrera\orcid{0000-0003-2796-2149}\inst{\ref{aff76},\ref{aff104}}
\and A.~Cappi\inst{\ref{aff1},\ref{aff73}}
\and C.~S.~Carvalho\inst{\ref{aff91}}
\and T.~Castro\orcid{0000-0002-6292-3228}\inst{\ref{aff6},\ref{aff7},\ref{aff2},\ref{aff4}}
\and K.~C.~Chambers\orcid{0000-0001-6965-7789}\inst{\ref{aff105}}
\and S.~Contarini\orcid{0000-0002-9843-723X}\inst{\ref{aff54},\ref{aff3}}
\and A.~R.~Cooray\orcid{0000-0002-3892-0190}\inst{\ref{aff106}}
\and M.~Costanzi\orcid{0000-0001-8158-1449}\inst{\ref{aff5},\ref{aff6},\ref{aff2}}
\and B.~De~Caro\inst{\ref{aff83},\ref{aff82}}
\and S.~de~la~Torre\inst{\ref{aff71}}
\and G.~Desprez\inst{\ref{aff107}}
\and A.~D\'iaz-S\'anchez\orcid{0000-0003-0748-4768}\inst{\ref{aff108}}
\and S.~Di~Domizio\orcid{0000-0003-2863-5895}\inst{\ref{aff28},\ref{aff29}}
\and H.~Dole\orcid{0000-0002-9767-3839}\inst{\ref{aff109}}
\and S.~Escoffier\orcid{0000-0002-2847-7498}\inst{\ref{aff63}}
\and A.~G.~Ferrari\orcid{0009-0005-5266-4110}\inst{\ref{aff41},\ref{aff16}}
\and P.~G.~Ferreira\orcid{0000-0002-3021-2851}\inst{\ref{aff110}}
\and I.~Ferrero\orcid{0000-0002-1295-1132}\inst{\ref{aff55}}
\and F.~Finelli\orcid{0000-0002-6694-3269}\inst{\ref{aff1},\ref{aff10}}
\and F.~Fornari\orcid{0000-0003-2979-6738}\inst{\ref{aff10}}
\and L.~Gabarra\orcid{0000-0002-8486-8856}\inst{\ref{aff110}}
\and K.~Ganga\orcid{0000-0001-8159-8208}\inst{\ref{aff74}}
\and J.~Garc\'ia-Bellido\orcid{0000-0002-9370-8360}\inst{\ref{aff98}}
\and E.~Gaztanaga\orcid{0000-0001-9632-0815}\inst{\ref{aff89},\ref{aff88},\ref{aff111}}
\and F.~Giacomini\orcid{0000-0002-3129-2814}\inst{\ref{aff16}}
\and G.~Gozaliasl\orcid{0000-0002-0236-919X}\inst{\ref{aff112},\ref{aff65}}
\and A.~Hall\orcid{0000-0002-3139-8651}\inst{\ref{aff43}}
\and H.~Hildebrandt\orcid{0000-0002-9814-3338}\inst{\ref{aff113}}
\and J.~Hjorth\orcid{0000-0002-4571-2306}\inst{\ref{aff114}}
\and A.~Jimenez~Mu\~noz\orcid{0009-0004-5252-185X}\inst{\ref{aff21}}
\and J.~J.~E.~Kajava\orcid{0000-0002-3010-8333}\inst{\ref{aff115},\ref{aff116}}
\and V.~Kansal\orcid{0000-0002-4008-6078}\inst{\ref{aff117},\ref{aff118}}
\and D.~Karagiannis\orcid{0000-0002-4927-0816}\inst{\ref{aff119},\ref{aff120}}
\and C.~C.~Kirkpatrick\inst{\ref{aff62}}
\and L.~Legrand\orcid{0000-0003-0610-5252}\inst{\ref{aff121}}
\and G.~Libet\inst{\ref{aff80}}
\and A.~Loureiro\orcid{0000-0002-4371-0876}\inst{\ref{aff122},\ref{aff123}}
\and G.~Maggio\orcid{0000-0003-4020-4836}\inst{\ref{aff6}}
\and M.~Magliocchetti\orcid{0000-0001-9158-4838}\inst{\ref{aff52}}
\and F.~Mannucci\orcid{0000-0002-4803-2381}\inst{\ref{aff124}}
\and R.~Maoli\orcid{0000-0002-6065-3025}\inst{\ref{aff19},\ref{aff40}}
\and C.~J.~A.~P.~Martins\orcid{0000-0002-4886-9261}\inst{\ref{aff125},\ref{aff33}}
\and S.~Matthew\orcid{0000-0001-8448-1697}\inst{\ref{aff43}}
\and L.~Maurin\orcid{0000-0002-8406-0857}\inst{\ref{aff109}}
\and R.~B.~Metcalf\orcid{0000-0003-3167-2574}\inst{\ref{aff3},\ref{aff1}}
\and P.~Monaco\orcid{0000-0003-2083-7564}\inst{\ref{aff5},\ref{aff6},\ref{aff7},\ref{aff2}}
\and C.~Moretti\orcid{0000-0003-3314-8936}\inst{\ref{aff25},\ref{aff4},\ref{aff6},\ref{aff2},\ref{aff7}}
\and G.~Morgante\inst{\ref{aff1}}
\and Nicholas~A.~Walton\orcid{0000-0003-3983-8778}\inst{\ref{aff126}}
\and L.~Patrizii\inst{\ref{aff16}}
\and A.~Pezzotta\orcid{0000-0003-0726-2268}\inst{\ref{aff54}}
\and M.~P\"ontinen\orcid{0000-0001-5442-2530}\inst{\ref{aff65}}
\and V.~Popa\inst{\ref{aff81}}
\and C.~Porciani\orcid{0000-0002-7797-2508}\inst{\ref{aff17}}
\and D.~Potter\orcid{0000-0002-0757-5195}\inst{\ref{aff127}}
\and I.~Risso\orcid{0000-0003-2525-7761}\inst{\ref{aff128}}
\and P.-F.~Rocci\inst{\ref{aff109}}
\and M.~Sahl\'en\orcid{0000-0003-0973-4804}\inst{\ref{aff129}}
\and A.~Schneider\orcid{0000-0001-7055-8104}\inst{\ref{aff127}}
\and M.~Schultheis\inst{\ref{aff73}}
\and P.~Simon\inst{\ref{aff17}}
\and A.~Spurio~Mancini\orcid{0000-0001-5698-0990}\inst{\ref{aff130},\ref{aff131}}
\and C.~Tao\orcid{0000-0001-7961-8177}\inst{\ref{aff63}}
\and G.~Testera\inst{\ref{aff29}}
\and R.~Teyssier\orcid{0000-0001-7689-0933}\inst{\ref{aff132}}
\and S.~Toft\orcid{0000-0003-3631-7176}\inst{\ref{aff60},\ref{aff133},\ref{aff134}}
\and S.~Tosi\orcid{0000-0002-7275-9193}\inst{\ref{aff28},\ref{aff29},\ref{aff8}}
\and A.~Troja\orcid{0000-0003-0239-4595}\inst{\ref{aff82},\ref{aff83}}
\and M.~Tucci\inst{\ref{aff20}}
\and C.~Valieri\inst{\ref{aff16}}
\and J.~Valiviita\orcid{0000-0001-6225-3693}\inst{\ref{aff65},\ref{aff66}}
\and D.~Vergani\orcid{0000-0003-0898-2216}\inst{\ref{aff1}}
\and G.~Verza\orcid{0000-0002-1886-8348}\inst{\ref{aff135},\ref{aff136}}}
										   
\institute{INAF-Osservatorio di Astrofisica e Scienza dello Spazio di Bologna, Via Piero Gobetti 93/3, 40129 Bologna, Italy\label{aff1}
\and
IFPU, Institute for Fundamental Physics of the Universe, via Beirut 2, 34151 Trieste, Italy\label{aff2}
\and
Dipartimento di Fisica e Astronomia "Augusto Righi" - Alma Mater Studiorum Universit\`a di Bologna, via Piero Gobetti 93/2, 40129 Bologna, Italy\label{aff3}
\and
ICSC - Centro Nazionale di Ricerca in High Performance Computing, Big Data e Quantum Computing, Via Magnanelli 2, Bologna, Italy\label{aff4}
\and
Dipartimento di Fisica - Sezione di Astronomia, Universit\`a di Trieste, Via Tiepolo 11, 34131 Trieste, Italy\label{aff5}
\and
INAF-Osservatorio Astronomico di Trieste, Via G. B. Tiepolo 11, 34143 Trieste, Italy\label{aff6}
\and
INFN, Sezione di Trieste, Via Valerio 2, 34127 Trieste TS, Italy\label{aff7}
\and
INAF-Osservatorio Astronomico di Brera, Via Brera 28, 20122 Milano, Italy\label{aff8}
\and
Universit\"ats-Sternwarte M\"unchen, Fakult\"at f\"ur Physik, Ludwig-Maximilians-Universit\"at M\"unchen, Scheinerstrasse 1, 81679 M\"unchen, Germany\label{aff9}
\and
INFN-Bologna, Via Irnerio 46, 40126 Bologna, Italy\label{aff10}
\and
Istituto Nazionale di Fisica Nucleare, Sezione di Bologna, Via Irnerio 46, 40126 Bologna, Italy\label{aff11}
\and
Laboratoire Univers et Th\'eorie, Observatoire de Paris, Universit\'e PSL, Universit\'e Paris Cit\'e, CNRS, 92190 Meudon, France\label{aff12}
\and
Institut d'Astrophysique de Paris, 98bis Boulevard Arago, 75014, Paris, France\label{aff13}
\and
Institut d'Astrophysique de Paris, UMR 7095, CNRS, and Sorbonne Universit\'e, 98 bis boulevard Arago, 75014 Paris, France\label{aff14}
\and
School of Physics, HH Wills Physics Laboratory, University of Bristol, Tyndall Avenue, Bristol, BS8 1TL, UK\label{aff15}
\and
INFN-Sezione di Bologna, Viale Berti Pichat 6/2, 40127 Bologna, Italy\label{aff16}
\and
Universit\"at Bonn, Argelander-Institut f\"ur Astronomie, Auf dem H\"ugel 71, 53121 Bonn, Germany\label{aff17}
\and
Universit\'e Paris-Saclay, Universit\'e Paris Cit\'e, CEA, CNRS, AIM, 91191, Gif-sur-Yvette, France\label{aff18}
\and
Dipartimento di Fisica, Sapienza Universit\`a di Roma, Piazzale Aldo Moro 2, 00185 Roma, Italy\label{aff19}
\and
Department of Astronomy, University of Geneva, ch. d'Ecogia 16, 1290 Versoix, Switzerland\label{aff20}
\and
Univ. Grenoble Alpes, CNRS, Grenoble INP, LPSC-IN2P3, 53, Avenue des Martyrs, 38000, Grenoble, France\label{aff21}
\and
Institut f\"ur Theoretische Physik, University of Heidelberg, Philosophenweg 16, 69120 Heidelberg, Germany\label{aff22}
\and
Zentrum f\"ur Astronomie, Universit\"at Heidelberg, Philosophenweg 12, 69120 Heidelberg, Germany\label{aff23}
\and
School of Mathematics and Physics, University of Surrey, Guildford, Surrey, GU2 7XH, UK\label{aff24}
\and
SISSA, International School for Advanced Studies, Via Bonomea 265, 34136 Trieste TS, Italy\label{aff25}
\and
Dipartimento di Fisica e Astronomia, Universit\`a di Bologna, Via Gobetti 93/2, 40129 Bologna, Italy\label{aff26}
\and
INAF-Osservatorio Astrofisico di Torino, Via Osservatorio 20, 10025 Pino Torinese (TO), Italy\label{aff27}
\and
Dipartimento di Fisica, Universit\`a di Genova, Via Dodecaneso 33, 16146, Genova, Italy\label{aff28}
\and
INFN-Sezione di Genova, Via Dodecaneso 33, 16146, Genova, Italy\label{aff29}
\and
Department of Physics "E. Pancini", University Federico II, Via Cinthia 6, 80126, Napoli, Italy\label{aff30}
\and
INAF-Osservatorio Astronomico di Capodimonte, Via Moiariello 16, 80131 Napoli, Italy\label{aff31}
\and
INFN section of Naples, Via Cinthia 6, 80126, Napoli, Italy\label{aff32}
\and
Instituto de Astrof\'isica e Ci\^encias do Espa\c{c}o, Universidade do Porto, CAUP, Rua das Estrelas, PT4150-762 Porto, Portugal\label{aff33}
\and
Dipartimento di Fisica, Universit\`a degli Studi di Torino, Via P. Giuria 1, 10125 Torino, Italy\label{aff34}
\and
INFN-Sezione di Torino, Via P. Giuria 1, 10125 Torino, Italy\label{aff35}
\and
INAF-IASF Milano, Via Alfonso Corti 12, 20133 Milano, Italy\label{aff36}
\and
Centro de Investigaciones Energ\'eticas, Medioambientales y Tecnol\'ogicas (CIEMAT), Avenida Complutense 40, 28040 Madrid, Spain\label{aff37}
\and
Port d'Informaci\'{o} Cient\'{i}fica, Campus UAB, C. Albareda s/n, 08193 Bellaterra (Barcelona), Spain\label{aff38}
\and
Institute for Theoretical Particle Physics and Cosmology (TTK), RWTH Aachen University, 52056 Aachen, Germany\label{aff39}
\and
INAF-Osservatorio Astronomico di Roma, Via Frascati 33, 00078 Monteporzio Catone, Italy\label{aff40}
\and
Dipartimento di Fisica e Astronomia "Augusto Righi" - Alma Mater Studiorum Universit\`a di Bologna, Viale Berti Pichat 6/2, 40127 Bologna, Italy\label{aff41}
\and
Instituto de Astrof\'isica de Canarias, Calle V\'ia L\'actea s/n, 38204, San Crist\'obal de La Laguna, Tenerife, Spain\label{aff42}
\and
Institute for Astronomy, University of Edinburgh, Royal Observatory, Blackford Hill, Edinburgh EH9 3HJ, UK\label{aff43}
\and
Jodrell Bank Centre for Astrophysics, Department of Physics and Astronomy, University of Manchester, Oxford Road, Manchester M13 9PL, UK\label{aff44}
\and
European Space Agency/ESRIN, Largo Galileo Galilei 1, 00044 Frascati, Roma, Italy\label{aff45}
\and
ESAC/ESA, Camino Bajo del Castillo, s/n., Urb. Villafranca del Castillo, 28692 Villanueva de la Ca\~nada, Madrid, Spain\label{aff46}
\and
Universit\'e Claude Bernard Lyon 1, CNRS/IN2P3, IP2I Lyon, UMR 5822, Villeurbanne, F-69100, France\label{aff47}
\and
Institute of Physics, Laboratory of Astrophysics, Ecole Polytechnique F\'ed\'erale de Lausanne (EPFL), Observatoire de Sauverny, 1290 Versoix, Switzerland\label{aff48}
\and
UCB Lyon 1, CNRS/IN2P3, IUF, IP2I Lyon, 4 rue Enrico Fermi, 69622 Villeurbanne, France\label{aff49}
\and
Departamento de F\'isica, Faculdade de Ci\^encias, Universidade de Lisboa, Edif\'icio C8, Campo Grande, PT1749-016 Lisboa, Portugal\label{aff50}
\and
Instituto de Astrof\'isica e Ci\^encias do Espa\c{c}o, Faculdade de Ci\^encias, Universidade de Lisboa, Campo Grande, 1749-016 Lisboa, Portugal\label{aff51}
\and
INAF-Istituto di Astrofisica e Planetologia Spaziali, via del Fosso del Cavaliere, 100, 00100 Roma, Italy\label{aff52}
\and
INAF-Osservatorio Astronomico di Padova, Via dell'Osservatorio 5, 35122 Padova, Italy\label{aff53}
\and
Max Planck Institute for Extraterrestrial Physics, Giessenbachstr. 1, 85748 Garching, Germany\label{aff54}
\and
Institute of Theoretical Astrophysics, University of Oslo, P.O. Box 1029 Blindern, 0315 Oslo, Norway\label{aff55}
\and
Jet Propulsion Laboratory, California Institute of Technology, 4800 Oak Grove Drive, Pasadena, CA, 91109, USA\label{aff56}
\and
Department of Physics, Lancaster University, Lancaster, LA1 4YB, UK\label{aff57}
\and
Felix Hormuth Engineering, Goethestr. 17, 69181 Leimen, Germany\label{aff58}
\and
Technical University of Denmark, Elektrovej 327, 2800 Kgs. Lyngby, Denmark\label{aff59}
\and
Cosmic Dawn Center (DAWN), Denmark\label{aff60}
\and
Max-Planck-Institut f\"ur Astronomie, K\"onigstuhl 17, 69117 Heidelberg, Germany\label{aff61}
\and
Department of Physics and Helsinki Institute of Physics, Gustaf H\"allstr\"omin katu 2, 00014 University of Helsinki, Finland\label{aff62}
\and
Aix-Marseille Universit\'e, CNRS/IN2P3, CPPM, Marseille, France\label{aff63}
\and
Universit\'e de Gen\`eve, D\'epartement de Physique Th\'eorique and Centre for Astroparticle Physics, 24 quai Ernest-Ansermet, CH-1211 Gen\`eve 4, Switzerland\label{aff64}
\and
Department of Physics, P.O. Box 64, 00014 University of Helsinki, Finland\label{aff65}
\and
Helsinki Institute of Physics, Gustaf H{\"a}llstr{\"o}min katu 2, University of Helsinki, Helsinki, Finland\label{aff66}
\and
NOVA optical infrared instrumentation group at ASTRON, Oude Hoogeveensedijk 4, 7991PD, Dwingeloo, The Netherlands\label{aff67}
\and
Dipartimento di Fisica "Aldo Pontremoli", Universit\`a degli Studi di Milano, Via Celoria 16, 20133 Milano, Italy\label{aff68}
\and
INFN-Sezione di Milano, Via Celoria 16, 20133 Milano, Italy\label{aff69}
\and
INFN-Sezione di Roma, Piazzale Aldo Moro, 2 - c/o Dipartimento di Fisica, Edificio G. Marconi, 00185 Roma, Italy\label{aff70}
\and
Aix-Marseille Universit\'e, CNRS, CNES, LAM, Marseille, France\label{aff71}
\and
Department of Physics, Institute for Computational Cosmology, Durham University, South Road, Durham, DH1 3LE, UK\label{aff72}
\and
Universit\'e C\^{o}te d'Azur, Observatoire de la C\^{o}te d'Azur, CNRS, Laboratoire Lagrange, Bd de l'Observatoire, CS 34229, 06304 Nice cedex 4, France\label{aff73}
\and
Universit\'e Paris Cit\'e, CNRS, Astroparticule et Cosmologie, 75013 Paris, France\label{aff74}
\and
Institut de F\'{i}sica d'Altes Energies (IFAE), The Barcelona Institute of Science and Technology, Campus UAB, 08193 Bellaterra (Barcelona), Spain\label{aff75}
\and
European Space Agency/ESTEC, Keplerlaan 1, 2201 AZ Noordwijk, The Netherlands\label{aff76}
\and
School of Mathematics, Statistics and Physics, Newcastle University, Herschel Building, Newcastle-upon-Tyne, NE1 7RU, UK\label{aff77}
\and
Department of Physics and Astronomy, University of Aarhus, Ny Munkegade 120, DK-8000 Aarhus C, Denmark\label{aff78}
\and
Space Science Data Center, Italian Space Agency, via del Politecnico snc, 00133 Roma, Italy\label{aff79}
\and
Centre National d'Etudes Spatiales -- Centre spatial de Toulouse, 18 avenue Edouard Belin, 31401 Toulouse Cedex 9, France\label{aff80}
\and
Institute of Space Science, Str. Atomistilor, nr. 409 M\u{a}gurele, Ilfov, 077125, Romania\label{aff81}
\and
Dipartimento di Fisica e Astronomia "G. Galilei", Universit\`a di Padova, Via Marzolo 8, 35131 Padova, Italy\label{aff82}
\and
INFN-Padova, Via Marzolo 8, 35131 Padova, Italy\label{aff83}
\and
Institut de Recherche en Astrophysique et Plan\'etologie (IRAP), Universit\'e de Toulouse, CNRS, UPS, CNES, 14 Av. Edouard Belin, 31400 Toulouse, France\label{aff84}
\and
Universit\'e St Joseph; Faculty of Sciences, Beirut, Lebanon\label{aff85}
\and
Departamento de F\'isica, FCFM, Universidad de Chile, Blanco Encalada 2008, Santiago, Chile\label{aff86}
\and
Universit\"at Innsbruck, Institut f\"ur Astro- und Teilchenphysik, Technikerstr. 25/8, 6020 Innsbruck, Austria\label{aff87}
\and
Institut d'Estudis Espacials de Catalunya (IEEC),  Edifici RDIT, Campus UPC, 08860 Castelldefels, Barcelona, Spain\label{aff88}
\and
Institute of Space Sciences (ICE, CSIC), Campus UAB, Carrer de Can Magrans, s/n, 08193 Barcelona, Spain\label{aff89}
\and
Satlantis, University Science Park, Sede Bld 48940, Leioa-Bilbao, Spain\label{aff90}
\and
Instituto de Astrof\'isica e Ci\^encias do Espa\c{c}o, Faculdade de Ci\^encias, Universidade de Lisboa, Tapada da Ajuda, 1349-018 Lisboa, Portugal\label{aff91}
\and
Universidad Polit\'ecnica de Cartagena, Departamento de Electr\'onica y Tecnolog\'ia de Computadoras,  Plaza del Hospital 1, 30202 Cartagena, Spain\label{aff92}
\and
Kapteyn Astronomical Institute, University of Groningen, PO Box 800, 9700 AV Groningen, The Netherlands\label{aff93}
\and
Infrared Processing and Analysis Center, California Institute of Technology, Pasadena, CA 91125, USA\label{aff94}
\and
INAF, Istituto di Radioastronomia, Via Piero Gobetti 101, 40129 Bologna, Italy\label{aff95}
\and
Astronomical Observatory of the Autonomous Region of the Aosta Valley (OAVdA), Loc. Lignan 39, I-11020, Nus (Aosta Valley), Italy\label{aff96}
\and
ICL, Junia, Universit\'e Catholique de Lille, LITL, 59000 Lille, France\label{aff97}
\and
Instituto de F\'isica Te\'orica UAM-CSIC, Campus de Cantoblanco, 28049 Madrid, Spain\label{aff98}
\and
CERCA/ISO, Department of Physics, Case Western Reserve University, 10900 Euclid Avenue, Cleveland, OH 44106, USA\label{aff99}
\and
Dipartimento di Fisica e Scienze della Terra, Universit\`a degli Studi di Ferrara, Via Giuseppe Saragat 1, 44122 Ferrara, Italy\label{aff100}
\and
Istituto Nazionale di Fisica Nucleare, Sezione di Ferrara, Via Giuseppe Saragat 1, 44122 Ferrara, Italy\label{aff101}
\and
Kavli Institute for the Physics and Mathematics of the Universe (WPI), University of Tokyo, Kashiwa, Chiba 277-8583, Japan\label{aff102}
\and
Minnesota Institute for Astrophysics, University of Minnesota, 116 Church St SE, Minneapolis, MN 55455, USA\label{aff103}
\and
Institute Lorentz, Leiden University, Niels Bohrweg 2, 2333 CA Leiden, The Netherlands\label{aff104}
\and
Institute for Astronomy, University of Hawaii, 2680 Woodlawn Drive, Honolulu, HI 96822, USA\label{aff105}
\and
Department of Physics \& Astronomy, University of California Irvine, Irvine CA 92697, USA\label{aff106}
\and
Department of Astronomy \& Physics and Institute for Computational Astrophysics, Saint Mary's University, 923 Robie Street, Halifax, Nova Scotia, B3H 3C3, Canada\label{aff107}
\and
Departamento F\'isica Aplicada, Universidad Polit\'ecnica de Cartagena, Campus Muralla del Mar, 30202 Cartagena, Murcia, Spain\label{aff108}
\and
Universit\'e Paris-Saclay, CNRS, Institut d'astrophysique spatiale, 91405, Orsay, France\label{aff109}
\and
Department of Physics, Oxford University, Keble Road, Oxford OX1 3RH, UK\label{aff110}
\and
Institute of Cosmology and Gravitation, University of Portsmouth, Portsmouth PO1 3FX, UK\label{aff111}
\and
Department of Computer Science, Aalto University, PO Box 15400, Espoo, FI-00 076, Finland\label{aff112}
\and
Ruhr University Bochum, Faculty of Physics and Astronomy, Astronomical Institute (AIRUB), German Centre for Cosmological Lensing (GCCL), 44780 Bochum, Germany\label{aff113}
\and
DARK, Niels Bohr Institute, University of Copenhagen, Jagtvej 155, 2200 Copenhagen, Denmark\label{aff114}
\and
Department of Physics and Astronomy, Vesilinnantie 5, 20014 University of Turku, Finland\label{aff115}
\and
Serco for European Space Agency (ESA), Camino bajo del Castillo, s/n, Urbanizacion Villafranca del Castillo, Villanueva de la Ca\~nada, 28692 Madrid, Spain\label{aff116}
\and
ARC Centre of Excellence for Dark Matter Particle Physics, Melbourne, Australia\label{aff117}
\and
Centre for Astrophysics \& Supercomputing, Swinburne University of Technology,  Hawthorn, Victoria 3122, Australia\label{aff118}
\and
School of Physics and Astronomy, Queen Mary University of London, Mile End Road, London E1 4NS, UK\label{aff119}
\and
Department of Physics and Astronomy, University of the Western Cape, Bellville, Cape Town, 7535, South Africa\label{aff120}
\and
ICTP South American Institute for Fundamental Research, Instituto de F\'{\i}sica Te\'orica, Universidade Estadual Paulista, S\~ao Paulo, Brazil\label{aff121}
\and
Oskar Klein Centre for Cosmoparticle Physics, Department of Physics, Stockholm University, Stockholm, SE-106 91, Sweden\label{aff122}
\and
Astrophysics Group, Blackett Laboratory, Imperial College London, London SW7 2AZ, UK\label{aff123}
\and
INAF-Osservatorio Astrofisico di Arcetri, Largo E. Fermi 5, 50125, Firenze, Italy\label{aff124}
\and
Centro de Astrof\'{\i}sica da Universidade do Porto, Rua das Estrelas, 4150-762 Porto, Portugal\label{aff125}
\and
Institute of Astronomy, University of Cambridge, Madingley Road, Cambridge CB3 0HA, UK\label{aff126}
\and
Department of Astrophysics, University of Zurich, Winterthurerstrasse 190, 8057 Zurich, Switzerland\label{aff127}
\and
Dipartimento di Fisica, Universit\`a degli studi di Genova, and INFN-Sezione di Genova, via Dodecaneso 33, 16146, Genova, Italy\label{aff128}
\and
Theoretical astrophysics, Department of Physics and Astronomy, Uppsala University, Box 515, 751 20 Uppsala, Sweden\label{aff129}
\and
Department of Physics, Royal Holloway, University of London, TW20 0EX, UK\label{aff130}
\and
Mullard Space Science Laboratory, University College London, Holmbury St Mary, Dorking, Surrey RH5 6NT, UK\label{aff131}
\and
Department of Astrophysical Sciences, Peyton Hall, Princeton University, Princeton, NJ 08544, USA\label{aff132}
\and
Cosmic Dawn Center (DAWN)\label{aff133}
\and
Niels Bohr Institute, University of Copenhagen, Jagtvej 128, 2200 Copenhagen, Denmark\label{aff134}
\and
Center for Cosmology and Particle Physics, Department of Physics, New York University, New York, NY 10003, USA\label{aff135}
\and
Center for Computational Astrophysics, Flatiron Institute, 162 5th Avenue, 10010, New York, NY, USA\label{aff136}}

%% file: AandA.bbl
\begin{thebibliography}{114}
\expandafter\ifx\csname natexlab\endcsname\relax\def\natexlab#1{#1}\fi

\bibitem[{{Abbott} {et~al.}(2020){Abbott}, {Aguena}, {Alarcon}, {Allam},
  {Allen}, {Annis}, {Avila}, {Bacon}, {Bechtol}, {Bermeo}, {Bernstein},
  {Bertin}, {Bhargava}, {Bocquet}, {Brooks}, {Brout}, {Buckley-Geer}, {Burke},
  {Carnero Rosell}, {Carrasco Kind}, {Carretero}, {Castander}, {Cawthon},
  {Chang}, {Chen}, {Choi}, {Costanzi}, {Crocce}, {da Costa}, {Davis}, {De
  Vicente}, {DeRose}, {Desai}, {Diehl}, {Dietrich}, {Dodelson}, {Doel},
  {Drlica-Wagner}, {Eckert}, {Eifler}, {Elvin-Poole}, {Estrada}, {Everett},
  {Evrard}, {Farahi}, {Ferrero}, {Flaugher}, {Fosalba}, {Frieman},
  {Garc{\'\i}a-Bellido}, {Gatti}, {Gaztanaga}, {Gerdes}, {Giannantonio},
  {Giles}, {Grandis}, {Gruen}, {Gruendl}, {Gschwend}, {Gutierrez}, {Hartley},
  {Hinton}, {Hollowood}, {Honscheid}, {Hoyle}, {Huterer}, {James}, {Jarvis},
  {Jeltema}, {Johnson}, {Johnson}, {Kent}, {Krause}, {Kron}, {Kuehn},
  {Kuropatkin}, {Lahav}, {Li}, {Lidman}, {Lima}, {Lin}, {MacCrann}, {Maia},
  {Mantz}, {Marshall}, {Martini}, {Mayers}, {Melchior}, {Mena-Fern{\'a}ndez},
  {Menanteau}, {Miquel}, {Mohr}, {Nichol}, {Nord}, {Ogando}, {Palmese},
  {Paz-Chinch{\'o}n}, {Plazas}, {Prat}, {Rau}, {Romer}, {Roodman}, {Rooney},
  {Rozo}, {Rykoff}, {Sako}, {Samuroff}, {S{\'a}nchez}, {Sanchez}, {Saro},
  {Scarpine}, {Schubnell}, {Scolnic}, {Serrano}, {Sevilla-Noarbe}, {Sheldon},
  {Smith}, {Smith}, {Suchyta}, {Swanson}, {Tarle}, {Thomas}, {To}, {Troxel},
  {Tucker}, {Varga}, {von der Linden}, {Walker}, {Wechsler}, {Weller},
  {Wilkinson}, {Wu}, {Yanny}, {Zhang}, {Zhang}, {Zuntz}, \& {DES
  Collaboration}}]{2020PhRvD.102b3509Abbott}
{Abbott}, T.~M.~C., {Aguena}, M., {Alarcon}, A., {et~al.} 2020, \prd, 102,
  023509

\bibitem[{{Allen} {et~al.}(2011){Allen}, {Evrard}, \&
  {Mantz}}]{2011ARA&A..49..409Allen}
{Allen}, S.~W., {Evrard}, A.~E., \& {Mantz}, A.~B. 2011, \araa, 49, 409

\bibitem[{{Anbajagane} {et~al.}(2020){Anbajagane}, {Evrard}, {Farahi},
  {Barnes}, {Dolag}, {McCarthy}, {Nelson}, \&
  {Pillepich}}]{2020MNRAS.495..686Anbajagane}
{Anbajagane}, D., {Evrard}, A.~E., {Farahi}, A., {et~al.} 2020, \mnras, 495,
  686

\bibitem[{{Andreon} {et~al.}(2016){Andreon}, {Dong}, \&
  {Raichoor}}]{2016A&A...593A...2Andreon}
{Andreon}, S., {Dong}, H., \& {Raichoor}, A. 2016, \aap, 593, A2

\bibitem[{{Andreon} \& {Moretti}(2011)}]{2011A&A...536A..37Andreon}
{Andreon}, S. \& {Moretti}, A. 2011, \aap, 536, A37

\bibitem[{{Angelinelli} {et~al.}(2022){Angelinelli}, {Ettori}, {Dolag},
  {Vazza}, \& {Ragagnin}}]{Angelinelli2022Mapping}
{Angelinelli}, M., {Ettori}, S., {Dolag}, K., {Vazza}, F., \& {Ragagnin}, A.
  2022, \aap, 663, L6

\bibitem[{{Angelinelli} {et~al.}(2023{\natexlab{a}}){Angelinelli}, {Ettori},
  {Dolag}, {Vazza}, \& {Ragagnin}}]{Angelinelli23Redshift}
{Angelinelli}, M., {Ettori}, S., {Dolag}, K., {Vazza}, F., \& {Ragagnin}, A.
  2023{\natexlab{a}}, \aap, 675, A188

\bibitem[{{Angelinelli} {et~al.}(2023{\natexlab{b}}){Angelinelli}, {Ettori},
  {Dolag}, {Vazza}, \& {Ragagnin}}]{Angelinelli2023Redshift}
{Angelinelli}, M., {Ettori}, S., {Dolag}, K., {Vazza}, F., \& {Ragagnin}, A.
  2023{\natexlab{b}}, \aap, 675, A188

\bibitem[{{Angulo} {et~al.}(2021){Angulo}, {Zennaro}, {Contreras}, {Aric{\`o}},
  {Pellejero-Iba{\~n}ez}, \& {St{\"u}cker}}]{Angulo2021Bacco}
{Angulo}, R.~E., {Zennaro}, M., {Contreras}, S., {et~al.} 2021, \mnras, 507,
  5869

\bibitem[{{Arnaud}(1996)}]{1996ASPC..101...17Arnaud}
{Arnaud}, K.~A. 1996, in Astronomical Society of the Pacific Conference Series,
  Vol. 101, Astronomical Data Analysis Software and Systems V, ed. G.~H.
  {Jacoby} \& J.~{Barnes}, 17

\bibitem[{{Balm{\`e}s} {et~al.}(2014){Balm{\`e}s}, {Rasera}, {Corasaniti}, \&
  {Alimi}}]{Balmes2014Sparsity}
{Balm{\`e}s}, I., {Rasera}, Y., {Corasaniti}, P.~S., \& {Alimi}, J.~M. 2014,
  \mnras, 437, 2328

\bibitem[{{Beck} {et~al.}(2016){Beck}, {Murante}, {Arth}, {Remus}, {Teklu},
  {Donnert}, {Planelles}, {Beck}, {F{\"o}rster}, {Imgrund}, {Dolag}, \&
  {Borgani}}]{2016MNRAS.455.2110Beck}
{Beck}, A.~M., {Murante}, G., {Arth}, A., {et~al.} 2016, \mnras, 455, 2110

\bibitem[{{Becker} \& {Kravtsov}(2011)}]{Becker11}
{Becker}, M.~R. \& {Kravtsov}, A.~V. 2011, \apj, 740, 25

\bibitem[{{Bellagamba} {et~al.}(2018{\natexlab{a}}){Bellagamba}, {Roncarelli},
  {Maturi}, \& {Moscardini}}]{amicoA}
{Bellagamba}, F., {Roncarelli}, M., {Maturi}, M., \& {Moscardini}, L.
  2018{\natexlab{a}}, \mnras, 473, 5221

\bibitem[{{Bellagamba} {et~al.}(2018{\natexlab{b}}){Bellagamba}, {Roncarelli},
  {Maturi}, \& {Moscardini}}]{Bellagamba2018Amico}
{Bellagamba}, F., {Roncarelli}, M., {Maturi}, M., \& {Moscardini}, L.
  2018{\natexlab{b}}, \mnras, 473, 5221

\bibitem[{{Bertocco} {et~al.}(2020){Bertocco}, {Goz}, {Tornatore}, {Ragagnin},
  {Maggio}, {Gasparo}, {Vuerli}, {Taffoni}, \&
  {Molinaro}}]{2020ASPC..527..303Bertocco}
{Bertocco}, S., {Goz}, D., {Tornatore}, L., {et~al.} 2020, in Astronomical
  Society of the Pacific Conference Series, Vol. 527, Astronomical Society of
  the Pacific Conference Series, ed. R.~{Pizzo}, E.~R. {Deul}, J.~D. {Mol},
  J.~{de Plaa}, \& H.~{Verkouter}, 303

\bibitem[{{Bhargava} {et~al.}(2023){Bhargava}, {Garrel}, {Koulouridis},
  {Pierre}, {Valtchanov}, {Cerardi}, {Maughan}, {Aguena}, {Benoist}, {Baguley},
  {Ramos-Ceja}, {Adami}, {Chiappetti}, {Vignali}, \&
  {Willis}}]{Bhargava2023AGNContamiantion}
{Bhargava}, S., {Garrel}, C., {Koulouridis}, E., {et~al.} 2023, \aap, 673, A92

\bibitem[{{Biffi} {et~al.}(2013){Biffi}, {Dolag}, \&
  {B{\"o}hringer}}]{2013MNRAS.428.1395Biffi}
{Biffi}, V., {Dolag}, K., \& {B{\"o}hringer}, H. 2013, \mnras, 428, 1395

\bibitem[{{Biffi} {et~al.}(2017){Biffi}, {Planelles}, {Borgani}, {Fabjan},
  {Rasia}, {Murante}, {Tornatore}, {Dolag}, {Granato}, {Gaspari}, \&
  {Beck}}]{Biffi17Xray}
{Biffi}, V., {Planelles}, S., {Borgani}, S., {et~al.} 2017, \mnras, 468, 531

\bibitem[{{Bocquet} {et~al.}(2019){Bocquet}, {Dietrich}, {Schrabback}, {Bleem},
  {Klein}, {Allen}, {Applegate}, {Ashby}, {Bautz}, {Bayliss}, {Benson},
  {Brodwin}, {Bulbul}, {Canning}, {Capasso}, {Carlstrom}, {Chang}, {Chiu},
  {Cho}, {Clocchiatti}, {Crawford}, {Crites}, {de Haan}, {Desai}, {Dobbs},
  {Foley}, {Forman}, {Garmire}, {George}, {Gladders}, {Gonzalez}, {Grandis},
  {Gupta}, {Halverson}, {Hlavacek-Larrondo}, {Hoekstra}, {Holder}, {Holzapfel},
  {Hou}, {Hrubes}, {Huang}, {Jones}, {Khullar}, {Knox}, {Kraft}, {Lee}, {von
  der Linden}, {Luong-Van}, {Mantz}, {Marrone}, {McDonald}, {McMahon}, {Meyer},
  {Mocanu}, {Mohr}, {Morris}, {Padin}, {Patil}, {Pryke}, {Rapetti},
  {Reichardt}, {Rest}, {Ruhl}, {Saliwanchik}, {Saro}, {Sayre}, {Schaffer},
  {Shirokoff}, {Stalder}, {Stanford}, {Staniszewski}, {Stark}, {Story},
  {Strazzullo}, {Stubbs}, {Vanderlinde}, {Vieira}, {Vikhlinin}, {Williamson},
  \& {Zenteno}}]{2019ApJ...878...55Bocquet}
{Bocquet}, S., {Dietrich}, J.~P., {Schrabback}, T., {et~al.} 2019, \apj, 878,
  55

\bibitem[{{Bocquet} {et~al.}(2024){Bocquet}, {Grandis}, {Bleem}, {Klein},
  {Mohr}, {Schrabback}, {Abbott}, {Ade}, {Aguena}, {Alarcon}, {Allam}, {Allen},
  {Alves}, {Amon}, {Anderson}, {Annis}, {Ansarinejad}, {Austermann}, {Avila},
  {Bacon}, {Bayliss}, {Beall}, {Bechtol}, {Becker}, {Bender}, {Benson},
  {Bernstein}, {Bhargava}, {Bianchini}, {Brodwin}, {Brooks}, {Bryant},
  {Campos}, {Canning}, {Carlstrom}, {Carnero Rosell}, {Carrasco Kind},
  {Carretero}, {Castander}, {Cawthon}, {Chang}, {Chang}, {Chaubal}, {Chen},
  {Chiang}, {Choi}, {Chou}, {Citron}, {Corbett Moran}, {Cordero}, {Costanzi},
  {Crawford}, {Crites}, {da Costa}, {Pereira}, {Davis}, {Davis}, {DeRose},
  {Desai}, {de Haan}, {Diehl}, {Dobbs}, {Dodelson}, {Doux}, {Drlica-Wagner},
  {Eckert}, {Elvin-Poole}, {Everett}, {Everett}, {Ferrero}, {Fert{\'e}},
  {Flores}, {Frieman}, {Gallicchio}, {Garc{\'\i}a-Bellido}, {Gatti}, {George},
  {Giannini}, {Gladders}, {Gruen}, {Gruendl}, {Gupta}, {Gutierrez},
  {Halverson}, {Harrison}, {Hartley}, {Herner}, {Hinton}, {Holder},
  {Hollowood}, {Holzapfel}, {Honscheid}, {Hrubes}, {Huang}, {Hubmayr}, {Huff},
  {Huterer}, {Irwin}, {James}, {Jarvis}, {Khullar}, {Kim}, {Knox}, {Kraft},
  {Krause}, {Kuehn}, {Kuropatkin}, {K{\'e}ruzor{\'e}}, {Lahav}, {Lee}, {Leget},
  {Li}, {Lin}, {Lowitz}, {MacCrann}, {Mahler}, {Mantz}, {Marshall},
  {McCullough}, {McDonald}, {McMahon}, {Mena-Fern{\'a}ndez}, {Menanteau},
  {Meyer}, {Miquel}, {Montgomery}, {Myles}, {Natoli}, {Navarro-Alsina},
  {Nibarger}, {Noble}, {Novosad}, {Ogando}, {Omori}, {Padin}, {Pandey},
  {Paschos}, {Patil}, {Pieres}, {Plazas Malag{\'o}n}, {Porredon}, {Prat},
  {Pryke}, {Raveri}, {Reichardt}, {Roberson}, {Rollins}, {Romero}, {Roodman},
  {Ruhl}, {Rykoff}, {Saliwanchik}, {Salvati}, {S{\'a}nchez}, {Sanchez},
  {Sanchez Cid}, {Saro}, {Schaffer}, {Secco}, {Sevilla-Noarbe}, {Sharon},
  {Sheldon}, {Shin}, {Sievers}, {Smecher}, {Smith}, {Somboonpanyakul},
  {Sommer}, {Stalder}, {Stark}, {Stephen}, {Strazzullo}, {Suchyta}, {Tarle},
  {To}, {Troxel}, {Tucker}, {Tutusaus}, {Varga}, {Veach}, {Vieira},
  {Vikhlinin}, {von der Linden}, {Wang}, {Weaverdyck}, {Weller}, {Whitehorn},
  {Wu}, {Yanny}, {Yefremenko}, {Yin}, {Young}, {Zebrowski}, {Zhang}, {Zohren},
  {Zuntz}, {(SPT}, \& {DES Collaborations)}}]{2024PhRvD.110h3510Bocquet}
{Bocquet}, S., {Grandis}, S., {Bleem}, L.~E., {et~al.} 2024, \prd, 110, 083510

\bibitem[{{Bocquet} {et~al.}(2020){Bocquet}, {Heitmann}, {Habib}, {Lawrence},
  {Uram}, {Frontiere}, {Pope}, \& {Finkel}}]{Bocquet2020MiraTitan}
{Bocquet}, S., {Heitmann}, K., {Habib}, S., {et~al.} 2020, \apj, 901, 5

\bibitem[{{Bocquet} {et~al.}(2016){Bocquet}, {Saro}, {Dolag}, \&
  {Mohr}}]{2016MNRAS.456.2361Bocquet}
{Bocquet}, S., {Saro}, A., {Dolag}, K., \& {Mohr}, J.~J. 2016, \mnras, 456,
  2361

\bibitem[{{Bose} {et~al.}(2019){Bose}, {Eisenstein}, {Hernquist}, {Pillepich},
  {Nelson}, {Marinacci}, {Springel}, \&
  {Vogelsberger}}]{2019MNRAS.490.5693Bose}
{Bose}, S., {Eisenstein}, D.~J., {Hernquist}, L., {et~al.} 2019, \mnras, 490,
  5693

\bibitem[{{Boylan-Kolchin} {et~al.}(2009){Boylan-Kolchin}, {Springel}, {White},
  {Jenkins}, \& {Lemson}}]{2009MNRAS.398.1150Boylan}
{Boylan-Kolchin}, M., {Springel}, V., {White}, S. D.~M., {Jenkins}, A., \&
  {Lemson}, G. 2009, \mnras, 398, 1150

\bibitem[{{Castignani} \& {Benoist}(2016)}]{Castignani2016richness}
{Castignani}, G. \& {Benoist}, C. 2016, \aap, 595, A111

\bibitem[{{Child} {et~al.}(2018){Child}, {Habib}, {Heitmann}, {Frontiere},
  {Finkel}, {Pope}, \& {Morozov}}]{Child2018MCscatter}
{Child}, H.~L., {Habib}, S., {Heitmann}, K., {et~al.} 2018, \apj, 859, 55

\bibitem[{{Contreras-Santos} {et~al.}(2024){Contreras-Santos}, {Buitrago},
  {Knebe}, {Rasia}, {Pearce}, {Cui}, {Power}, \&
  {Winstanley}}]{ContrerasSantos24NoExotic}
{Contreras-Santos}, A., {Buitrago}, F., {Knebe}, A., {et~al.} 2024, \aap, 690,
  A109

\bibitem[{{Corasaniti} {et~al.}(2022){Corasaniti}, {Le Brun}, {Richardson},
  {Rasera}, {Ettori}, {Arnaud}, \& {Pratt}}]{Corasaniti2022Sparsity}
{Corasaniti}, P.~S., {Le Brun}, A.~M.~C., {Richardson}, T.~R.~G., {et~al.}
  2022, \mnras, 516, 437

\bibitem[{{Costanzi} {et~al.}(2019){Costanzi}, {Rozo}, {Simet}, {Zhang},
  {Evrard}, {Mantz}, {Rykoff}, {Jeltema}, {Gruen}, {Allen}, {McClintock},
  {Romer}, {von der Linden}, {Farahi}, {DeRose}, {Varga}, {Weller}, {Giles},
  {Hollowood}, {Bhargava}, {Bermeo-Hernandez}, {Chen}, {Abbott}, {Abdalla},
  {Avila}, {Bechtol}, {Brooks}, {Buckley-Geer}, {Burke}, {Rosell}, {Kind},
  {Carretero}, {Crocce}, {Cunha}, {da Costa}, {Davis}, {De Vicente}, {Diehl},
  {Dietrich}, {Doel}, {Eifler}, {Estrada}, {Flaugher}, {Fosalba}, {Frieman},
  {Garc{\'\i}a-Bellido}, {Gaztanaga}, {Gerdes}, {Giannantonio}, {Gruendl},
  {Gschwend}, {Gutierrez}, {Hartley}, {Honscheid}, {Hoyle}, {James}, {Krause},
  {Kuehn}, {Kuropatkin}, {Lima}, {Lin}, {Maia}, {March}, {Marshall}, {Martini},
  {Menanteau}, {Miller}, {Miquel}, {Mohr}, {Ogando}, {Plazas}, {Roodman},
  {Sanchez}, {Scarpine}, {Schindler}, {Schubnell}, {Serrano}, {Sevilla-Noarbe},
  {Sheldon}, {Smith}, {Soares-Santos}, {Sobreira}, {Suchyta}, {Swanson},
  {Tarle}, {Thomas}, \& {Wechsler}}]{2019MNRAS.488.4779Costanzi}
{Costanzi}, M., {Rozo}, E., {Simet}, M., {et~al.} 2019, \mnras, 488, 4779

\bibitem[{{Cui} {et~al.}(2022){Cui}, {Dave}, {Knebe}, {Rasia}, {Gray},
  {Pearce}, {Power}, {Yepes}, {Anbajagane}, {Ceverino}, {Contreras-Santos}, {de
  Andres}, {De Petris}, {Ettori}, {Haggar}, {Li}, {Wang}, {Yang}, {Borgani},
  {Dolag}, {Zu}, {Kuchner}, {Ca{\~n}as}, {Ferragamo}, \&
  {Gianfagna}}]{Cui22GizmoRun}
{Cui}, W., {Dave}, R., {Knebe}, A., {et~al.} 2022, \mnras, 514, 977

\bibitem[{{Davies} {et~al.}(2020){Davies}, {Crain}, {Oppenheimer}, \&
  {Schaye}}]{2020MNRAS.491.4462Davies}
{Davies}, J.~J., {Crain}, R.~A., {Oppenheimer}, B.~D., \& {Schaye}, J. 2020,
  \mnras, 491, 4462

\bibitem[{{Davis} {et~al.}(1985){Davis}, {Efstathiou}, {Frenk}, \&
  {White}}]{1985ApJ...292..371Davis}
{Davis}, M., {Efstathiou}, G., {Frenk}, C.~S., \& {White}, S.~D.~M. 1985, \apj,
  292, 371

\bibitem[{{Diemer}(2018)}]{2018ApJS..239...35Diemer}
{Diemer}, B. 2018, \apjs, 239, 35

\bibitem[{{Dolag} {et~al.}(2009){Dolag}, {Borgani}, {Murante}, \&
  {Springel}}]{2009MNRAS.399..497Dolag}
{Dolag}, K., {Borgani}, S., {Murante}, G., \& {Springel}, V. 2009, \mnras, 399,
  497

\bibitem[{{Dolag} {et~al.}(2015){Dolag}, {Gaensler}, {Beck}, \&
  {Beck}}]{2015MNRAS.451.4277Dolag}
{Dolag}, K., {Gaensler}, B.~M., {Beck}, A.~M., \& {Beck}, M.~C. 2015, \mnras,
  451, 4277

\bibitem[{{Dolag} {et~al.}(2016){Dolag}, {Komatsu}, \&
  {Sunyaev}}]{2016MNRAS.463.1797Dolag}
{Dolag}, K., {Komatsu}, E., \& {Sunyaev}, R. 2016, \mnras, 463, 1797

\bibitem[{{Ettori} {et~al.}(2013){Ettori}, {Donnarumma}, {Pointecouteau},
  {Reiprich}, {Giodini}, {Lovisari}, \& {Schmidt}}]{Ettori2013MassProfiles}
{Ettori}, S., {Donnarumma}, A., {Pointecouteau}, E., {et~al.} 2013, \ssr, 177,
  119

\bibitem[{{Euclid Collaboration: Adam} {et~al.}(2019){Euclid Collaboration:
  Adam}, {Vannier}, {Maurogordato}, {Biviano}, {Adami}, {Ascaso}, {Bellagamba},
  {Benoist}, {Cappi}, {D{\'\i}az-S{\'a}nchez}, {Durret}, {Farrens}, {Gonzalez},
  {Iovino}, {Licitra}, {Maturi}, {Mei}, {Merson}, {Munari}, {Pell{\'o}},
  {Ricci}, {Rocci}, {Roncarelli}, {Sarron}, {Amoura}, {Andreon}, {Apostolakos},
  {Arnaud}, {Bardelli}, {Bartlett}, {Baugh}, {Borgani}, {Brodwin}, {Castander},
  {Castignani}, {Cucciati}, {De Lucia}, {Dubath}, {Fosalba}, {Giocoli},
  {Hoekstra}, {Mamon}, {Melin}, {Moscardini}, {Paltani}, {Radovich},
  {Sartoris}, {Schultheis}, {Sereno}, {Weller}, {Burigana}, {Carvalho},
  {Corcione}, {Kurki-Suonio}, {Lilje}, {Sirri}, {Toledo-Moreo}, \&
  {Zamorani}}]{Adam-EP3}
{Euclid Collaboration: Adam}, R., {Vannier}, M., {Maurogordato}, S., {et~al.}
  2019, \aap, 627, A23

\bibitem[{{Euclid Collaboration: Ajani} {et~al.}(2023){Euclid Collaboration:
  Ajani}, {Baldi}, {Barthelemy}, {Boyle}, {Burger}, {Cardone}, {Cheng},
  {Codis}, {Giocoli}, {Harnois-D{\'e}raps}, {Heydenreich}, {Kansal},
  {Kilbinger}, {Linke}, {Llinares}, {Martinet}, {Parroni}, {Peel}, {Pires},
  {Porth}, {Tereno}, {Uhlemann}, {Vicinanza}, {Vinciguerra}, {Aghanim},
  {Auricchio}, {Bonino}, {Branchini}, {Brescia}, {Brinchmann}, {Camera},
  {Capobianco}, {Carbone}, {Carretero}, {Castander}, {Castellano}, {Cavuoti},
  {Cimatti}, {Cledassou}, {Congedo}, {Conselice}, {Conversi}, {Corcione},
  {Courbin}, {Cropper}, {Da Silva}, {Degaudenzi}, {Di Giorgio}, {Dinis},
  {Douspis}, {Dubath}, {Dupac}, {Farrens}, {Ferriol}, {Fosalba}, {Frailis},
  {Franceschi}, {Galeotta}, {Garilli}, {Gillis}, {Grazian}, {Grupp},
  {Hoekstra}, {Holmes}, {Hornstrup}, {Hudelot}, {Jahnke}, {Jhabvala},
  {K{\"u}mmel}, {Kitching}, {Kunz}, {Kurki-Suonio}, {Lilje}, {Lloro},
  {Maiorano}, {Mansutti}, {Marggraf}, {Markovic}, {Marulli}, {Massey}, {Mei},
  {Mellier}, {Meneghetti}, {Moresco}, {Moscardini}, {Niemi}, {Nightingale},
  {Nutma}, {Padilla}, {Paltani}, {Pedersen}, {Pettorino}, {Polenta}, {Poncet},
  {Popa}, {Raison}, {Renzi}, {Rhodes}, {Riccio}, {Romelli}, {Roncarelli},
  {Rossetti}, {Saglia}, {Sapone}, {Sartoris}, {Schneider}, {Schrabback},
  {Secroun}, {Seidel}, {Serrano}, {Sirignano}, {Stanco}, {Starck},
  {Tallada-Cresp{\'\i}}, {Taylor}, {Toledo-Moreo}, {Torradeflot}, {Tutusaus},
  {Valentijn}, {Valenziano}, {Vassallo}, {Wang}, {Weller}, {Zamorani},
  {Zoubian}, {Andreon}, {Bardelli}, {Boucaud}, {Bozzo}, {Colodro-Conde}, {Di
  Ferdinando}, {Fabbian}, {Farina}, {Graci{\'a}-Carpio}, {Keih{\"a}nen},
  {Lindholm}, {Maino}, {Mauri}, {Neissner}, {Schirmer}, {Scottez}, {Zucca},
  {Akrami}, {Baccigalupi}, {Balaguera-Antol{\'\i}nez}, {Ballardini},
  {Bernardeau}, {Biviano}, {Blanchard}, {Borgani}, {Borlaff}, {Burigana},
  {Cabanac}, {Cappi}, {Carvalho}, {Casas}, {Castignani}, {Castro}, {Chambers},
  {Cooray}, {Coupon}, {Courtois}, {Davini}, {de la Torre}, {De Lucia},
  {Desprez}, {Dole}, {Escartin}, {Escoffier}, {Ferrero}, {Finelli}, {Ganga},
  {Garcia-Bellido}, {George}, {Giacomini}, {Gozaliasl}, {Hildebrandt}, {Jimenez
  Mu{\~n}oz}, {Joachimi}, {Kajava}, {Kirkpatrick}, {Legrand}, {Loureiro},
  {Magliocchetti}, {Maoli}, {Marcin}, {Martinelli}, {Martins}, {Matthew},
  {Maurin}, {Metcalf}, {Monaco}, {Morgante}, {Nadathur}, {Nucita}, {Popa},
  {Potter}, {Pourtsidou}, {P{\"o}ntinen}, {Reimberg}, {S{\'a}nchez}, {Sakr},
  {Schneider}, {Sefusatti}, {Sereno}, {Shulevski}, {Spurio Mancini},
  {Steinwagner}, {Teyssier}, {Valiviita}, {Veropalumbo}, {Viel}, \&
  {Zinchenko}}]{Ajani-EP29}
{Euclid Collaboration: Ajani}, V., {Baldi}, M., {Barthelemy}, A., {et~al.}
  2023, \aap, 675, A120

\bibitem[{{Euclid Collaboration: Blanchard} {et~al.}(2020){Euclid
  Collaboration: Blanchard}, {Camera}, {Carbone}, {Cardone}, {Casas}, {Clesse},
  {Ili{\'c}}, {Kilbinger}, {Kitching}, {Kunz}, {Lacasa}, {Linder}, {Majerotto},
  {Markovi{\v{c}}}, {Martinelli}, {Pettorino}, {Pourtsidou}, {Sakr},
  {S{\'a}nchez}, {Sapone}, {Tutusaus}, {Yahia-Cherif}, {Yankelevich},
  {Andreon}, {Aussel}, {Balaguera-Antol{\'\i}nez}, {Baldi}, {Bardelli},
  {Bender}, {Biviano}, {Bonino}, {Boucaud}, {Bozzo}, {Branchini}, {Brau-Nogue},
  {Brescia}, {Brinchmann}, {Burigana}, {Cabanac}, {Capobianco}, {Cappi},
  {Carretero}, {Carvalho}, {Casas}, {Castander}, {Castellano}, {Cavuoti},
  {Cimatti}, {Cledassou}, {Colodro-Conde}, {Congedo}, {Conselice}, {Conversi},
  {Copin}, {Corcione}, {Coupon}, {Courtois}, {Cropper}, {Da Silva}, {de la
  Torre}, {Di Ferdinando}, {Dubath}, {Ducret}, {Duncan}, {Dupac}, {Dusini},
  {Fabbian}, {Fabricius}, {Farrens}, {Fosalba}, {Fotopoulou}, {Fourmanoit},
  {Frailis}, {Franceschi}, {Franzetti}, {Fumana}, {Galeotta}, {Gillard},
  {Gillis}, {Giocoli}, {G{\'o}mez-Alvarez}, {Graci{\'a}-Carpio}, {Grupp},
  {Guzzo}, {Hoekstra}, {Hormuth}, {Israel}, {Jahnke}, {Keihanen}, {Kermiche},
  {Kirkpatrick}, {Kohley}, {Kubik}, {Kurki-Suonio}, {Ligori}, {Lilje}, {Lloro},
  {Maino}, {Maiorano}, {Marggraf}, {Martinet}, {Marulli}, {Massey},
  {Medinaceli}, {Mei}, {Mellier}, {Metcalf}, {Metge}, {Meylan}, {Moresco},
  {Moscardini}, {Munari}, {Nichol}, {Niemi}, {Nucita}, {Padilla}, {Paltani},
  {Pasian}, {Percival}, {Pires}, {Polenta}, {Poncet}, {Pozzetti}, {Racca},
  {Raison}, {Renzi}, {Rhodes}, {Romelli}, {Roncarelli}, {Rossetti}, {Saglia},
  {Schneider}, {Scottez}, {Secroun}, {Sirri}, {Stanco}, {Starck}, {Sureau},
  {Tallada-Cresp{\'\i}}, {Tavagnacco}, {Taylor}, {Tenti}, {Tereno},
  {Toledo-Moreo}, {Torradeflot}, {Valenziano}, {Vassallo}, {Verdoes Kleijn},
  {Viel}, {Wang}, {Zacchei}, {Zoubian}, \& {Zucca}}]{Blanchard-EP7}
{Euclid Collaboration: Blanchard}, A., {Camera}, S., {Carbone}, C., {et~al.}
  2020, \aap, 642, A191

\bibitem[{{Euclid Collaboration: Desprez} {et~al.}(2020){Euclid Collaboration:
  Desprez}, {Paltani}, {Coupon}, {Almosallam}, {Alvarez-Ayllon}, {Amaro},
  {Brescia}, {Brodwin}, {Cavuoti}, {De Vicente-Albendea}, {Fotopoulou},
  {Hatfield}, {Hartley}, {Ilbert}, {Jarvis}, {Longo}, {Rau}, {Saha}, {Speagle},
  {Tramacere}, {Castellano}, {Dubath}, {Galametz}, {Kuemmel}, {Laigle},
  {Merlin}, {Mohr}, {Pilo}, {Salvato}, {Andreon}, {Auricchio}, {Baccigalupi},
  {Balaguera-Antol{\'\i}nez}, {Baldi}, {Bardelli}, {Bender}, {Biviano},
  {Bodendorf}, {Bonino}, {Bozzo}, {Branchini}, {Brinchmann}, {Burigana},
  {Cabanac}, {Camera}, {Capobianco}, {Cappi}, {Carbone}, {Carretero},
  {Carvalho}, {Casas}, {Casas}, {Castander}, {Castignani}, {Cimatti},
  {Cledassou}, {Colodro-Conde}, {Congedo}, {Conselice}, {Conversi}, {Copin},
  {Corcione}, {Courtois}, {Cuby}, {Da Silva}, {de la Torre}, {Degaudenzi}, {Di
  Ferdinando}, {Douspis}, {Duncan}, {Dupac}, {Ealet}, {Fabbian}, {Fabricius},
  {Farrens}, {Ferreira}, {Finelli}, {Fosalba}, {Fourmanoit}, {Frailis},
  {Franceschi}, {Fumana}, {Galeotta}, {Garilli}, {Gillard}, {Gillis},
  {Giocoli}, {Gozaliasl}, {Graci{\'a}-Carpio}, {Grupp}, {Guzzo}, {Hailey},
  {Haugan}, {Holmes}, {Hormuth}, {Humphrey}, {Jahnke}, {Keihanen}, {Kermiche},
  {Kilbinger}, {Kirkpatrick}, {Kitching}, {Kohley}, {Kubik}, {Kunz},
  {Kurki-Suonio}, {Ligori}, {Lilje}, {Lloro}, {Maino}, {Maiorano}, {Marggraf},
  {Markovic}, {Martinet}, {Marulli}, {Massey}, {Maturi}, {Mauri},
  {Maurogordato}, {Medinaceli}, {Mei}, {Meneghetti}, {Metcalf}, {Meylan},
  {Moresco}, {Moscardini}, {Munari}, {Niemi}, {Padilla}, {Pasian}, {Patrizii},
  {Pettorino}, {Pires}, {Polenta}, {Poncet}, {Popa}, {Potter}, {Pozzetti},
  {Raison}, {Renzi}, {Rhodes}, {Riccio}, {Rossetti}, {Saglia}, {Sapone},
  {Schneider}, {Scottez}, {Secroun}, {Serrano}, {Sirignano}, {Sirri}, {Stanco},
  {Stern}, {Sureau}, {Tallada Cresp{\'\i}}, {Tavagnacco}, {Taylor}, {Tenti},
  {Tereno}, {Toledo-Moreo}, {Torradeflot}, {Valenziano}, {Valiviita},
  {Vassallo}, {Viel}, {Wang}, {Welikala}, {Whittaker}, {Zacchei}, {Zamorani},
  {Zoubian}, \& {Zucca}}]{Desprez-EP10}
{Euclid Collaboration: Desprez}, G., {Paltani}, S., {Coupon}, J., {et~al.}
  2020, \aap, 644, A31

\bibitem[{{Euclid Collaboration: Giocoli} {et~al.}(2024){Euclid Collaboration:
  Giocoli}, {Meneghetti}, {Rasia}, {Borgani}, {Despali}, {Lesci}, {Marulli},
  {Moscardini}, {Sereno}, {Cui}, {Knebe}, {Yepes}, {Castro}, {Corasaniti},
  {Pires}, {Castignani}, {Schrabback}, {Pratt}, {Le Brun}, {Aghanim},
  {Amendola}, {Auricchio}, {Baldi}, {Bodendorf}, {Bonino}, {Branchini},
  {Brescia}, {Brinchmann}, {Camera}, {Capobianco}, {Carbone}, {Carretero},
  {Castander}, {Castellano}, {Cavuoti}, {Cledassou}, {Congedo}, {Conselice},
  {Conversi}, {Copin}, {Corcione}, {Courbin}, {Cropper}, {Da Silva},
  {Degaudenzi}, {Dinis}, {Dubath}, {Dupac}, {Dusini}, {Farrens}, {Ferriol},
  {Fosalba}, {Frailis}, {Franceschi}, {Fumana}, {Galeotta}, {Garilli},
  {Gillis}, {Grazian}, {Grupp}, {Haugan}, {Holmes}, {Hornstrup}, {Jahnke},
  {K{\"u}mmel}, {Kermiche}, {Kilbinger}, {Kunz}, {Kurki-Suonio}, {Ligori},
  {Lilje}, {Lloro}, {Maiorano}, {Mansutti}, {Marggraf}, {Markovic}, {Massey},
  {Maurogordato}, {Mei}, {Merlin}, {Meylan}, {Moresco}, {Munari}, {Niemi},
  {Nightingale}, {Nutma}, {Padilla}, {Paltani}, {Pasian}, {Pedersen},
  {Pettorino}, {Polenta}, {Poncet}, {Popa}, {Raison}, {Renzi}, {Rhodes},
  {Riccio}, {Romelli}, {Roncarelli}, {Rossetti}, {Saglia}, {Sapone},
  {Sartoris}, {Schneider}, {Secroun}, {Serrano}, {Sirignano}, {Sirri},
  {Stanco}, {Starck}, {Tallada-Cresp{\'\i}}, {Taylor}, {Tereno},
  {Toledo-Moreo}, {Torradeflot}, {Tutusaus}, {Valentijn}, {Valenziano},
  {Vassallo}, {Wang}, {Weller}, {Zamorani}, {Zoubian}, {Andreon}, {Bardelli},
  {Boucaud}, {Bozzo}, {Colodro-Conde}, {Di Ferdinando}, {Fabbian}, {Farina},
  {Israel}, {Keih{\"a}nen}, {Lindholm}, {Mauri}, {Neissner}, {Schirmer},
  {Scottez}, {Tenti}, {Zucca}, {Akrami}, {Baccigalupi}, {Ballardini},
  {Bernardeau}, {Biviano}, {Borlaff}, {Burigana}, {Cabanac}, {Cappi},
  {Carvalho}, {Casas}, {Chambers}, {Cooray}, {Courtois}, {Davini}, {de la
  Torre}, {De Lucia}, {Desprez}, {Dole}, {Escartin}, {Escoffier}, {Ferrero},
  {Finelli}, {Gabarra}, {Ganga}, {Garcia-Bellido}, {George}, {Giacomini},
  {Gozaliasl}, {Hildebrandt}, {Hook}, {Jimenez Mu{\~n}oz}, {Joachimi},
  {Kajava}, {Kansal}, {Kirkpatrick}, {Legrand}, {Loureiro}, {Macias-Perez},
  {Magliocchetti}, {Mainetti}, {Maoli}, {Marcin}, {Martinelli}, {Martinet},
  {Martins}, {Matthew}, {Maurin}, {Metcalf}, {Monaco}, {Morgante}, {Nadathur},
  {Nucita}, {Patrizii}, {Peel}, {Pollack}, {Popa}, {Porciani}, {Potter},
  {P{\"o}ntinen}, {Reimberg}, {S{\'a}nchez}, {Sakr}, {Schneider}, {Sefusatti},
  {Shulevski}, {Spurio Mancini}, {Stadel}, {Steinwagner}, {Valiviita},
  {Veropalumbo}, {Viel}, \& {Zinchenko}}]{Giocoli-EP30}
{Euclid Collaboration: Giocoli}, C., {Meneghetti}, M., {Rasia}, E., {et~al.}
  2024, \aap, 681, A67

\bibitem[{{Euclid Collaboration: Mellier} {et~al.}(2024){Euclid Collaboration:
  Mellier}, {Abdurro'uf}, {Acevedo~Barroso}, {Ach\'ucarro},
  {et~al.}}]{EuclidSkyOverview}
{Euclid Collaboration: Mellier}, Y., {Abdurro'uf}, {Acevedo~Barroso}, J.,
  {Ach\'ucarro}, A., {et~al.} 2024, \aap, submitted, arXiv:2405.13491

\bibitem[{{Fabjan} {et~al.}(2010){Fabjan}, {Borgani}, {Tornatore}, {Saro},
  {Murante}, \& {Dolag}}]{2010MNRAS.401.1670Fabjan}
{Fabjan}, D., {Borgani}, S., {Tornatore}, L., {et~al.} 2010, \mnras, 401, 1670

\bibitem[{{Farahi} {et~al.}(2019){Farahi}, {Mulroy}, {Evrard}, {Smith},
  {Finoguenov}, {Bourdin}, {Carlstrom}, {Haines}, {Marrone}, {Martino},
  {Mazzotta}, {O'Donnell}, \& {Okabe}}]{2019NatCo..10.2504Farahi}
{Farahi}, A., {Mulroy}, S.~L., {Evrard}, A.~E., {et~al.} 2019, Nature
  Communications, 10, 2504

\bibitem[{{Ferland} {et~al.}(1998){Ferland}, {Korista}, {Verner}, {Ferguson},
  {Kingdon}, \& {Verner}}]{1998PASP..110..761Ferland}
{Ferland}, G.~J., {Korista}, K.~T., {Verner}, D.~A., {et~al.} 1998, \pasp, 110,
  761

\bibitem[{{Fischer} {et~al.}(2024){Fischer}, {Kasselmann}, {Br{\"u}ggen},
  {Dolag}, {Kahlhoefer}, {Ragagnin}, {Robertson}, \&
  {Schmidt-Hoberg}}]{Fischer24Mergers}
{Fischer}, M.~S., {Kasselmann}, L., {Br{\"u}ggen}, M., {et~al.} 2024, \mnras,
  529, 2327

\bibitem[{{Ghirardini} {et~al.}(2024){Ghirardini}, {Bulbul}, {Artis}, {Clerc},
  {Garrel}, {Grandis}, {Kluge}, {Liu}, {Bahar}, {Balzer}, {Chiu}, {Comparat},
  {Gruen}, {Kleinebreil}, {Krippendorf}, {Merloni}, {Nandra}, {Okabe},
  {Pacaud}, {Predehl}, {Ramos-Ceja}, {Reiprich}, {Sanders}, {Schrabback},
  {Seppi}, {Zelmer}, {Zhang}, {Bornemann}, {Brunner}, {Burwitz}, {Coutinho},
  {Dennerl}, {Freyberg}, {Friedrich}, {Gaida}, {Gueguen}, {Haberl}, {Kink},
  {Lamer}, {Li}, {Liu}, {Maitra}, {Meidinger}, {Mueller}, {Miyatake},
  {Miyazaki}, {Robrade}, {Schwope}, \& {Stewart}}]{Ghirardini2024arXivCosmo}
{Ghirardini}, V., {Bulbul}, E., {Artis}, E., {et~al.} 2024, arXiv e-prints,
  arXiv:2402.08458

\bibitem[{{Giocoli} {et~al.}(2008){Giocoli}, {Tormen}, \& {van den
  Bosch}}]{Giocoli2008mass-loss}
{Giocoli}, C., {Tormen}, G., \& {van den Bosch}, F.~C. 2008, \mnras, 386, 2135

\bibitem[{{Giodini} {et~al.}(2013){Giodini}, {Lovisari}, {Pointecouteau},
  {Ettori}, {Reiprich}, \& {Hoekstra}}]{2013SSRv..177..247Giodini}
{Giodini}, S., {Lovisari}, L., {Pointecouteau}, E., {et~al.} 2013, \ssr, 177,
  247

\bibitem[{{Hirschmann} {et~al.}(2014){Hirschmann}, {Dolag}, {Saro}, {Bachmann},
  {Borgani}, \& {Burkert}}]{2014MNRAS.442.2304Hirschmann}
{Hirschmann}, M., {Dolag}, K., {Saro}, A., {et~al.} 2014, \mnras, 442, 2304

\bibitem[{{Hoekstra}(2003)}]{2003MNRAS.339.1155Hoekstra}
{Hoekstra}, H. 2003, \mnras, 339, 1155

\bibitem[{{Hoekstra} {et~al.}(2012){Hoekstra}, {Mahdavi}, {Babul}, \&
  {Bildfell}}]{2012MNRAS.427.1298Hoekstra}
{Hoekstra}, H., {Mahdavi}, A., {Babul}, A., \& {Bildfell}, C. 2012, \mnras,
  427, 1298

\bibitem[{{Hudson} {et~al.}(2010){Hudson}, {Mittal}, {Reiprich}, {Nulsen},
  {Andernach}, \& {Sarazin}}]{2010A&A...513A..37Hudson}
{Hudson}, D.~S., {Mittal}, R., {Reiprich}, T.~H., {et~al.} 2010, \aap, 513, A37

\bibitem[{{Komatsu} {et~al.}(2011){Komatsu}, {Smith}, {Dunkley}, {Bennett},
  {Gold}, {Hinshaw}, {Jarosik}, {Larson}, {Nolta}, {Page}, {Spergel},
  {Halpern}, {Hill}, {Kogut}, {Limon}, {Meyer}, {Odegard}, {Tucker}, {Weiland},
  {Wollack}, \& {Wright}}]{2011ApJS..192...18Komatsu}
{Komatsu}, E., {Smith}, K.~M., {Dunkley}, J., {et~al.} 2011, \apjs, 192, 18

\bibitem[{{Kravtsov} \& {Borgani}(2012)}]{2012ARA&A..50..353Kravtsov}
{Kravtsov}, A.~V. \& {Borgani}, S. 2012, \araa, 50, 353

\bibitem[{{Lima} \& {Hu}(2005)}]{2005PhRvD..72d3006Lima}
{Lima}, M. \& {Hu}, W. 2005, \prd, 72, 043006

\bibitem[{{Ludlow} {et~al.}(2012){Ludlow}, {Navarro}, {Li}, {Angulo},
  {Boylan-Kolchin}, \& {Bett}}]{2012MNRAS.427.1322Ludlow}
{Ludlow}, A.~D., {Navarro}, J.~F., {Li}, M., {et~al.} 2012, \mnras, 427, 1322

\bibitem[{{Macci{\`o}} {et~al.}(2007){Macci{\`o}}, {Dutton}, {van den Bosch},
  {Moore}, {Potter}, \& {Stadel}}]{Maccio07}
{Macci{\`o}}, A.~V., {Dutton}, A.~A., {van den Bosch}, F.~C., {et~al.} 2007,
  \mnras, 378, 55

\bibitem[{{Maturi} {et~al.}(2019){Maturi}, {Bellagamba}, {Radovich},
  {Roncarelli}, {Sereno}, {Moscardini}, {Bardelli}, \& {Puddu}}]{amicoB}
{Maturi}, M., {Bellagamba}, F., {Radovich}, M., {et~al.} 2019, \mnras, 485, 498

\bibitem[{{McClintock} {et~al.}(2019){McClintock}, {Varga}, {Gruen}, {Rozo},
  {Rykoff}, {Shin}, {Melchior}, {DeRose}, {Seitz}, {Dietrich}, {Sheldon},
  {Zhang}, {von der Linden}, {Jeltema}, {Mantz}, {Romer}, {Allen}, {Becker},
  {Bermeo}, {Bhargava}, {Costanzi}, {Everett}, {Farahi}, {Hamaus}, {Hartley},
  {Hollowood}, {Hoyle}, {Israel}, {Li}, {MacCrann}, {Morris}, {Palmese},
  {Plazas}, {Pollina}, {Rau}, {Simet}, {Soares-Santos}, {Troxel}, {Vergara
  Cervantes}, {Wechsler}, {Zuntz}, {Abbott}, {Abdalla}, {Allam}, {Annis},
  {Avila}, {Bridle}, {Brooks}, {Burke}, {Carnero Rosell}, {Carrasco Kind},
  {Carretero}, {Castander}, {Crocce}, {Cunha}, {D'Andrea}, {da Costa}, {Davis},
  {De Vicente}, {Diehl}, {Doel}, {Drlica-Wagner}, {Evrard}, {Flaugher},
  {Fosalba}, {Frieman}, {Garc{\'\i}a-Bellido}, {Gaztanaga}, {Gerdes},
  {Giannantonio}, {Gruendl}, {Gutierrez}, {Honscheid}, {James}, {Kirk},
  {Krause}, {Kuehn}, {Lahav}, {Li}, {Lima}, {March}, {Marshall}, {Menanteau},
  {Miquel}, {Mohr}, {Nord}, {Ogando}, {Roodman}, {Sanchez}, {Scarpine},
  {Schindler}, {Sevilla-Noarbe}, {Smith}, {Smith}, {Sobreira}, {Suchyta},
  {Swanson}, {Tarle}, {Tucker}, {Vikram}, {Walker}, {Weller}, \& {DES
  Collaboration}}]{McClintock2019}
{McClintock}, T., {Varga}, T.~N., {Gruen}, D., {et~al.} 2019, \mnras, 482, 1352

\bibitem[{{Melchior} {et~al.}(2015){Melchior}, {Suchyta}, {Huff}, {Hirsch},
  {Kacprzak}, {Rykoff}, {Gruen}, {Armstrong}, {Bacon}, {Bechtol}, {Bernstein},
  {Bridle}, {Clampitt}, {Honscheid}, {Jain}, {Jouvel}, {Krause}, {Lin},
  {MacCrann}, {Patton}, {Plazas}, {Rowe}, {Vikram}, {Wilcox}, {Young}, {Zuntz},
  {Abbott}, {Abdalla}, {Allam}, {Banerji}, {Bernstein}, {Bernstein}, {Bertin},
  {Buckley-Geer}, {Burke}, {Castander}, {da Costa}, {Cunha}, {Depoy}, {Desai},
  {Diehl}, {Doel}, {Estrada}, {Evrard}, {Neto}, {Fernandez}, {Finley},
  {Flaugher}, {Frieman}, {Gaztanaga}, {Gerdes}, {Gruendl}, {Gutierrez},
  {Jarvis}, {Karliner}, {Kent}, {Kuehn}, {Kuropatkin}, {Lahav}, {Maia},
  {Makler}, {Marriner}, {Marshall}, {Merritt}, {Miller}, {Miquel}, {Mohr},
  {Neilsen}, {Nichol}, {Nord}, {Reil}, {Roe}, {Roodman}, {Sako}, {Sanchez},
  {Santiago}, {Schindler}, {Schubnell}, {Sevilla-Noarbe}, {Sheldon}, {Smith},
  {Soares-Santos}, {Swanson}, {Sypniewski}, {Tarle}, {Thaler}, {Thomas},
  {Tucker}, {Walker}, {Wechsler}, {Weller}, \&
  {Wester}}]{2015MNRAS.449.2219Melchior}
{Melchior}, P., {Suchyta}, E., {Huff}, E., {et~al.} 2015, \mnras, 449, 2219

\bibitem[{{Meneghetti} {et~al.}(2010){Meneghetti}, {Rasia}, {Merten},
  {Bellagamba}, {Ettori}, {Mazzotta}, {Dolag}, \& {Marri}}]{meneghetti10}
{Meneghetti}, M., {Rasia}, E., {Merten}, J., {et~al.} 2010, \aap, 514, A93

\bibitem[{{Meneghetti} {et~al.}(2014){Meneghetti}, {Rasia}, {Vega}, {Merten},
  {Postman}, {Yepes}, {Sembolini}, {Donahue}, {Ettori}, {Umetsu}, {Balestra},
  {Bartelmann}, {Ben{\'\i}tez}, {Biviano}, {Bouwens}, {Bradley}, {Broadhurst},
  {Coe}, {Czakon}, {De Petris}, {Ford}, {Giocoli}, {Gottl{\"o}ber}, {Grillo},
  {Infante}, {Jouvel}, {Kelson}, {Koekemoer}, {Lahav}, {Lemze}, {Medezinski},
  {Melchior}, {Mercurio}, {Molino}, {Moscardini}, {Monna}, {Moustakas},
  {Moustakas}, {Nonino}, {Rhodes}, {Rosati}, {Sayers}, {Seitz}, {Zheng}, \&
  {Zitrin}}]{2014ApJ...797...34Meneghetti}
{Meneghetti}, M., {Rasia}, E., {Vega}, J., {et~al.} 2014, \apj, 797, 34

\bibitem[{{Nagai} {et~al.}(2007){Nagai}, {Kravtsov}, \&
  {Vikhlinin}}]{Nagai2007gNFW}
{Nagai}, D., {Kravtsov}, A.~V., \& {Vikhlinin}, A. 2007, \apj, 668, 1

\bibitem[{{Navarro} {et~al.}(1997){Navarro}, {Frenk}, \&
  {White}}]{1997ApJ...490..493Navarro}
{Navarro}, J.~F., {Frenk}, C.~S., \& {White}, S. D.~M. 1997, \apj, 490, 493

\bibitem[{{Oguri} \& {Hamana}(2011)}]{Oguri+2011truncatedNFW}
{Oguri}, M. \& {Hamana}, T. 2011, \mnras, 414, 1851

\bibitem[{{Okabe} {et~al.}(2010){Okabe}, {Zhang}, {Finoguenov}, {Takada},
  {Smith}, {Umetsu}, \& {Futamase}}]{2010ApJ...721..875Okabe}
{Okabe}, N., {Zhang}, Y.~Y., {Finoguenov}, A., {et~al.} 2010, \apj, 721, 875

\bibitem[{{Pacaud} {et~al.}(2007){Pacaud}, {Pierre}, {Adami}, {Altieri},
  {Andreon}, {Chiappetti}, {Detal}, {Duc}, {Galaz}, {Gueguen}, {Le F{\`e}vre},
  {Hertling}, {Libbrecht}, {Melin}, {Ponman}, {Quintana}, {Refregier},
  {Sprimont}, {Surdej}, {Valtchanov}, {Willis}, {Alloin}, {Birkinshaw},
  {Bremer}, {Garcet}, {Jean}, {Jones}, {Le F{\`e}vre}, {Maccagni}, {Mazure},
  {Proust}, {R{\"o}ttgering}, \& {Trinchieri}}]{2007MNRAS.382.1289Pacaud}
{Pacaud}, F., {Pierre}, M., {Adami}, C., {et~al.} 2007, \mnras, 382, 1289

\bibitem[{{Pires} {et~al.}(2020){Pires}, {Vandenbussche}, {Kansal}, {Bender},
  {Blot}, {Bonino}, {Boucaud}, {Brinchmann}, {Capobianco}, {Carretero},
  {Castellano}, {Cavuoti}, {Cl{\'e}dassou}, {Congedo}, {Conversi}, {Corcione},
  {Dubath}, {Fosalba}, {Frailis}, {Franceschi}, {Fumana}, {Grupp}, {Hormuth},
  {Kermiche}, {Knabenhans}, {Kohley}, {Kubik}, {Kunz}, {Ligori}, {Lilje},
  {Lloro}, {Maiorano}, {Marggraf}, {Massey}, {Meylan}, {Padilla}, {Paltani},
  {Pasian}, {Poncet}, {Potter}, {Raison}, {Rhodes}, {Roncarelli}, {Saglia},
  {Schneider}, {Secroun}, {Serrano}, {Stadel}, {Tallada Cresp{\'\i}}, {Tereno},
  {Toledo-Moreo}, \& {Wang}}]{2020A&A...638A.141Pires}
{Pires}, S., {Vandenbussche}, V., {Kansal}, V., {et~al.} 2020, \aap, 638, A141

\bibitem[{{Pratt} {et~al.}(2019){Pratt}, {Arnaud}, {Biviano}, {Eckert},
  {Ettori}, {Nagai}, {Okabe}, \& {Reiprich}}]{pratt19}
{Pratt}, G.~W., {Arnaud}, M., {Biviano}, A., {et~al.} 2019, \ssr, 215, 25

\bibitem[{{Puddu} \& {Andreon}(2022)}]{2022MNRAS.511.2968Puddu}
{Puddu}, E. \& {Andreon}, S. 2022, \mnras, 511, 2968

\bibitem[{Pugh \& Winslow(1966)}]{pugh1966analysis}
Pugh, E. \& Winslow, G. 1966, The Analysis of Physical Measurements (By>
  Emerson M. Pugh (And> George H. Winslow, Addison-Wesley Series in Physics

\bibitem[{{Ragagnin} {et~al.}(2022){Ragagnin}, {Andreon}, \&
  {Puddu}}]{2022A&A...666A..22Ragagnin}
{Ragagnin}, A., {Andreon}, S., \& {Puddu}, E. 2022, \aap, 666, A22

\bibitem[{{Ragagnin} {et~al.}(2017){Ragagnin}, {Dolag}, {Biffi}, {Cadolle Bel},
  {Hammer}, {Krukau}, {Petkova}, \& {Steinborn}}]{2017A&C....20...52Ragagnin}
{Ragagnin}, A., {Dolag}, K., {Biffi}, V., {et~al.} 2017, Astronomy and
  Computing, 20, 52

\bibitem[{{Ragagnin} {et~al.}(2019){Ragagnin}, {Dolag}, {Moscardini},
  {Biviano}, \& {D'Onofrio}}]{Ragagnin2019MC}
{Ragagnin}, A., {Dolag}, K., {Moscardini}, L., {Biviano}, A., \& {D'Onofrio},
  M. 2019, \mnras, 486, 4001

\bibitem[{{Ragagnin} {et~al.}(2023){Ragagnin}, {Fumagalli}, {Castro}, {Dolag},
  {Saro}, {Costanzi}, \& {Bocquet}}]{Ragagnin2023HOD}
{Ragagnin}, A., {Fumagalli}, A., {Castro}, T., {et~al.} 2023, \aap, 675, A77

\bibitem[{{Ragagnin} {et~al.}(2024){Ragagnin}, {Meneghetti}, {Calura},
  {Despali}, {Dolag}, {Fischer}, {Giocoli}, \& {Moscardini}}]{Ragagnin24SIDM}
{Ragagnin}, A., {Meneghetti}, M., {Calura}, F., {et~al.} 2024, \aap, 687, A270

\bibitem[{{Ragagnin} {et~al.}(2021){Ragagnin}, {Saro}, {Singh}, \&
  {Dolag}}]{Ragagnin2021MC}
{Ragagnin}, A., {Saro}, A., {Singh}, P., \& {Dolag}, K. 2021, \mnras, 500, 5056

\bibitem[{{Ragagnin} {et~al.}(2016){Ragagnin}, {Tchipev}, {Bader}, {Dolag}, \&
  {Hammer}}]{2016pcre.conf..411Ragagnin}
{Ragagnin}, A., {Tchipev}, N., {Bader}, M., {Dolag}, K., \& {Hammer}, N.~J.
  2016, in Advances in Parallel Computing, Volume 27: Parallel Computing: On
  the Road to Exascale, Edited by Gerhard R. Joubert, Hugh Leather, Mark
  Parsons, Frans Peters, Mark Sawyer. IOP Ebook, ISBN: 978-1-61499-621-7, pages
  411-420

\bibitem[{{Rozo} {et~al.}(2010){Rozo}, {Wechsler}, {Rykoff}, {Annis}, {Becker},
  {Evrard}, {Frieman}, {Hansen}, {Hao}, {Johnston}, {Koester}, {McKay},
  {Sheldon}, \& {Weinberg}}]{2010ApJ...708..645Rozo}
{Rozo}, E., {Wechsler}, R.~H., {Rykoff}, E.~S., {et~al.} 2010, \apj, 708, 645

\bibitem[{{Saro} {et~al.}(2014){Saro}, {Liu}, {Mohr}, {Aird}, {Ashby},
  {Bayliss}, {Benson}, {Bleem}, {Bocquet}, {Brodwin}, {Carlstrom}, {Chang},
  {Chiu}, {Cho}, {Clocchiatti}, {Crawford}, {Crites}, {de Haan}, {Desai},
  {Dietrich}, {Dobbs}, {Dolag}, {Dudley}, {Foley}, {Gangkofner}, {George},
  {Gladders}, {Gonzalez}, {Halverson}, {Hennig}, {Hlavacek-Larrondo},
  {Holzapfel}, {Hrubes}, {Jones}, {Keisler}, {Lee}, {Leitch}, {Lueker},
  {Luong-Van}, {Mantz}, {Marrone}, {McDonald}, {McMahon}, {Mehl}, {Meyer},
  {Mocanu}, {Montroy}, {Murray}, {Nurgaliev}, {Padin}, {Patej}, {Pryke},
  {Reichardt}, {Rest}, {Ruel}, {Ruhl}, {Saliwanchik}, {Sayre}, {Schaffer},
  {Shirokoff}, {Spieler}, {Stalder}, {Staniszewski}, {Stark}, {Story}, {van
  Engelen}, {Vanderlinde}, {Vieira}, {Vikhlinin}, {Williamson}, {Zahn}, \&
  {Zenteno}}]{2014MNRAS.440.2610Saro}
{Saro}, A., {Liu}, J., {Mohr}, J.~J., {et~al.} 2014, \mnras, 440, 2610

\bibitem[{{Sartoris} {et~al.}(2016){Sartoris}, {Biviano}, {Fedeli}, {Bartlett},
  {Borgani}, {Costanzi}, {Giocoli}, {Moscardini}, {Weller}, {Ascaso},
  {Bardelli}, {Maurogordato}, \& {Viana}}]{Sartoris2016}
{Sartoris}, B., {Biviano}, A., {Fedeli}, C., {et~al.} 2016, \mnras, 459, 1764

\bibitem[{{Schrabback} {et~al.}(2021){Schrabback}, {Bocquet}, {Sommer},
  {Zohren}, {van den Busch}, {Hern{\'a}ndez-Mart{\'\i}n}, {Hoekstra}, {Raihan},
  {Schirmer}, {Applegate}, {Bayliss}, {Benson}, {Bleem}, {Dietrich}, {Floyd},
  {Hilbert}, {Hlavacek-Larrondo}, {McDonald}, {Saro}, {Stark}, \&
  {Weissgerber}}]{2021MNRAS.505.3923Schrabback}
{Schrabback}, T., {Bocquet}, S., {Sommer}, M., {et~al.} 2021, \mnras, 505, 3923

\bibitem[{{Sereno} {et~al.}(2016){Sereno}, {Fedeli}, \&
  {Moscardini}}]{2016JCAP...01..042Sereno}
{Sereno}, M., {Fedeli}, C., \& {Moscardini}, L. 2016, \jcap, 2016, 042

\bibitem[{{Sereno} {et~al.}(2020){Sereno}, {Umetsu}, {Ettori}, {Eckert},
  {Gastaldello}, {Giles}, {Lieu}, {Maughan}, {Okabe}, {Birkinshaw}, {Chiu},
  {Fujita}, {Miyazaki}, {Rapetti}, {Koulouridis}, \&
  {Pierre}}]{2020MNRAS.492.4528Sereno}
{Sereno}, M., {Umetsu}, K., {Ettori}, S., {et~al.} 2020, \mnras, 492, 4528

\bibitem[{{Singh} {et~al.}(2020){Singh}, {Saro}, {Costanzi}, \&
  {Dolag}}]{2020MNRAS.494.3728Singh}
{Singh}, P., {Saro}, A., {Costanzi}, M., \& {Dolag}, K. 2020, \mnras, 494, 3728

\bibitem[{{Smith} {et~al.}(2001){Smith}, {Brickhouse}, {Liedahl}, \&
  {Raymond}}]{2001ApJ...556L..91Smith}
{Smith}, R.~K., {Brickhouse}, N.~S., {Liedahl}, D.~A., \& {Raymond}, J.~C.
  2001, \apjl, 556, L91

\bibitem[{{Sommer} {et~al.}(2022){Sommer}, {Schrabback}, {Applegate},
  {Hilbert}, {Ansarinejad}, {Floyd}, \& {Grandis}}]{2022MNRAS.509.1127Sommer}
{Sommer}, M.~W., {Schrabback}, T., {Applegate}, D.~E., {et~al.} 2022, \mnras,
  509, 1127

\bibitem[{{Sommer} {et~al.}(2023){Sommer}, {Schrabback}, {Ragagnin}, \&
  {Rockenfeller}}]{Sommer+23subm}
{Sommer}, M.~W., {Schrabback}, T., {Ragagnin}, A., \& {Rockenfeller}, R. 2023,
  arXiv, arXiv:2306.13187

\bibitem[{{Springel}(2005)}]{2005MNRAS.364.1105Springel}
{Springel}, V. 2005, \mnras, 364, 1105

\bibitem[{{Springel} {et~al.}(2005{\natexlab{a}}){Springel}, {Di Matteo}, \&
  {Hernquist}}]{2005MNRAS.361..776Springel}
{Springel}, V., {Di Matteo}, T., \& {Hernquist}, L. 2005{\natexlab{a}}, \mnras,
  361, 776

\bibitem[{{Springel} {et~al.}(2005{\natexlab{b}}){Springel}, {White},
  {Jenkins}, {Frenk}, {Yoshida}, {Gao}, {Navarro}, {Thacker}, {Croton},
  {Helly}, {Peacock}, {Cole}, {Thomas}, {Couchman}, {Evrard}, {Colberg}, \&
  {Pearce}}]{2005Natur.435..629Springel}
{Springel}, V., {White}, S.~D.~M., {Jenkins}, A., {et~al.} 2005{\natexlab{b}},
  \nat, 435, 629

\bibitem[{{Springel} {et~al.}(2001){Springel}, {White}, {Tormen}, \&
  {Kauffmann}}]{2001MNRAS.328..726Springel}
{Springel}, V., {White}, S.~D.~M., {Tormen}, G., \& {Kauffmann}, G. 2001,
  \mnras, 328, 726

\bibitem[{{Stanek} {et~al.}(2010){Stanek}, {Rasia}, {Evrard}, {Pearce}, \&
  {Gazzola}}]{2010ApJ...715.1508Stanek}
{Stanek}, R., {Rasia}, E., {Evrard}, A.~E., {Pearce}, F., \& {Gazzola}, L.
  2010, \apj, 715, 1508

\bibitem[{{Steinborn} {et~al.}(2016){Steinborn}, {Dolag}, {Comerford},
  {Hirschmann}, {Remus}, \& {Teklu}}]{2016MNRAS.458.1013Steinborn}
{Steinborn}, L.~K., {Dolag}, K., {Comerford}, J.~M., {et~al.} 2016, \mnras,
  458, 1013

\bibitem[{{Steinborn} {et~al.}(2015){Steinborn}, {Dolag}, {Hirschmann},
  {Prieto}, \& {Remus}}]{2015MNRAS.448.1504Steinborn}
{Steinborn}, L.~K., {Dolag}, K., {Hirschmann}, M., {Prieto}, M.~A., \& {Remus},
  R.-S. 2015, \mnras, 448, 1504

\bibitem[{{Stern} {et~al.}(2019){Stern}, {Dietrich}, {Bocquet}, {Applegate},
  {Mohr}, {Bridle}, {Carrasco Kind}, {Gruen}, {Jarvis}, {Kacprzak}, {Saro},
  {Sheldon}, {Troxel}, {Zuntz}, {Benson}, {Capasso}, {Chiu}, {Desai},
  {Rapetti}, {Reichardt}, {Saliwanchik}, {Schrabback}, {Gupta}, {Abbott},
  {Abdalla}, {Avila}, {Bertin}, {Brooks}, {Burke}, {Carnero Rosell},
  {Carretero}, {Castander}, {D'Andrea}, {da Costa}, {Davis}, {De Vicente},
  {Diehl}, {Doel}, {Estrada}, {Evrard}, {Flaugher}, {Fosalba}, {Frieman},
  {Garc{\'\i}a-Bellido}, {Gaztanaga}, {Gruendl}, {Gschwend}, {Gutierrez},
  {Hollowood}, {Jeltema}, {Kirk}, {Kuehn}, {Kuropatkin}, {Lahav}, {Lima},
  {Maia}, {March}, {Melchior}, {Menanteau}, {Miquel}, {Plazas}, {Romer},
  {Sanchez}, {Schindler}, {Schubnell}, {Sevilla-Noarbe}, {Smith}, {Smith},
  {Sobreira}, {Suchyta}, {Swanson}, {Tarle}, {Walker}, {DES Collaboration}, \&
  {SPT Collaboration}}]{2019MNRAS.485...69Stern}
{Stern}, C., {Dietrich}, J.~P., {Bocquet}, S., {et~al.} 2019, \mnras, 485, 69

\bibitem[{{Sugiyama} {et~al.}(2023){Sugiyama}, {Miyatake}, {More}, {Li},
  {Shirasaki}, {Takada}, {Kobayashi}, {Takahashi}, {Nishimichi}, {Nishizawa},
  {Rau}, {Zhang}, {Dalal}, {Mandelbaum}, {Strauss}, {Hamana}, {Oguri}, {Osato},
  {Kannawadi}, {Hsieh}, {Luo}, {Armstrong}, {Bosch}, {Komiyama}, {Lupton},
  {Lust}, {Miyazaki}, {Murayama}, {Okura}, {Price}, {Tait}, {Tanaka}, \&
  {Wang}}]{2023PhRvD.108l3521Sugiyama}
{Sugiyama}, S., {Miyatake}, H., {More}, S., {et~al.} 2023, \prd, 108, 123521

\bibitem[{{Sun} {et~al.}(2009){Sun}, {Voit}, {Donahue}, {Jones}, {Forman}, \&
  {Vikhlinin}}]{2009ApJ...693.1142Sun}
{Sun}, M., {Voit}, G.~M., {Donahue}, M., {et~al.} 2009, \apj, 693, 1142

\bibitem[{{Sunayama} {et~al.}(2020){Sunayama}, {Park}, {Takada}, {Kobayashi},
  {Nishimichi}, {Kurita}, {More}, {Oguri}, \& {Osato}}]{Sunayama2020ImpactProj}
{Sunayama}, T., {Park}, Y., {Takada}, M., {et~al.} 2020, \mnras, 496, 4468

\bibitem[{{Sunyaev} \& {Zeldovich}(1972)}]{1972CoASP...4..173Sunyaev}
{Sunyaev}, R.~A. \& {Zeldovich}, Y.~B. 1972, Comments on Astrophysics and Space
  Physics, 4, 173

\bibitem[{{Taffoni} {et~al.}(2020){Taffoni}, {Becciani}, {Garilli}, {Maggio},
  {Pasian}, {Umana}, {Smareglia}, \& {Vitello}}]{2020ASPC..527..307Taffoni}
{Taffoni}, G., {Becciani}, U., {Garilli}, B., {et~al.} 2020, in Astronomical
  Society of the Pacific Conference Series, Vol. 527, Astronomical Society of
  the Pacific Conference Series, ed. R.~{Pizzo}, E.~R. {Deul}, J.~D. {Mol},
  J.~{de Plaa}, \& H.~{Verkouter}, 307

\bibitem[{{Teklu} {et~al.}(2015){Teklu}, {Remus}, {Dolag}, {Beck}, {Burkert},
  {Schmidt}, {Schulze}, \& {Steinborn}}]{2015ApJ...812...29Teklu}
{Teklu}, A.~F., {Remus}, R.-S., {Dolag}, K., {et~al.} 2015, \apj, 812, 29

\bibitem[{Tornatore {et~al.}(2007)Tornatore, Borgani, Dolag, \&
  Matteucci}]{2007MNRAS.382.1050Tornatore}
Tornatore, L., Borgani, S., Dolag, K., \& Matteucci, F. 2007, \mnras, 382, 1050

\bibitem[{{Truong} {et~al.}(2018){Truong}, {Rasia}, {Mazzotta}, {Planelles},
  {Biffi}, {Fabjan}, {Beck}, {Borgani}, {Dolag}, {Gaspari}, {Granato},
  {Murante}, {Ragone-Figueroa}, \& {Steinborn}}]{2018MNRAS.474.4089Truong}
{Truong}, N., {Rasia}, E., {Mazzotta}, P., {et~al.} 2018, \mnras, 474, 4089

\bibitem[{{van den Bosch} {et~al.}(2005){van den Bosch}, {Yang}, {Mo}, \&
  {Norberg}}]{VanDenBosch2005Satellites}
{van den Bosch}, F.~C., {Yang}, X., {Mo}, H.~J., \& {Norberg}, P. 2005, \mnras,
  356, 1233

\bibitem[{{van Uitert} {et~al.}(2016){van Uitert}, {Cacciato}, {Hoekstra},
  {Brouwer}, {Sif{\'o}n}, {Viola}, {Baldry}, {Bland-Hawthorn}, {Brough},
  {Brown}, {Choi}, {Driver}, {Erben}, {Heymans}, {Hildebrandt}, {Joachimi},
  {Kuijken}, {Liske}, {Loveday}, {McFarland}, {Miller}, {Nakajima}, {Peacock},
  {Radovich}, {Robotham}, {Schneider}, {Sikkema}, {Taylor}, \& {Verdoes
  Kleijn}}]{2016MNRAS.459.3251VanUitert}
{van Uitert}, E., {Cacciato}, M., {Hoekstra}, H., {et~al.} 2016, \mnras, 459,
  3251

\bibitem[{{Vikhlinin} {et~al.}(2006){Vikhlinin}, {Kravtsov}, {Forman}, {Jones},
  {Markevitch}, {Murray}, \& {Van Speybroeck}}]{2006ApJ...640..691Vikhlinin}
{Vikhlinin}, A., {Kravtsov}, A., {Forman}, W., {et~al.} 2006, \apj, 640, 691

\bibitem[{{Villaescusa-Navarro} {et~al.}(2022){Villaescusa-Navarro}, {Ding},
  {Genel}, {Tonnesen}, {La Torre}, {Spergel}, {Teyssier}, {Li}, {Heneka},
  {Lemos}, {Angl{\'e}s-Alc{\'a}zar}, {Nagai}, \&
  {Vogelsberger}}]{VillaescusaNavarro22Cosmology1Galaxy}
{Villaescusa-Navarro}, F., {Ding}, J., {Genel}, S., {et~al.} 2022, \apj, 929,
  132

\bibitem[{{Wiersma} {et~al.}(2009){Wiersma}, {Schaye}, {Theuns}, {Dalla
  Vecchia}, \& {Tornatore}}]{Wiersma2009Cooling}
{Wiersma}, R. P.~C., {Schaye}, J., {Theuns}, T., {Dalla Vecchia}, C., \&
  {Tornatore}, L. 2009, \mnras, 399, 574

\bibitem[{{Wu} {et~al.}(2022){Wu}, {Costanzi}, {To}, {Salcedo}, {Weinberg},
  {Annis}, {Bocquet}, {da Silva Pereira}, {DeRose}, {Esteves}, {Farahi},
  {Grandis}, {Rozo}, {Rykoff}, {Varga}, {Wechsler}, {Zeng}, {Zhang}, {Zhang},
  \& {DES Collaboration}}]{Wu2022OpticalBias}
{Wu}, H.-Y., {Costanzi}, M., {To}, C.-H., {et~al.} 2022, \mnras, 515, 4471

\bibitem[{{Xu} {et~al.}(2018){Xu}, {Ramos-Ceja}, {Pacaud}, {Reiprich}, \&
  {Erben}}]{2018A&A...619A.162Xu}
{Xu}, W., {Ramos-Ceja}, M.~E., {Pacaud}, F., {Reiprich}, T.~H., \& {Erben}, T.
  2018, \aap, 619, A162

\end{thebibliography}
